\documentclass[draft]{agujournal2019}
\usepackage{url} 
\usepackage[finalnew]{trackchanges} 
\usepackage{soul}

\usepackage[flushleft]{threeparttable}
\usepackage{amssymb}
\usepackage{textcomp, gensymb}  
\usepackage{commath}
\usepackage{dsfont}
\usepackage{multirow}
\usepackage{multicol}
\usepackage{subcaption}
\usepackage{graphicx}

\draftfalse
\journalname{Journal of Advances in Modeling Earth Systems (JAMES)}

\begin{document}

\title{A Generative Deep Learning Approach to Stochastic Downscaling of Precipitation Forecasts}

\authors{Lucy Harris\affil{1}, Andrew T. T. McRae\affil{1}, Matthew Chantry\affil{2}, Peter D. Dueben\affil{2}, Tim N. Palmer\affil{1}}

\affiliation{1}{Department of Physics, University of Oxford}
\affiliation{2}{European Centre for Medium-Range Weather Forecasts}

\correspondingauthor{Lucy Harris}{lucy.harris2@physics.ox.ac.uk}

\begin{keypoints}
\item We use generative adversarial neural networks to post-process global weather forecast model output over the UK.
\item We produce more realistic precipitation forecasts \add{than the input forecast data}, at 10x resolution, with excellent statistical properties.
\item We match or outperform a state-of-the-art pointwise downscaling scheme, while also producing spatially coherent images.
\end{keypoints}

\begin{abstract}
Despite continuous improvements, precipitation forecasts are still not as accurate and reliable as those of other meteorological variables. A major contributing factor to this is that several key processes affecting precipitation distribution and intensity occur below the resolved scale of global weather models. Generative adversarial networks (GANs) have been demonstrated by the computer vision community to be successful at super-resolution problems, i.e., learning to add fine-scale structure to coarse images. Leinonen et al.~(2020) previously applied a GAN to produce ensembles of reconstructed high-resolution atmospheric fields, given coarsened input data. In this paper, we demonstrate this approach can be extended to the more challenging problem of increasing the accuracy and resolution of comparatively low-resolution input from a weather forecasting model, using high-resolution radar measurements as a ``ground truth". The neural network must learn to add resolution and structure whilst accounting for non-negligible forecast error. We show that GANs and VAE-GANs can match the statistical properties of state-of-the-art pointwise post-processing methods whilst creating high-resolution, spatially coherent precipitation maps. Our model compares favourably to the best existing downscaling methods in both pixel-wise and pooled CRPS scores, power spectrum information and rank histograms (used to assess calibration). We test our models and show that they perform in a range of scenarios, including heavy rainfall.
\end{abstract}

\section*{Plain Language Summary}
The processes that lead to precipitation (rainfall) happen on a very small scale. Weather forecast computer models work on much larger scales, so rainfall is often poorly predicted. In this paper, we develop a method that enhances the resolution of rainfall forecasts by a factor of 10, and makes the forecasts more accurate. We generate many samples of what the rainfall pattern could be, which gives an idea of the uncertainty in the forecast. Our method is based on machine learning and neural networks, which means that we use many past examples of weather forecasts, together with the rainfall that actually happened, and our method `automatically' learns how the forecasts can be improved. We use an existing idea called ``Generative Adversarial Networks'', which has been used very successfully in image-related tasks, such as producing realistic higher-resolution images from low-resolution ones. Our task is similar to producing a high-resolution image from a low-resolution one, hence this approach is promising. Our method outperforms a variety of existing approaches, and even produces good predictions for the most extreme rainfall situations in our data set. These are the scenarios that cause the most real-world disruption, the most useful events to produce good forecasts for.

\section{Introduction}
Weather prediction and climate models are constantly evolving, and are generally considered to perform well for most applications. However, it is a well-recognised problem that precipitation events are imperfectly predicted \cite{Sha2020,Gascon2018,Berrocal2008,Applequist2002}. This is in part due to the low spatial resolution of the outputs of most models: global weather and climate models are produced on a much larger spatial scale than is typically required to accurately predict the finer structures and extremes of rainfall events \cite{Adewoyin2021TRU-Net}. Limitations of computational resources, numerical stability, and knowledge of initial conditions lead to constraints on model resolution such that most global numerical forecast models operate at roughly 10--80 km grid spacings, and are consequently only capable of resolving large-scale weather phenomena, as well as a limited representation of small mesoscale atmospheric processes, topography, and land-sea distribution \cite{Feser2011,DOnofrio2014}. \add{When a major weather event hits some part of the world, devastating the local population, it is only days later that emergency relief is distributed} \cite{PalmerarXiv,PalmerWhitePaper}. The direct application of weather and climate model outputs to precipitation impact assessment is therefore inadequate \cite{Yu2016}, particularly for extreme rainfall and situations with significant small-scale variability, for example, in the presence of heterogeneous orography and along coastlines.

In weather and climate sciences, downscaling refers to an operation that infers high-resolution information from lower-resolution data\footnote{Confusingly, downsampling (upsampling) in machine learning refers to a reduction (increase) in the image resolution. In this paper we will discuss downscaling in a weather-related context, and upsampling in a computer vision or machine learning context.}. Downscaling is particularly important in precipitation forecasting: the intensity of precipitation can vary considerably over short spatial scales (1 km or less), which is much lower than the typical resolution of global weather models. Increasing the resolution of precipitation forecasts is essential for assessing the potential impacts, particularly for extreme rainfall scenarios. Increasingly, stochastic downscaling techniques are applied to generate ensembles of possible small-scale rainfall fields from an initial large-scale distribution, as a way to introduce rainfall variability at scales not resolved by physical models, since full, high-resolution, deterministic models are computationally intractable \cite{DOnofrio2014}. \citeA{PalmerarXiv,PalmerWhitePaper} \add{advocates for the use of stochastic neural network approaches to post-process and downscale global forecast model output, perhaps in place of traditional limited-area models.} In stochastic downscaling, the goal is to produce an ensemble of possible realisations where the small-scale fields are consistent with the large scale features of the low-resolution data, as well as any smaller-scale information, such as the terrain geometry or land-sea distribution. Downscaling is inherently an under-determined problem, where one low-resolution forecast state could be valid for a multitude of high-resolution truths. This low-resolution state will generally contain errors when compared to a coarsened version of the truth data. By employing a stochastic method we can sample these high-resolution states to capture the uncertainty of both the mapping between data sources and the downscaling. We can condition this mapping further by including additional model fields and surface descriptors.

Downscaling precipitation using convolutional neural networks is a very active area of research. Many authors have approached the problem as a pure super-resolution task by coarsening their `truth' data and inputting this to their model \add{(sometimes alongside other fields)}, then trying to retrieve the lost resolution. Papers that take this approach include \citeA{Sha2020}, \citeA{Wang2021}, and \citeA{Kumar2021}. However, we argue that this is not sufficient to tackle the full downscaling problem, since it does not account for the inevitable errors in the input forecast data. Two papers that use independent input and truth datasets are \citeA{Huang2020} and \citeA{Adewoyin2021TRU-Net}. However, both of these are motivated by climate models rather than weather prediction and so operate at much coarser scales in space and time than we do. This changes the flavour of the problem substantially -- for example, both papers prioritise standard metrics like RMSE, which is inappropriate on higher resolutions and shorter timescales \cite{double_penalty}.  \citeA{Hess2022} \add{uses independent datasets, and has a particular focus on heavy rainfall events, but does not increase resolution.} Finally, many authors have used convolutional neural networks for nowcasting: forecasting precipitation events over short lead times (typically 0--6 hours). This problem differs from the downscaling task examined here, with a focus on evolving fields forward in time instead of enhancing and increasing resolution of input data. Nevertheless, many network architecture elements are shared across these domains. Recent work in this area includes \citeA{ShiChen2015}, \citeA{Agrawal2019}, \citeA{Sonderby2020}, \citeA{deepmind_2021}, \citeA{Klocek2021}, and \citeA{Espeholt2021}.

In digital image processing, super-resolution refers to enhancing the spatial resolution of an image by estimating a high resolution image from its low-resolution counterpart. This has clear parallels with the downscaling problem from weather and climate science. Super-resolution is a highly challenging task, receives substantial attention within the computer vision research community, and has a wide range of applications. Recent developments in this field have led to the application of convolutional neural networks (CNNs), and subsequently, generative adversarial networks (GANs) \cite{Goodfellow2014} to super-resolution problems \cite{Dong2015,Lin2017-DCSR,Ledig2017}. The purpose of a GAN model is to generate realistic artificial samples similar to those encountered during training. GANs differ from typical neural network approaches -- in place of a standard `loss function', a second network (the discriminator) is used to evaluate generated samples. The generator network is hence trained to produce outputs that the discriminator considers to be realistic, while the discriminator is trained to better differentiate between real and artificial data. This approach has found great success in super-resolution applications.

\add{Generative adversarial network approaches have started to appear in post-processing/downscaling and forecasting/nowcasting applications.} \citeA{Bihlo2020} \add{produced 24-hour large-scale predictions, trained on ERA5 reanalysis data. Promising results were obtained for 500 hPa geopotential height and 2m temperature, but not for precipitation.} \citeA{Watson2020} \add{performs precipitation downscaling, trained to map between different configurations of the forecast model WRF. The results are promising, although only a preliminary analysis is presented.} \citeA{deepmind_2021} \add{successfully tackled the precipitation nowcasting problem, producing high-resolution 90-minute forecasts over the UK.} \citeA{Gong2022} \add{forecasts the evolution of 2m temperature over 12 hours, trained on ERA5 data, using an existing adversarial video-prediction architecture.}

Previously, \citeA{Leinonen2020} successfully applied a GAN to stochastically downscale time-series of atmospheric fields, including precipitation. However, this took the pure super-resolution approach of first coarsening the high-resolution `truth' data and then recovering the lost resolution. The absence of future radar truth images means that any application of Leinonen's model would have to infer from a future forecast model state, which is somewhat different to the task for which it was trained. In this paper, we work on an extension of this problem. We do not just learn the mapping from coarse- to fine-resolution representations of the same data. Instead, our models learn the mapping from (multiple) low-resolution atmospheric fields, originating from a weather forecast model, to high-resolution `truth' radar data. Thus, we are aiming to both increase resolution of the original forecast and provide error correction in a probabilistic sense. The neural networks are also supplied with high-resolution orography data and a land-sea mask, which are expected to affect local precipitation due to physical principles \cite{Holden2011}. We are therefore tackling the complete downscaling problem: using the predictive power of atmospheric model fields and surface properties to match an observation of Earth’s weather. Following the excellent performance of the Leinonen approach we closely follow their model architecture. However, due to computational constraints we have removed the time-series aspect of Leinonen's approach.

Shortly before completion of this work, \citeA{Price2022} appeared in the literature, which builds upon Leinonen's GAN model in a similar way to us. Like us, they map low-resolution weather forecast data to a higher-resolution precipitation truth dataset. Their downscaling factor is comparable to ours, but they work at coarser resolutions in space and time. They make several different choices to us regarding network inputs and training, and an optimal solution may well combine strengths from both approaches.

\section{Data}
\label{sec:data}

We trained our model to map hourly data from the Integrated Forecast System (IFS) to hourly accumulated rainfall based on the NIMROD radar network~\cite{NIMROD}. Our domain of interest covers latitudes 49.5\degree{} to 59\degree{} and longitudes -7.5\degree{} to 2\degree{}, covering mainland UK.

\subsection{IFS data}
Our input data is the ECMWF's IFS operational forecast dataset, using years 2016--2020. During training we use 7--17h lead time forecasts, initialised at 00Z and 12Z. Earlier lead times are discarded to ensure any artefacts from data assimilation do not affect training. Later lead times are discarded as the chaotic nature of the atmosphere means the predicted cloud locations within the IFS will be increasingly poorly aligned with real world observations. We did not test the sensitivity of these two lead time thresholds, but we later evaluate the model on lead times out to 72h and find the model performs well despite only being trained on short-term forecasts.

From the IFS model we use 9 fields:
\begin{itemize}
    \item Total precipitation
    \item Convective precipitation
    \item Surface pressure
    \item TOA incident solar radiation
    \item Convective available potential energy
    \item Total column cloud liquid water
    \item Total column water vapour
    \item u \& v (horizontal wind) velocities at 700hPa
\end{itemize}

The choice of these fields was motivated by the ecPoint model \cite{ecPoint}, and domain knowledge. IFS data is linearly interpolated to a 0.1\degree{} grid (approximately 10km), resulting in images of size $94 \times 94$ pixels. To normalise the precipitation fields in the IFS data (total and convective precipitation) for input into the neural networks, we use the transformation $\log_{10}\left( 1 + x \right )$ on the mm/hr rate. The surface pressure field is normalised by subtracting the mean and dividing by the standard deviation, where these values are scalars, calculated from all grid points in 2018. Each of the other fields are normalised by calculating the (absolute) maximum value observed in 2018 (across all grid points) and dividing by this value. Winds u \& v are normalised independently from one another. The inputs to the neural network are hence all $\mathcal{O}(1)$.

\subsection{NIMROD data}
As a \change{truth}{`truth'} dataset, we use the 1km Resolution UK Composite Rainfall Data from the Met Office NIMROD System~\cite{NIMROD}. This system delivers radar-derived precipitation maps every 5 minutes, covering 2004 to present day. As with the IFS we use 2016--2020 data and aggregate to hourly precipitation. Calendar days with more than 30 minutes of missing data were removed from the dataset. On average this results in 330 days per year, or approximately 8000 hours. For ease of grid alignment with the IFS the data is re-gridded to a 0.01\degree{} grid, resulting in images of size $940 \times 940$ pixels. \remove{The NIMROD dataset occasionally contains artefacts resulting from the radar system. Here, we make the decision to make no further steps of data cleaning or processing to our truth dataset.} We again use a $\log_{10}\left( 1 + x \right)$ transformation for the NIMROD precipitation.

\add{The NIMROD dataset itself naturally has inherent errors, and even contains obvious artefacts resulting from the radar system. However, we believe the NIMROD data is different enough to the IFS input, and close enough to the genuine `truth', that training a successful model on the NIMROD data is of equivalent difficulty to training a model on any more accurate precipitation dataset that may be available in the future. Furthermore, even though the data is imperfect, the trained models will still provide significant value over the IFS input. As a result, we make no further steps of data cleaning or processing to the NIMROD dataset.}

\subsection{Geographic data}
To improve model performance we augment our model input data with high resolution surface geopotential and land-sea mask data, which depend only on location and do not vary with time and date. These may help the network add meaningful information on length scales smaller than the input data. These fields are derived from 1.25km input data (originally generated for high resolution IFS simulations) and are re-gridded to the same 0.01\degree{} grid as the NIMROD dataset. The surface geopotential is normalised by dividing by the global maximum value. Before this, values less than $5 \textrm{m}^2 \textrm{s}^{-2}$ are clipped to this value. This is done to remove artefacts stemming from the spectral origin of our data. The land-sea mask already takes fractional values between 0 (no land in grid box) to 1 (grid box only comprised of land).

\subsection{Data subsets}
\label{sec:datasubset}
The model was trained on data from 2016--2018. Data from 2019 was used for validation, and data from 2020 was held out for final testing. All quantitative evaluation in this paper is performed solely on 2020 data. Some interesting synoptical situations from 2019 have also been included as case studies.

Contrary to popular opinion, it is not raining in the UK for the overwhelming majority of the time. We were therefore concerned that training the model on randomly-sampled input data would cause significant under-prediction of rainfall during high-rainfall events. Furthermore, although we could use full-sized low- and high-resolution images during inference ($94 \times 94$ and $940 \times 940$, respectively), we did not have the computational resources to use such large images during model training. The data across the UK were therefore split into smaller sub-images of $20 \times 20$ (low-resolution) and $200 \times 200$ (high-resolution), by randomly sampling patches from the full-sized images. Each sub-image was scored on \emph{``how rainy"} it was in that image and categorised into one of four bins, depending on what fraction of pixels contained rainfall ($>$0.1mm/hour) -- 0--25\%, 25--50\%, 50--75\%, or 75--100\%. This allowed us to select the distribution with which we sample from the different bins during model training, and we treated this as a hyperparameter to be optimised. The results of varying the distribution of images shown to the network are discussed in the appendix. We remark that \citeA{deepmind_2021} also increased the prevalence of rainy images in their training data, although their weighting was based on both spatial coverage and intensity.

\section{Methods}
\label{sec:method}

We use two generative deep learning approaches; both post-process lower-resolution atmospheric field forecast data and aim to produce well-calibrated ensembles of high-resolution precipitation forecasts.

\subsection{Model 1: GAN}
\begin{figure}
    \centering
    \includegraphics[width=\textwidth]{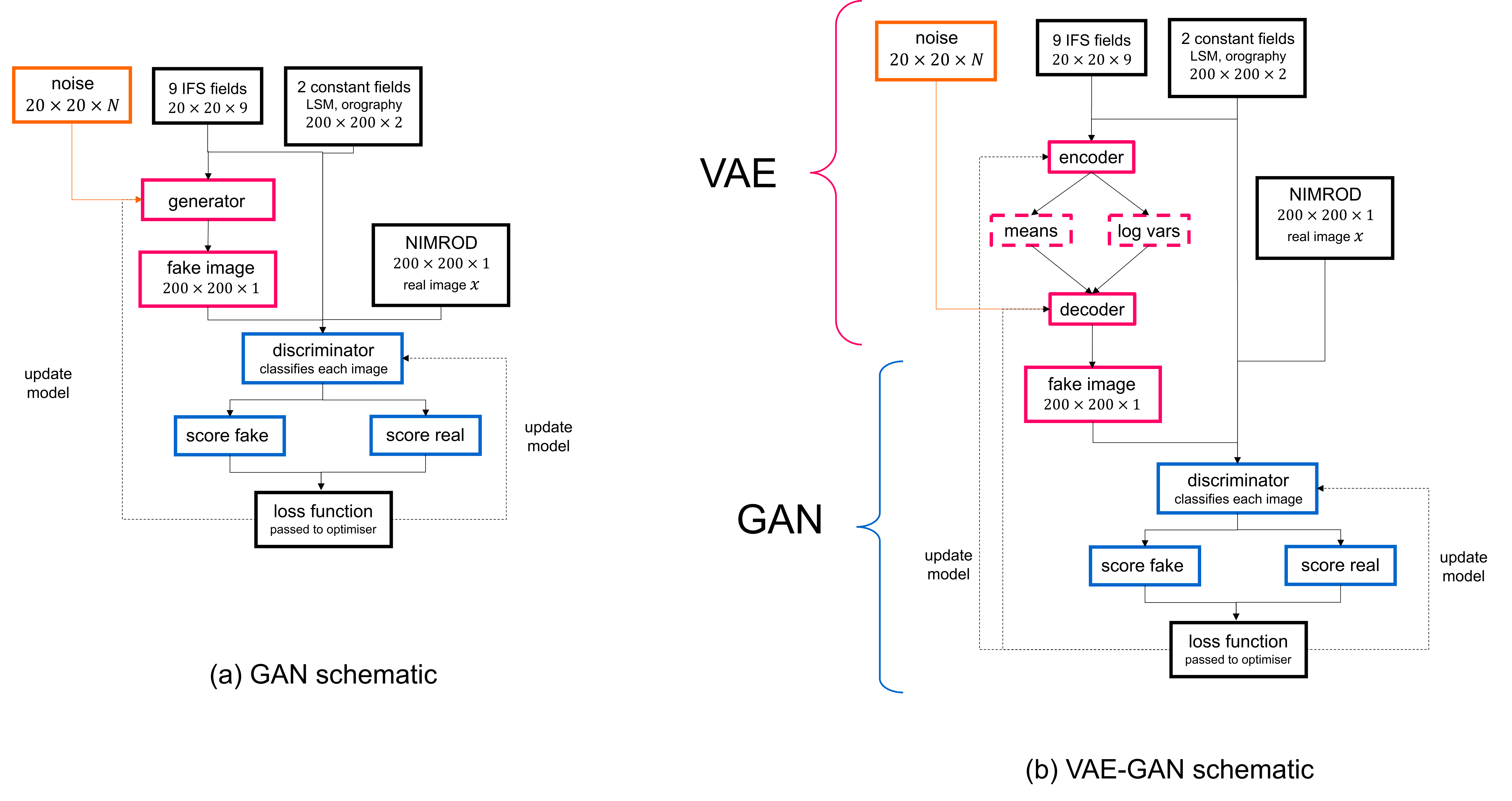}
    \caption{Schematic of the information flow through (a) the conditional GAN model and (b) the conditional VAE-GAN model}
    \label{fig:schematic}
\end{figure}

The first model is a conditional GAN \cite{mirza2014conditional}, where both the generator and the discriminator are conditioned on additional information: in this case, lower-resolution atmospheric fields and full-resolution orography and land-sea mask data. The generator has an explicit noise input, which allows multiple samples to be generated for a given forecast state. The discriminator is trained to distinguish between the high-resolution predictions from the generator and corresponding ``ground-truth'' high-resolution rainfall data. We follow the work of \citeA{Arjovsky2017} and \citeA{Gulrajani2017} by using a Wasserstein-GAN with a gradient penalty to enable stable GAN training. A high-level schematic of our conditional GAN is shown in Figure~\ref{fig:schematic}\add{(a)}.

\subsection{Model 2: VAE-GAN}
We initially explored using a variational auto-encoder (VAE) as an alternative approach to GANs. This is a model consisting of an encoder network and a decoder network. The encoder network maps from the input to some latent space representation of the input data, encoded in the means and (log-)variances of normal random variables. The decoder network then samples from the normal distributions described by these variables, via an external noise input, and attempts to recreate the higher resolution `truth' data. This required us to define a `content loss' term that penalises deviations between the network output and the truth data. Despite trying numerous content loss terms, results were uniformly disappointing -- the resulting ensemble was greatly under-dispersive, and the predictions `blurry'.

We therefore developed a hybrid VAE-GAN model which substituted the GAN generator with a VAE. This effectively employs a full discriminator network as the VAE content loss function, and produced much sharper and better-calibrated results. A high-level schematic of our hybrid conditional VAE-GAN is shown in Figure~\ref{fig:schematic}\add{(b)}.

\subsection{Model Architecture}
The generator of the GAN, encoder and decoder of the VAE-GAN, and discriminator are all deep, convolutional neural networks which make heavy use of residual blocks \cite{He2015}. The architecture is closely based on that used in \citeA{Leinonen2020}, modified for our downscaling factor of 10, and with blocks facilitating the temporal component of their problem removed. The generator networks in both models are fully convolutional, without any dense layers. This allows them to be size-agnostic, and hence we can train the network on $20 \times 20$ input images but use full-size $94 \times 94$ input images during inference. Due to this restriction, the latent variables in the VAE-GAN model will only represent local rather than global information.

The inputs to the models are:
\begin{itemize}
    \item Low-resolution conditioning fields (weather forecasts), with dimensions $(b \times h_l \times w_l \times N_i)$,
    \item High-resolution geographic fields (land-sea mask and orography), with dimensions $(b \times h_h \times w_h \times 2)$,
    \item A noise input, with dimensions $(b \times h_l \times w_l \times n)$,
\end{itemize}
where $b$ is the batch size, $h_l$ and $w_l$ are the low-resolution input image dimensions, $N_i$ is the number of input conditioning fields (for us, typically 9 IFS variables), $h_h$ and $w_h$ are the high-resolution target image dimensions, and $n$ is the number of noise channels per input image pixel. The ratio between the high- and low-resolution image dimensions is the \emph{downscaling factor}, $K$; in this paper, we use a downscaling factor $K = 10$ throughout. In the GAN generator, the number of noise channels, $n$ is a parameter that can be varied. In the VAE-based models, there is one noise input for each latent variable, where the number of latent variables per pixel is a parameter that can be varied. We did not attempt to use a more sophisticated `conditioning stack' to further process the IID noise input, as was done in \citeA{deepmind_2021}.

\subsection{GAN architecture}

\begin{figure*}
    \centering
    \includegraphics[width=\textwidth]{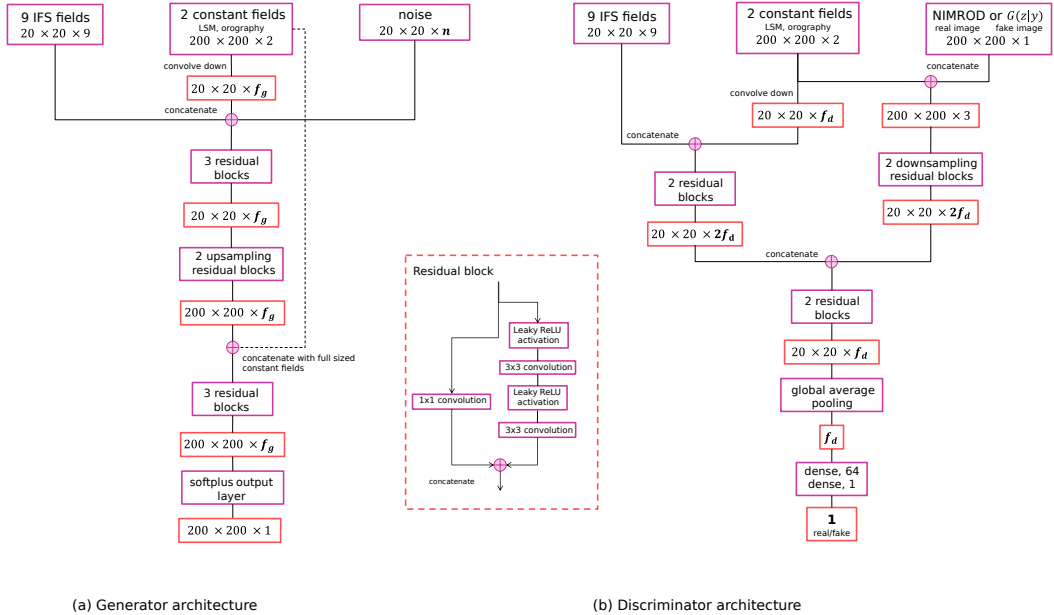}
    \caption{Network architecture for the conditional GAN: (a) the generator model and (b) the discriminator model}
    \label{fig:architecture}
\end{figure*}

The GAN architecture is displayed in Figure~\ref{fig:architecture}. The number of trainable parameters in the generator depends on the number of filters, $f_g$. When $f_g = 128$, the value we use throughout this paper, the generator network has approximately 3.2 million trainable parameters. The number of trainable parameters in the discriminator depends on the number of filters, $f_d$. When $f_d = 512$, the value we use, the discriminator network has approximately 64 million trainable parameters. The networks were designed by assessing the overall performance of the architecture with different hyperparameter choices. Since a Wasserstein GAN can be trained to optimality \cite{Gulrajani2017}, we deliberately choose $f_d > f_g$ so that the discriminator network is more powerful than the generator network. This helps to avoid mode collapse from occurring. We were limited to a maximum value of $f_d = 512$ by the hardware available (\add{initially, a} V100 GPU with 16GB RAM, \add{although we later gained access to an A100 GPU}). Increasing the number of channels in the generator had a much smaller impact on the model performance.

\subsection{VAE-GAN architecture}

\begin{figure}
    \centering
    \includegraphics[width=0.4\textwidth]{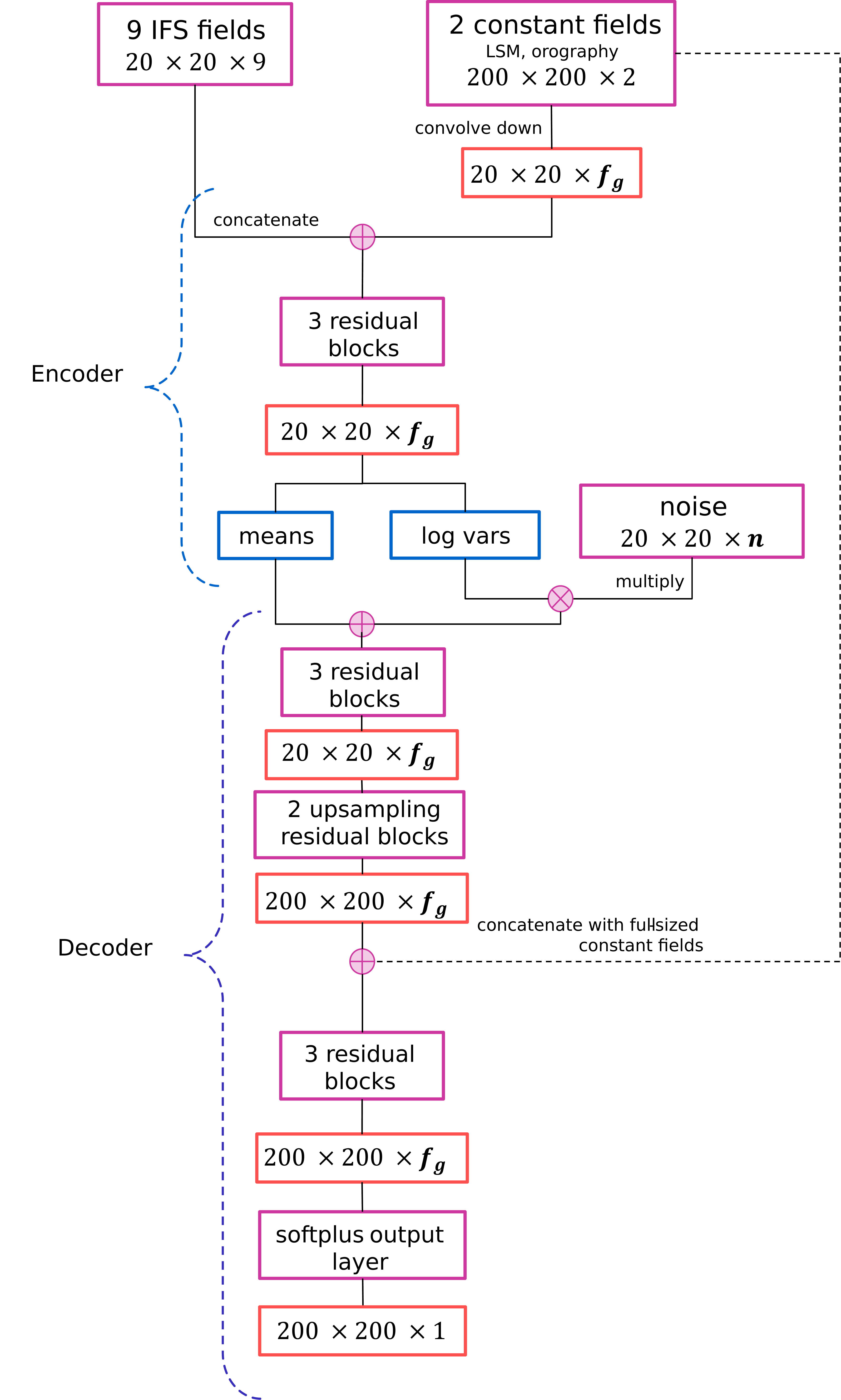}
    \caption{Network architecture for the VAE-GAN generator model. The discriminator model is identical to that shown in Figure~\ref{fig:architecture}}
    \label{fig:VAEGAN-architecture}
\end{figure}

The VAE-GAN model has a \remove{very} similar architecture to the GAN model, with the key difference in where the noise input is passed to the model. In the VAE-GAN generator, the noise is introduced at an intermediate stage when the latent variable distributions are sampled. This is performed after three low-resolution residual \change{blocks, before upsampling occurs.}{blocks. A further three residual blocks are used, before upsampling occurs\footnote{\add{Our initial work performed upsampling immediately after sampling from the latent variable distributions. The resulting ensemble members had overly similar large-scale features. The extra network layers are hence crucial for allowing the sampled latent variables to develop into coherent, larger-scale spatial variations.}}.} The rest of the \change{generator -- the \emph{decoder} -- and}{generator, and} the entire discriminator is identical to the architecture used in the pure GAN model. The network architecture of the VAE-GAN generator model is shown in Figure~\ref{fig:VAEGAN-architecture}. The discriminator model remains the same as before, shown in Figure~\ref{fig:architecture}.

We use a specific number of latent variables per pixel of the low-resolution image. The results shown in this paper use \change{100}{50} latent variables per pixel. In early trials, we used far fewer, corresponding to a significant network bottleneck compared to the network width of 128 in other layers. However, the results were rather worse than the pure GAN, which does not have such a bottleneck. This led us to increase the number dramatically.

\subsection{Remarks}

Leaky rectified linear unit (ReLU) activations \cite{Maas13} with a negative slope of 0.2 are used in the residual blocks in both the generator and the discriminator. Regular ReLU activations are used in the upscaling (dimension-reducing) convolutions of the high-resolution input pathways of the discriminator. The final activation function used is a softplus layer on the output of the generator, leading to precipitation values (in the transformed variable). Using a softplus activation (instead of, e.g., a sigmoid) prevents the output from having an artificially-constrained maximum value, which we originally considered desirable. However, we found that in extreme convective scenarios, the network could produce ensemble members with unphysical values of localised precipitation, of order 1000mm/hr\footnote{Recall we use a $\log_{10}(1 + x)$ variable transform for precipitation, hence the network output $y$ is converted to a precipitation value $(10^{y} - 1)$ mm/hr. As a result, $\mathcal{O}(1)$ errors in extremes of $y$ lead to order-of-magnitude errors in extremes of precipitation.} We therefore clip values above 500 mm/hr, although this could be lowered somewhat further.

Formally, the network can only predict precipitation values in the half-open interval $(0, 500]$ mm/hr, in contrast to methods which explicitly assign a probability to zero precipitation (e.g., \citeA{vaughan2022convolutional}). However, the precipitation values can be arbitrarily small, and it would be a trivial post-processing step to flush values below some threshold to zero. The final activation function of the discriminator is linear, a standard choice in a WGAN. In terms of computational resources, evaluating the GAN or VAE-GAN generator on a single full-size image, mapping a $94 \times 94$ input to a $940 \times 940$ output, takes approximately \change{0.4s}{0.13s} per ensemble member on an NVIDIA \change{V100}{A100} GPU.

\section{Training and Validation}
\label{sec:training-and-validation}

\subsection{Training}
\label{sec:training}

The standard training objective for a GAN is a \emph{minimax} game: the generator, $G$, tries to minimise a loss function, whilst the discriminator, $D$, tries to maximise it. This loss function represents the ability of the discriminator to tell a real sample from a fake one. In a Wasserstein GAN \cite{Gulrajani2017}, the loss function is constructed from the Wasserstein metric, or earth-mover distance.

In our setting, of a conditional GAN \cite{mirza2014conditional}, both the generator and discriminator receive common inputs -- the IFS data, and geographic data -- which we represent by $y$. The GAN generator also takes in a noise input, $z$, while the discriminator takes in either `truth' data $x_\mathrm{true}$, or generated data $G(z | y)$. The loss functions themselves are simple; the discriminator has a loss function:
\begin{equation}
    \mathcal{L}_D = D \left(x_\mathrm{true} | y \right) - D \left(G \left(z | y \right) | y \right)
\end{equation}
The term $D(x_\mathrm{true}|y)$ therefore represents the discriminator's `score' that the real data instance is real, while $D(G(z|y)|y)$ is the discriminator's `score' that the generated, fake instance is real. The discriminator tries to maximise this function, i.e., it tries to maximise the difference between its output on real instances and its output on fake instances.
The generator has loss function:
\begin{equation}
    \mathcal{L}_G = D \left(G \left(z | y \right) | y \right)
\end{equation}
The generator tries to maximise this function, i.e., it tries to maximise the output of the discriminator for the generated fake instances. Intuitively, it tries to `trick' the discriminator into thinking the generated output is real.

Wasserstein GANs (WGANs) have various theoretical advantages over traditional GANs: They avoid problems with vanishing gradients, and the earth-mover distance is a true metric: a measure of distance in a space of probability distributions. Since it is continuous and differentiable, the discriminator can be trained to optimality \cite{Arjovsky2017}. Furthermore, in practice, WGANs are less vulnerable to getting stuck during training than traditional GANs. We follow \citeA{Gulrajani2017} by using a WGAN with a gradient penalty term, as shown in equation~\ref{eq:loss-function-1}.

Motivated by \citeA{deepmind_2021}, we added a further `content loss' term to the generator loss function. We implement this as the mean squared error between the truth and an \emph{ensemble mean} prediction \change{We typically use}{over} $8$ ensemble members\footnote{\add{We also experimented with a content loss term based on the pixel-wise CRPS. This produced similar results, and allowed a smaller ensemble size to be used. However, we felt we saw more instability during training, perhaps because CRPS penalises large errors less than ensemble-mean MSE.}}. This calculation is performed in the \emph{transformed} precipitation variable. Mean squared error terms often penalise a model from making bold predictions and result in `blurry' images; this effect is far less pronounced here since the loss function is applied to an ensemble mean, rather than an individual prediction.

The loss functions for a conditional WGAN-GP, as employed in this paper, take the form:
\begin{align}
  \mathcal{L}_{D} \left( x_\mathrm{true}, y, z; \theta_D \right) &= \underbrace{D \left(x_\mathrm{true}|y \right) - D \left(G \left( z|y \right) |y \right) }_{\textrm{original discriminator loss}} + \underbrace{\gamma \left( \norm{\nabla_{\hat{x}}D \left( \hat{x}|y \right) }_{2}-1 \right) ^{2}}_{\textrm{gradient penalty}}, \label{eq:loss-function-1} \\
  \mathcal{L}_{G} \left( x_\mathrm{true}, y, z; \theta_G \right) &= \underbrace{D \left( G \left( z|y \right) | y \right)}_{\textrm{original generator loss}}
  + \underbrace{\frac{\lambda}{N}  \norm{x_\textrm{true} - \frac{1}{P}\sum_{i=1}^P G \left( z_i|y \right) }_{2} ^{2}}_{\textrm{content loss term}},
\label{eq:loss-function-2}
\end{align}
where $L_D$ and $L_G$ are the loss functions for the discriminator and the generator, respectively, and $\theta_D$ and and $\theta_G$ are the corresponding trainable weights, with a gradient penalty weight $\gamma = 10$, after \citeA{Gulrajani2017}, and content loss weight $\lambda = 1000$ from experimentation. The samples $\hat{x}$, used to calculate the gradient penalty term, are randomly weighted averages of the real and generated terms:
\begin{equation}
    \hat{x} = \epsilon x + \left( 1 - \epsilon \right) G \left(z|y \right),
\end{equation}
where $\epsilon$ is randomly sampled from a uniform distribution between 0 and 1.

For the VAE-GAN generator, the generator loss contains an additional term based on the Kullback--Leibler divergence:
\begin{equation}
    \frac{1}{2}\left [ \left( \sum_{j=1}^M \mu_j^2 + \sum_{j=1}^M \sigma_j^2 \right) - \sum_{j=1}^M(\log(\sigma_j^2) + 1) \right ].
\end{equation}
The sums are taken over the $M$ intermediate latent variables, whose distributions are $\left\{\mathcal{N}(\mu_j, \sigma_j)\right\}$. This term must be weighted against the original generator loss and the content loss term; we use a multiplicative factor of $10^{-5}$. However, the results did not seem especially sensitive to this choice.

The generator and discriminator are trained adversarially, with the model alternating between training the discriminator for five iterations and the generator for one, after \citeA{Kurach2018}. The Adam optimiser \cite{Kingma2014} is used for both the generator and the discriminator, with a constant learning rate of $10^{-5}$ \add{for the pure GAN, and $5 \times 10^{-6}$ for the VAE-GAN}; we found larger learning rates resulted in unstable training. The model was trained with a batch size of 2 (limited by GPU memory) for 320,000 batches. The discriminator is trained on five times as many samples. Model weights are written to disk at 100 intermediate `checkpoints' in order to facilitate model selection, as described in section~\ref{sec:model-select}. Training a single model took approximately \change{six days}{three days}, using a single NVIDIA \change{V100}{A100} GPU.

\subsection{Validation}
A number of metrics are used to assess the performance of the networks. We describe them here.

\subsubsection{CRPS}
\label{subsubsec:crps}

A commonly used distance metric in the field of weather and climate forecasting is the \change{continuously}{continuous} ranked probability score (CRPS) \cite{MathesonWinkler1976,Hersbach2000,Gneiting2007}. The CRPS uses the entire ensemble of predictions to score the forecast. For each pixel in a predicted image, the CRPS is the integral of the squared difference of the cumulative distribution function (CDF) of the ensemble members, $F$, to the CDF of the observations. The observation CDF is a Heaviside step function H at the point $x_{\textrm{true},i}$. The CRPS for the pixel $i$ is therefore:
\begin{equation}
    \textrm{CRPS} = \int_{-\infty}^{\infty} \left( F \left(x' \right) - H \left(x' - x_{\textrm{true},i} \right) \right) ^{2} dx'
\end{equation}
The CRPS for the entire image is the mean of the pixel-wise CRPS scores. The CRPS can therefore be understood as a generalisation of the mean absolute error, and in the case of only one ensemble member it reduces to this metric.

Pixel-wise CRPS scores reward well-calibrated local forecasts, but do not promote spatially-coherent forecasts. We hence also calculate CRPS on spatially-pooled forecasts, per \citeA{deepmind_2021}. We use both \emph{average-pooling} and \emph{max-pooling}, in which we consider \emph{average} and \emph{maximum} values over local neighbourhoods. The former can be motivated by flood forecasts, in which rainfall accumulations over larger spatial regions are relevant. The latter is perhaps relevant for extreme localised rainfall events, whose location is unlikely to be forecast precisely. We follow \citeA{deepmind_2021} by using neighbourhood sizes of $4 \times 4$ (stride 2) and $16 \times 16$ (stride 4).

\subsubsection{Rank histograms}
\label{subsubsec:rankhist}
These aim to assess the amount of variability in the images produced by the network. For each low-resolution sample passed into the network, we have a ground truth image and an ensemble of predictions. For each pixel in each truth image, we can therefore determine the \emph{normalised rank} of the actual value compared to all $N_p$ predictions: $r = \frac{N_s}{N_p}$, where $N_s$ is the number of ensemble members below the truth. If the ensemble is perfectly calibrated, $r$ would be uniformly distributed across the range $0 \leq r \leq 1$ when sampled enough times. The shape of the distribution of $r$ can therefore be used as an evaluation metric to assess the variability of the generated images. We examine this distribution of $r$ visually by plotting a histogram, after \citeA{Hamill2000}. Since our networks cannot explicitly predict zero rainfall, our histograms would be distorted by the presence of zero rainfall values in the truth image. We therefore add a meteorologically-insignificant amount of noise, of order $10^{-3}$ mm/hr, to both the model-generated images and the ground-truth images before performing rank calculations.

\add{Since heavy rainfall events are particularly important, we produce separate rank histogram plots that only consider events corresponding to the top 0.01\% of IFS precipitation predictions within the evaluation sample. While it may sound more natural to condition on the top fraction of pixels in the `truth' NIMROD dataset, we found that these pixels were generally higher than all 100 ensemble members, whether using our approach or a strong baseline method} (described in Section~\ref{sec:altmeth}). \add{This is likely because there is no way to reliably predict the precise pixels that will experience heavy localised rainfall events, at least at the high spatial resolution we are working at. Our `thresholded' rank histograms are therefore conditioned on the most extreme \emph{forecast} values.}

\subsubsection{Image quality metrics}
\label{subsubsec:image-quality-metrics}

The simplest measure of image accuracy is the root-mean-squared error. However, we found this metric to be unsuitable for assessing our model performance since we are in a regime where the well-known ``double penalty problem'' applies \cite{double_penalty}. The uncertainty in small scale spatial variations is beyond what we can reliably infer from the input data, and \add{hence} predictions that \remove{hence} forecast correct amounts of rain in slightly incorrect locations often score worse than forecasts of no rain at all. Similarly, we found that metrics like the multi-scale structural similarity index (MS-SSIM) \cite{Wang2003MSSSIM}, which is popular in the computer vision community, were not particularly useful for our problem. \add{We do report the ensemble mean RMSE: the root-mean-squared error of the mean of an ensemble of generated predictions.}

\citeA{Leinonen2020} used a log spectral distance metric (LSD) to compute a root mean square error in the 2D power spectra, in decibels (dB). However, we also found little correlation between good scores from this metric and good model predictions, perhaps because using the full 2D power spectrum is overly stringent. Instead, we compute the radially averaged power spectral density (RAPSD), which was also used in \citeA{deepmind_2021}. This involves calculating the 2D power spectrum, then collapsing over all angular directions (with binning) to form a 1D power spectrum. We then compute a log spectral distance of this. We are unaware of an established name for this metric, so we label this a Radially Averaged Log Spectral Distance (RALSD):
\begin{equation}
    \textrm{RALSD} = \sqrt{ \frac{1}{N} \sum^{N}_{i=1} \left( 10 \log_{10} \frac{\overline{P}_{\textrm{true},i}}{\overline{P}_{\textrm{gen},i}}\right)^2}
    \label{eq:ralsd}
\end{equation}
where $\overline{P}_\textrm{true}$ and $\overline{P}_\textrm{gen}$ are the radially averaged power spectra of the true and generated images, respectively\add{, and $N$ is the number of bins}. We calculate the spectra in accordance with \citeA{ruzanski_scale_2011}, using the pySTEPS implementation \cite{pysteps}. Due to the logarithm, we found that this metric can produce distorted results in cases with very low rainfall nationwide. Since these cases are of little physical interest for our application, we exclude cases where the average rainfall over the entire image is less than 0.002 mm/h when calculating the RALSD.

\subsubsection{ROC and Precision-Recall curves}
\label{subsubsec:rocpr}
Receiver Operating Characteristic (ROC) curves are a standard diagnostic in machine learning applications. The ROC curve assesses the skill of a binary classifier by plotting the true positive rate (sensitivity) against the false positive rate (1 - specificity), across the range of probability thresholds. To construct ROC curves for a particular precipitation intensity, we make an ensemble of neural-network predictions for each forecast event. For each pixel, we look at what fraction of the ensemble members predicted rainfall above the prescribed intensity. We interpret this as the probability that our system outputs for the event taking place. Each point on the ROC curve then indicates the performance of our system when a specific probability threshold is used to separate positive predictions from negative ones. The ROC curve then indicates the performance of our system across all probability thresholds from 0 to 1, over $\mathcal{O}(10^8)$ individual predictions (i.e., each image pixel, for several hundred forecast events). We produce these curves for a range of precipitation intensities, from 0.1mm/hr (common) to 5.0mm/hr (rare). The ROC curve is often reduced further to a single number, the area under the curve (AUC), which would ideally be 1. This metric is also shown in the plots.

Precision-Recall curves are a closely-related diagnostic, which plot precision against sensitivity (recall). These are often considered better suited for low-probability events \cite{saito2015precision}, such as high rainfall intensities within our application. Precision-Recall curves for our models are shown in the supporting information document.

\subsubsection{Fractions Skill Score}
\label{subsubsec:fss}
Many of the preceding metrics were point-wise, i.e., defined by comparing the predictions with the truth image at a pixel-by-pixel level. The fractions skill score (FSS) \cite{RobertsLean2008,Roberts2008} is a popular verification method that takes spatial consistency into account. For a given precipitation threshold, the prediction and truth images are binarised according to whether the rainfall is above the prescribed intensity. The neighbourhood of each forecast pixel is then compared with the neighbourhood of each truth pixel, based on the fraction of pixels meeting the criteria. A skill score is calculated from this, representing the forecast performance at a particular spatial scale. When the score exceeds a certain number, the forecast is said to have useful skill at that spatial scale.

The basic FSS compares multiple individual ensemble members with the truth sequentially. We found that this metric can be artificially inflated (at intermediate spatial scales) by forecasts with small-scale noise, \change{and this}{so this metric should be interpreted with caution. This} behaviour has been observed independently\footnote{Suman Ravuri, private communication}. We also use the `ensemble FSS' concept described in \citeA{Duc2013}, in which the binarised prediction is replaced by probabilities in $[0, 1]$, representing the proportion of ensemble members predicting rainfall above the prescribed threshold. This metric does not seem to suffer the same flaw.

\subsection{Model selection}
\label{sec:model-select}

\begin{figure}
    \centering
    \includegraphics[width=\textwidth]{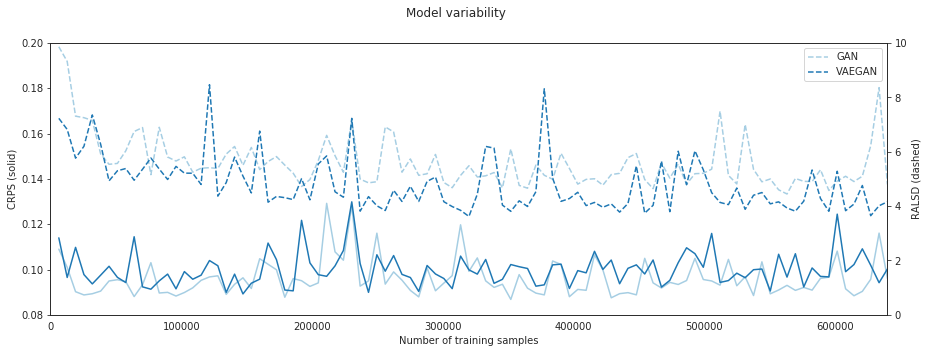}
    \caption{\change{Plot displaying}{A representative example of} the highly variable CRPS \add{and RALSD} of GAN and VAE-GAN generated samples as a function of number of generator training samples. Note that this plot is from the validation data, i.e., 2019. \add{The final one-third of these checkpoints is examined in more detail, and a model checkpoint that scores the best across multiple metrics (on the validation data) is selected for final evaluation.}}
    \label{fig:variability}
\end{figure}

We found that the GAN and VAE-GAN models did not improve monotonically with training, and using the final trained model would generally be far from optimal. Figure~\ref{fig:variability} shows \add{a typical example of} the variability of pixel-wise CRPS scores, \add{and RALSD scores}, for the GAN and VAE-GAN as a function of generator training samples. Each new checkpoint represents the generator seeing an additional 6,400 $20\times20$ images. The training instability may have been alleviated with a more sophisticated treatment of the learning rate. Instead, we adopted the simple strategy of saving the generator model weights at 100 intermediate checkpoints during training, and evaluating \change{these models}{the final one-third of these models} on the 2019 validation data.

We selected a `best' model by looking at the evaluation results for \change{pixel-wise CRPS scores, and cross-referencing with models that also produced good RALSD scores}{checkpoints that produced the best pixel-wise CRPS scores, then looking manually at the full and thresholded rank histograms of these}. A final, best model checkpoint was selected and then evaluated on the hold-out 2020 dataset. The final models that we selected were the GAN generator saved after \change{$563,200$}{460,800} training samples, and the VAE-GAN generator saved after 550,400 training samples.

\section{Results}
\label{sec:results}

\add{Unless stated otherwise, all quantitative results in this section are produced using 256 randomly-chosen examples from the unseen 2020 dataset. The same examples are used for all models, and for all the different metrics assessed. For all stochastic models, including our own neural networks, we draw 100 ensemble members for each example.}

\subsection{Description of alternative methods}
\label{sec:altmeth}

We compare our approach to a number of different methods, which we describe here briefly. These include very simple methods, such as naïve upsampling and Lanczos filtering~\cite{Turkowski1990}, intermediate methods such as \change{RainFARM}{Rainfall Filtered Auto Regressive Modelling (RainFARM)}~\cite{rebora2006rainfarm}, and sophisticated methods such as the ecPoint approach \cite{ecPoint}, and a deterministic neural network trained on mean squared error.

ecPoint \cite{ecPoint} is an ECMWF statistical post-processing technique that gives a probabilistic prediction for rainfall intensity at a specific point, accounting for sub-grid variability and model biases. This is done by assigning the parent grid-cell into one of over 100 bins (categories), based on the atmospheric conditions predicted by the model. Mapping functions, which scale the precipitation multiplicatively, are pre-calculated for each bin based upon the percentiles of the training data. The input fields for the ecPoint approach are similar to, but not exactly the same as, the input fields to our generative model. The ecPoint model was originally trained with global station observations, using precipitation accumulated over 12 hours. For our application, we use the same atmospheric variable decision tree ``break-points'' as the standard ecPoint implementation, but naïvely converted to work with hourly data -- accumulated quantities are divided by 12, other quantities are left unchanged. We then retrain the probability mapping functions on our output dataset: NIMROD, at hourly intervals. However we appreciate that this implementation of the ecPoint approach is somewhat flawed, and ideally optimal break-points would be re-derived for hourly data.

While ecPoint was designed to sample gridbox uncertainty, the method was first designed as a post-processing tool and as such only describes how to get a point-wise probabilistic prediction of the possible sub-grid values within an IFS gridbox. To use it as a downscaling tool requires a choice on how to sample multiple high-resolution grid points within the same IFS gridbox. We use two example approaches that maintain the correct pixel-wise statistics in the high-resolution output. \emph{ecPoint no-corr} refers to a ``no-correlation'' method of sampling every pixel independently from the parent distribution. Visually, this would lead to a very noisy image. Quantitatively, accumulated rainfall forecasts over larger regions will have insufficient variation, since there is no spatial coherence to the forecast. \emph{ecPoint part-corr} refers to a ``part-correlation'' method in which the same sampled value is used for every pixel within one low-resolution parent grid cell. This is equivalent to the standard use of ecPoint, where the input image is post-processed but not upsampled. Visually, this would produce a blocky image, but accumulated rainfall over larger regions may vary more realistically. \add{We always use 100 ensemble members, which allows us to \emph{permute} the 100 candidate ecPoint predictions at each pixel, i.e., sample without replacement. This improves the CRPS very slightly compared to sampling \emph{with} replacement.} Again, we emphasise that these steps are not part of the core ecPoint approach, but are merely simple methods for combining single-pixel probabilistic predictions into complete images.

\remove{Rainfall Filtered Auto Regressive Modelling (RainFARM)} \add{RainFARM} is a downscaling method which has been developed specifically for rainfall. It is based on non-linearly filtering the output of a linear auto-regressive process, whose properties are derived from the information available at the large scales. This process extrapolates the large-scale spatio-temporal power spectrum of the meteorological predictions to the small, unresolved scales. The basic concept is to preserve the amplitude and phases of the original field at the scales with high confidence in original model prediction, and to reconstruct the Fourier spectrum at the smaller (unreliable, unresolved) scales. RainFARM can be used stochastically to generate multiple predictions for the same input.

Lanczos filtering is a traditional, interpolation-based widely used image scaling method. It is perhaps better than `constant upsampling' of the input by repetition, although in our application the two approaches are very similar. We also use a deterministic convolutional neural network method that has the same architecture as the GAN generator, with the noise input removed, and is trained on a mean squared error loss function. Lanczos filtering, constant upsampling, and this CNN are all deterministic.

\subsection{Model evaluation}
\label{sec:model-evaluation}

\begin{threeparttable}[t]
    \centering
    \caption{Table showing evaluation results for different models, on previously unseen 2020 data, for CRPS, power spectra error (RALSD), and RMSE. The best score for each metric is highlighted in bold.}
    \label{table:model-eval-results}
    \begin{tabular}{ccccccccc}
        \hline
        \multirow{2}{*}{Model} & \multicolumn{8}{c}{Evaluation Metric}\\
        \cline{2-9}
         & \multicolumn{5}{c}{CRPS (mm/hr)} & \multirow{2}{*}{RALSD (dB)} & \multicolumn{2}{c}{RMSE (mm/hr)}\\
         & pixelwise\tnote{1} & avg 4\tnote{1} & max 4\tnote{1} & avg 16\tnote{1} & max 16\tnote{1} & & ens-mean & individual\\
         \hline
         GAN   & 0.0856 & 0.0844 & 0.1151 & 0.0806 & 0.2117 & \textbf{4.88} & \textbf{0.404} & 0.528 \\
         VAE-GAN & \textbf{0.0852}  & \textbf{0.0840} & \textbf{0.1147} & \textbf{0.0802} & \textbf{0.2104} & 5.34 & 0.405 & 0.499 \\
         ecPoint no-corr\tnote{2} & 0.0895 & 0.1075 & 0.3987 & 0.1195 & 1.5948 & 16.35 & 0.423 & 0.644 \\
         ecPoint part-corr\tnote{2} & 0.0895 & 0.0889 & 0.1255 & 0.0883 & 0.2485 & 9.78 & 0.423 & 0.644 \\
         RainFARM & 0.1331 & 0.1332 & 0.1697 & 0.1286 & 0.2888 & 9.95 & 0.442 & \textbf{0.444} \\
         Lanczos\tnote{3} & 0.1412 & 0.1392 & 0.1731 & 0.1309 & 0.2923 & 15.38 & \multicolumn{2}{c}{0.447} \\
         Det CNN\tnote{3} & 0.1347 & 0.1325 & 0.1644 & 0.1250 & 0.2817 & 16.74 & \multicolumn{2}{c}{0.404} \\
         \hline
    \end{tabular}
    \begin{tablenotes}
    \item[1] \change{The different CRPS metrics}{These} correspond to different methods of spatial pooling, as described in Section~\ref{subsubsec:crps}.
    \item[2] \add{The two ecPoint variants have identical pixel-wise statistics, by construction.}
    \item[3] \add{These are deterministic methods, hence the CRPS reduces to the mean absolute error, and there is no separate ensemble-mean RMSE.}
    \end{tablenotes}
\end{threeparttable}

Table~\ref{table:model-eval-results} shows numerical results for the pixel-wise and pooled CRPS scores, the RALSD score, and RMSE, for the GAN and VAE-GAN models we developed, compared to existing models/approaches. For each of the CRPS metrics, the best score is obtained by the \change{GAN}{VAE-GAN} model, \add{marginally ahead of the GAN}. Notably, the \change{GAN compares}{GAN and VAE-GAN compare} favourably to ecPoint on pixel-wise CRPS, despite the latter being a very well-calibrated point-wise method. We consider this a strong result. \remove{The VAE-GAN model also scores consistently well on CRPS.} The deterministic model scores poorly, showing the added value of the stochastic nature of the generative approaches. The RALSD scores show that only the GAN and VAE-GAN produce images with realistic power spectra. The RALSD figures for both ecPoint variations are included, but ecPoint is not designed to produce coherent spatial forecasts so these scores are unimportant. \add{Ensemble-mean RMSE values are given too: the GAN produces the best score here, marginally better than a deterministic CNN trained to minimise mean squared error. The VAE-GAN ensemble-mean RMSE is only slightly worse, and these all score notably better than the other methods.} \add{Individual prediction} RMSE values are given for completeness, but, as discussed in Section~\ref{subsubsec:image-quality-metrics}, it is a poor metric to optimise for due to the double penalty effect.

\begin{figure*}[hp!]
    \centering
    \includegraphics[width=0.85\textwidth]{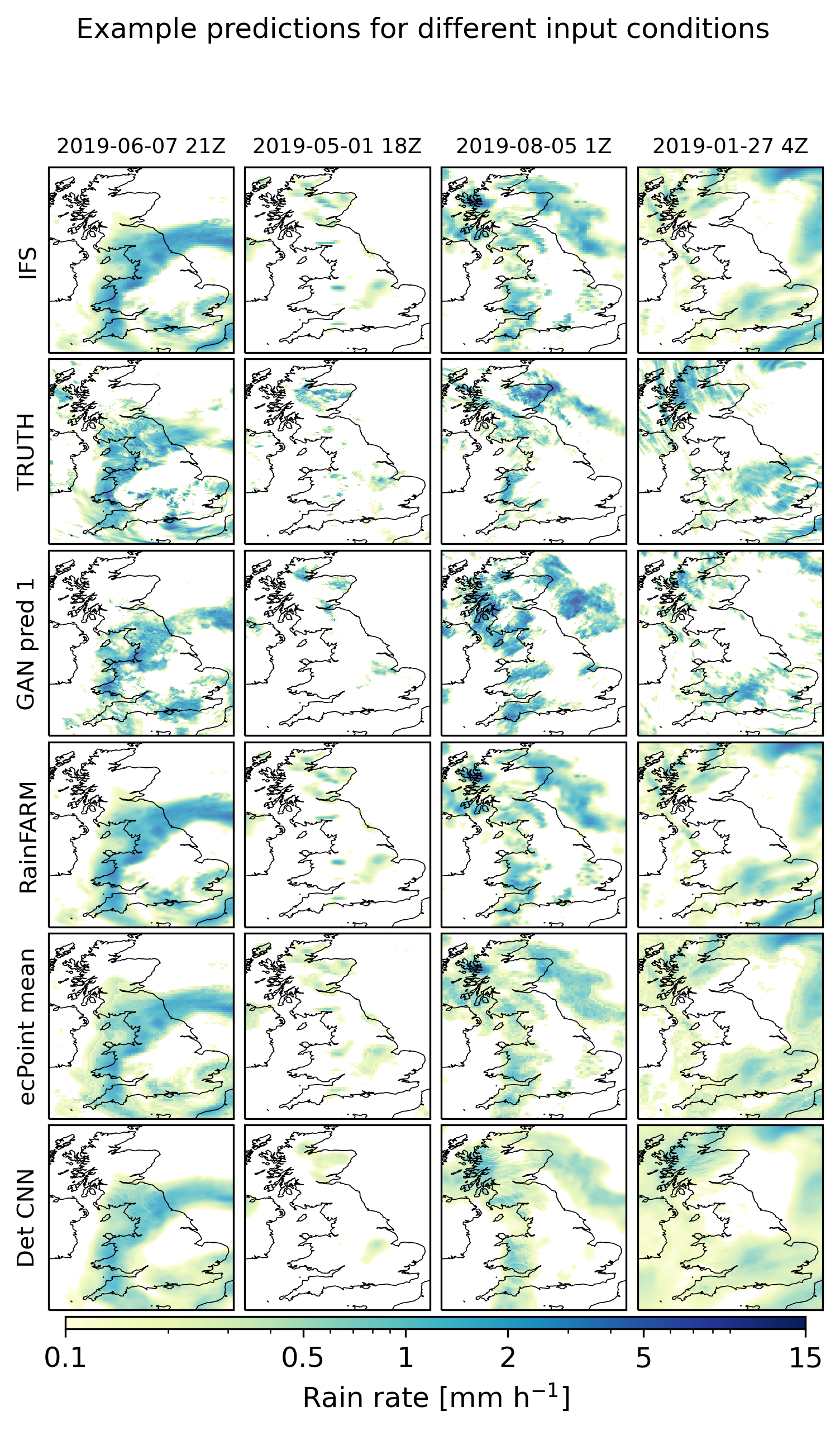}
    \caption{Comparison of predictions generated by the GAN with those produced by existing methods, for four randomly-chosen cases. The following fields are shown: the IFS forecast low-resolution input data, the NIMROD high-resolution ground truth data, a single GAN ensemble member prediction, a RainFARM prediction, the mean ecPoint prediction, and a deterministic neural network applied to the IFS data.}
    \label{fig:GAN-comparison}
\end{figure*}

Figure~\ref{fig:GAN-comparison} shows plots for four example cases. \add{More detailed descriptions of these meteorological scenarios can be found in the supporting information document.} The examples give a clear indication of how our GAN model produces \remove{significantly} more detailed and more visually realistic images than any other method, as well as being more robust at forecasting more intense rainfall. The RainFARM algorithm does produce some small-scale detail compared to the IFS input, but it is limited to producing the same texture everywhere in the image and it does not reproduce the overall structure of the high-resolution truth as well as the GAN, nor does it predict extremes of rainfall missed by the IFS forecast. The ecPoint mean prediction is shown, for completeness, and is effectively a bias-corrected IFS. However, for clarity, none of our quantitative methods use this ecPoint mean; instead they use ecPoint ensemble members constructed via the \emph{no-corr} and \emph{part-corr} methods described previously. Finally, the deterministic CNN trained on mean squared error produces very `blurry' predictions. This model often greatly over-predicts the spatial extent of very light rainfall, and is incapable of predicting extremes. In general, the smoother plots with less variance and fine-scale structure are rewarded by the RMSE metric, but punished by the RALSD metric (further details in section~\ref{subsubsec:image-quality-metrics}, with results displayed in Table~\ref{table:model-eval-results}).

\begin{figure*}
    \centering
    \includegraphics[width=0.9\textwidth]{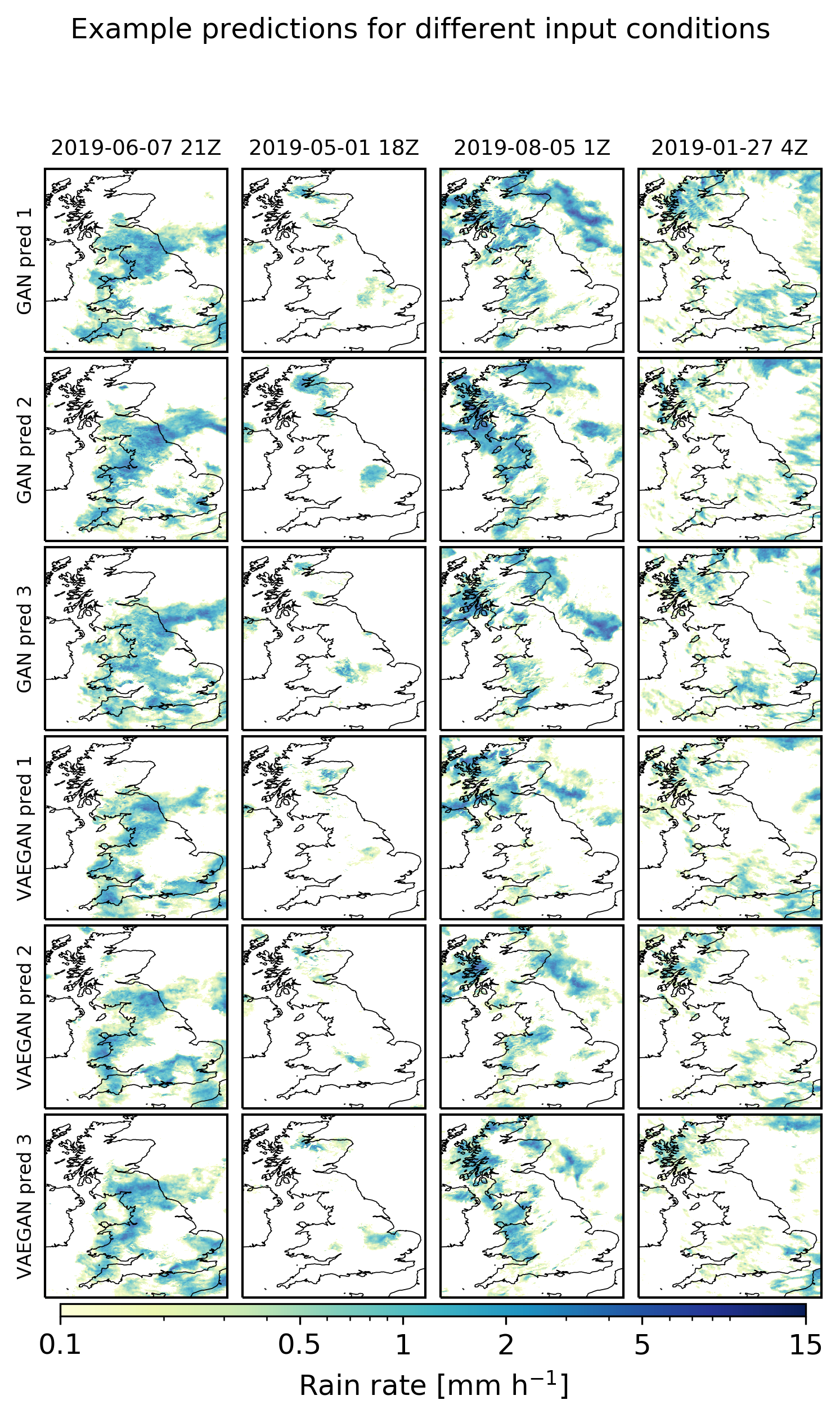}
    \caption{Examples of multiple GAN and VAE-GAN ensemble predictions for four different input examples. The examples used are the same as in Figure~\ref{fig:GAN-comparison}.}
    \label{fig:predictions}
\end{figure*}

A set of example predictions for the best GAN and VAE-GAN models are shown in Figure~\ref{fig:predictions}. For each example, Figure~\ref{fig:predictions} shows three different ensemble predictions from each model. The same randomly selected cases are shown in this example, encompassing a range of precipitation conditions. The predictions produced by these models provide a high-quality solution set to a range of different meteorological conditions. There is sharply varying spatial structure in the predictions that is reminiscent of the true conditions, and not produced by any of the existing approaches.

We can clearly see from these examples that the GAN and VAE-GAN models are very capable of improving on the IFS forecast and bringing the predictions closer to the truth. Further, both models produce multiple realisations for the same situation, giving a clearer idea of the uncertainty. The main \change{improvements}{improvement} offered by the GAN over the VAE-GAN \change{are}{is} an improved tendency to predict more intense rainfall.\remove{, and a capability of producing much more realistic structure to the larger-scale events.}

\subsection{Model predictions for extreme events}

Since machine-learnt models are trained on historic data, they often struggle with extreme events. However, these are some of the most important situations to forecast accurately and reliably, so we were particularly interested in assessing our models' performances on extreme events.

\begin{figure*}
\centering
    \includegraphics[width=\textwidth]{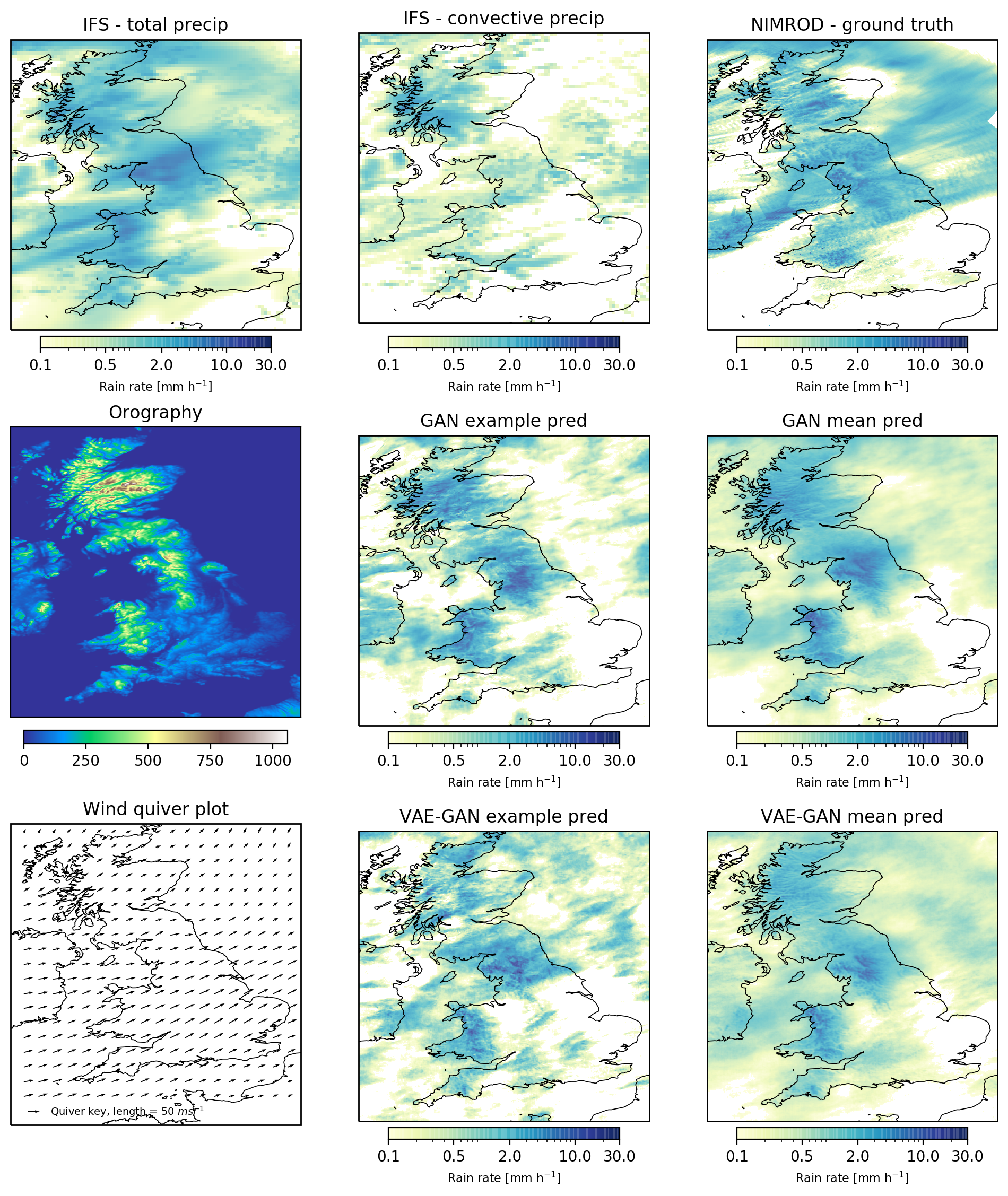}
    \caption{GAN and VAE-GAN model output example predictions and means for one of the most extreme examples in our dataset, from 09:00-10:00 UTC of the 9\textsuperscript{th} February 2020, showing the total precipitation and wind direction and strength from the IFS forecast, orography, and the ground truth NIMROD data. Note that the colourbar has been changed from the previous set of examples and now ranges from \change{1mm-30mm}{0.1--30mm}.}
    \label{fig:extremes}
\end{figure*}

Figure~\ref{fig:extremes} shows the GAN and VAE-GAN model responses to one of the most extreme rainfall events in our dataset. These data points are taken from 09:00-10:00 UTC of the 9\textsuperscript{th} February 2020, during which there was a significant rainstorm across the UK, named Storm Ciara. Figure~\ref{fig:extremes} shows input fields including total precipitation, convective precipitation, orography and a wind quiver plot, as well as the truth data, a single example prediction and the mean prediction. The GAN and VAE-GAN models capture the peak intensities and fine-scale structure of the rainfall event better than the IFS forecast.

\subsection{Rank statistics}
\label{subsec:rank-statistics}

\begin{figure}
    \centering
    \includegraphics[width=\textwidth]{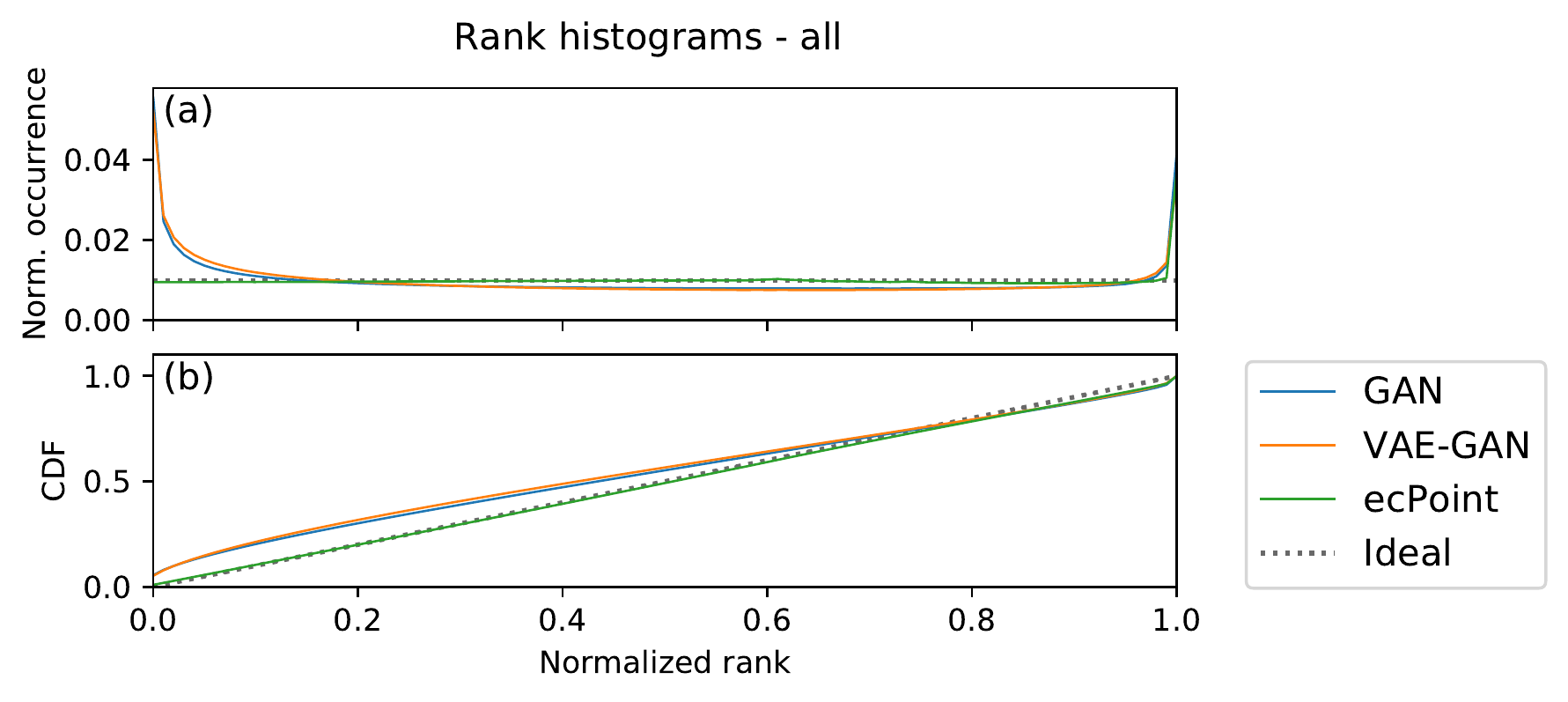}
    \caption{Calibration plot \change{for GAN and VAE-GAN models}{across all events}: (a) shows the frequency of per-pixel normalised ranks for the trained GAN and VAE-GAN models evaluated on the hold-out dataset (2020), \add{compared to the ecPoint approach}. The dotted grey line shows the ideal distribution for comparison. (b) shows the same as panel a, except displaying the CDFs of the distributions.}
    \label{fig:GAN-histograms}
\end{figure}

\begin{figure}
    \centering
    \includegraphics[width=\textwidth]{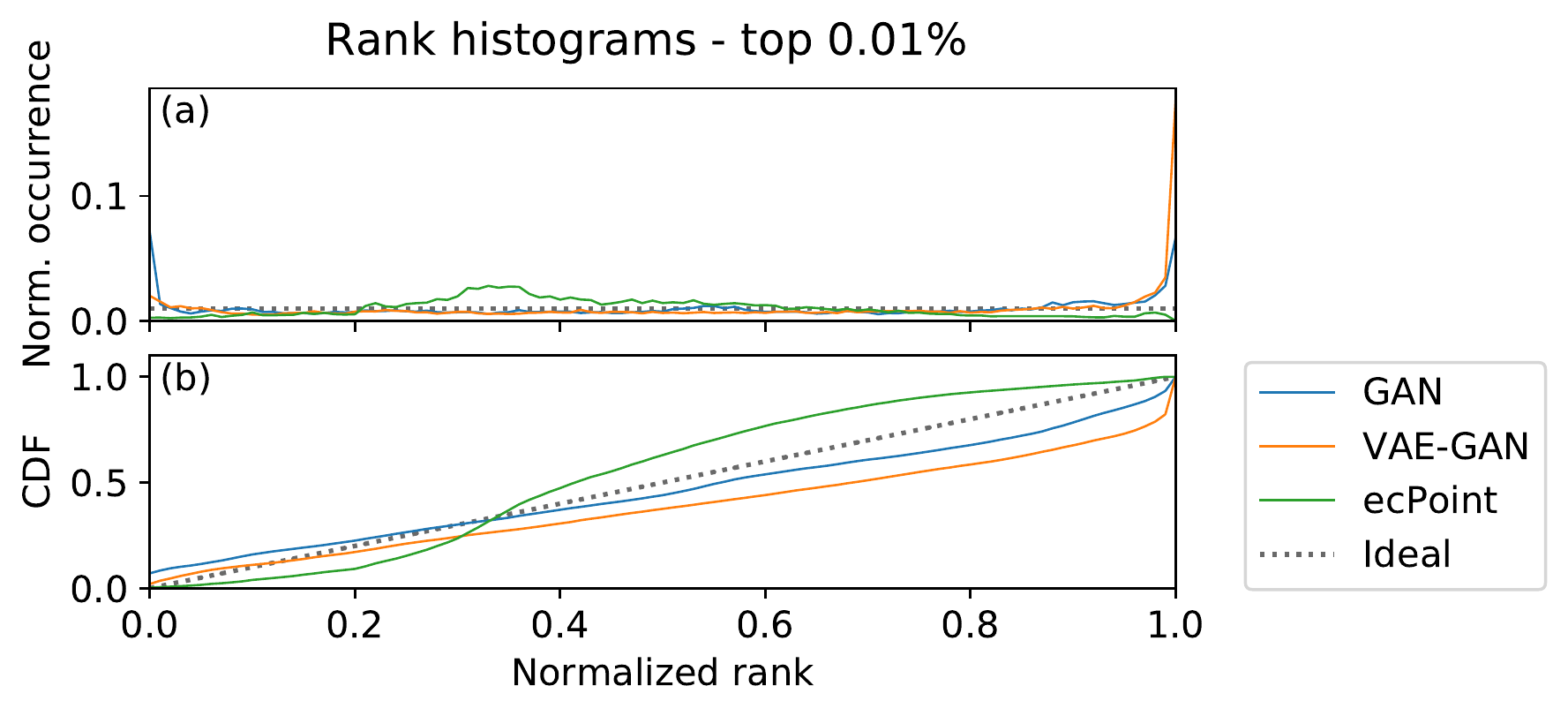}
    \caption{\add{Calibration plot; similar to} Figure~\ref{fig:GAN-histograms}, \add{but only for the top 0.01\% of IFS forecasted precipitation events. This corresponds to IFS predictions above 5.7mm/hr of precipitation; 226 of these are present in the 256 $94 \times 94$ input images used.}}
    \label{fig:GAN-histograms-thresh}
\end{figure}

Figure~\ref{fig:GAN-histograms} shows the pixel-wise rank distributions of the GAN and VAE-GAN ensembles, based on 100-member ensembles. \remove{for 256 full-image examples} These plots show that the majority of the outlier ranks are in the tail ends of the distribution, where $r$ is either close to 0 or 1, \add{implying that the GAN and VAE-GAN are slightly underdispersive}. \remove{There are few outliers in the middle of the distribution, and overall these outlier samples are only a small part of the overall distribution. In the majority of cases, the real sample falls within the ensemble set of predictions.} \change{This is another metric where the GAN outperforms the VAE-GAN -- the rank histograms for the GAN are consistently better calibrated across the distribution.}{The GAN marginally out-performs the VAE-GAN on this metric.} The ecPoint approach outperforms our networks on this metric considerably; however, ecPoint is essentially optimised for this metric, as its \emph{raison d'être} is to produce well-calibrated pointwise forecasts. Our approaches, on the other hand, also try to produce realistic larger-scale spatial structures. \remove{Interestingly, the GAN model does outperform the ecPoint approach with respect to heavy precipitation outliers: the truth measurement exceeds all ecPoint ensemble members far more than the expected $\frac{1}{101}$ frequency, while the GAN model is only slightly mis-calibrated here.} \add{The ecPoint approach is still somewhat underdispersive on the right-hand tail, though.}

Figure~\ref{fig:GAN-histograms-thresh} \add{shows the same analysis but now restricted to `extreme events' -- the top 0.01\% of forecasted precipitation events seen in the IFS input. The GAN now outperforms the VAE-GAN, but both are under-dispersive, particularly on the right-hand tail. The ecPoint approach is now over-dispersive, with the real sample rarely falling in the bottom or top 20\% of predictions. This is perhaps related to the multiplicative ansatz of the ecPoint approach. Among the three methods, the GAN perhaps performs best on extreme events, but none of the three methods are particularly reliable on these.}

\subsection{Power spectra}
\label{subsubsec:power-spectra}

\begin{figure}
     \centering
     \begin{subfigure}[b]{0.495\textwidth}
         \centering
         \includegraphics[width=\textwidth]{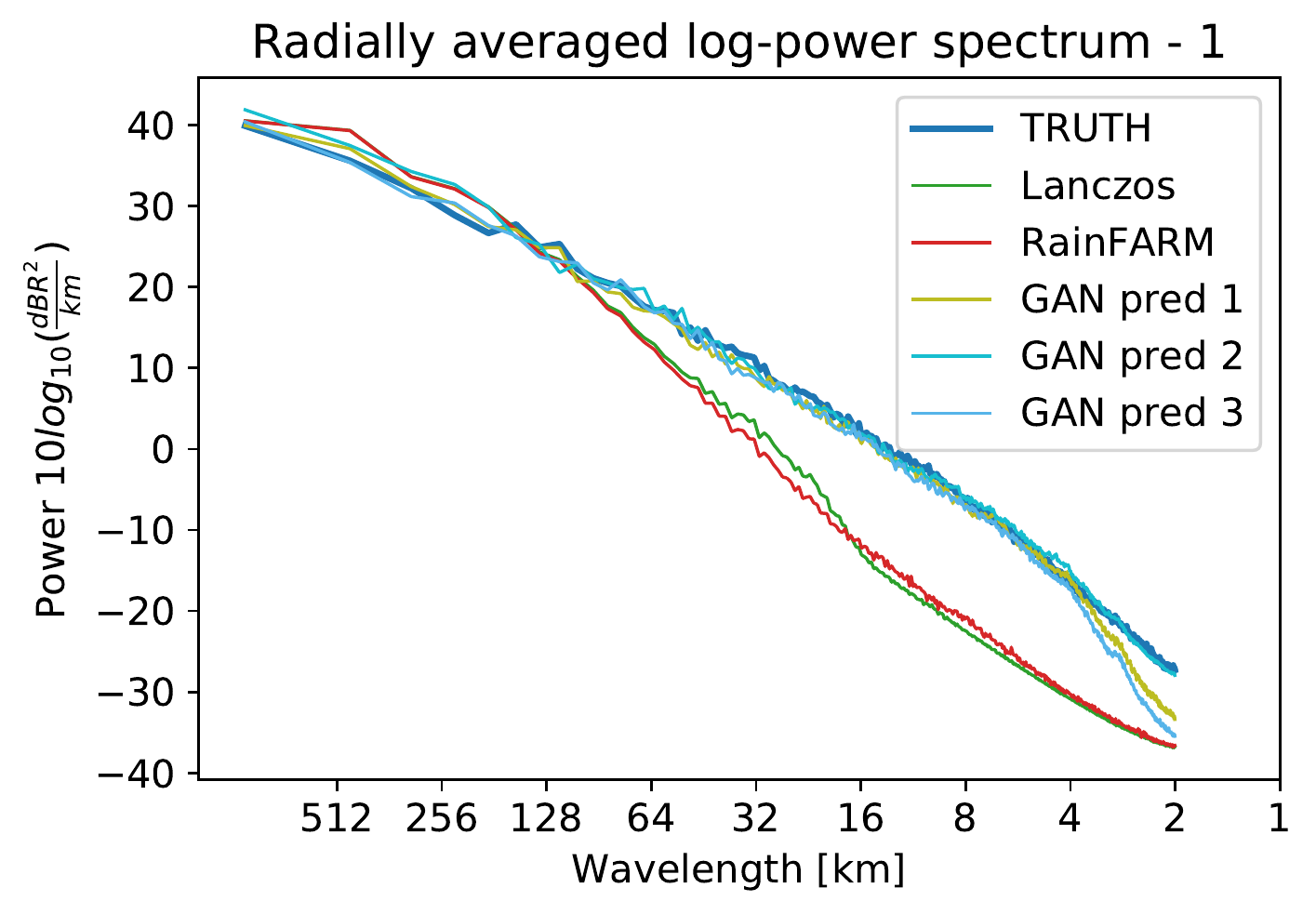}
         \caption{GAN model}
         \label{fig:GAN-RAPSD}
     \end{subfigure}
     \hfill
     \begin{subfigure}[b]{0.495\textwidth}
         \centering
         \includegraphics[width=\textwidth]{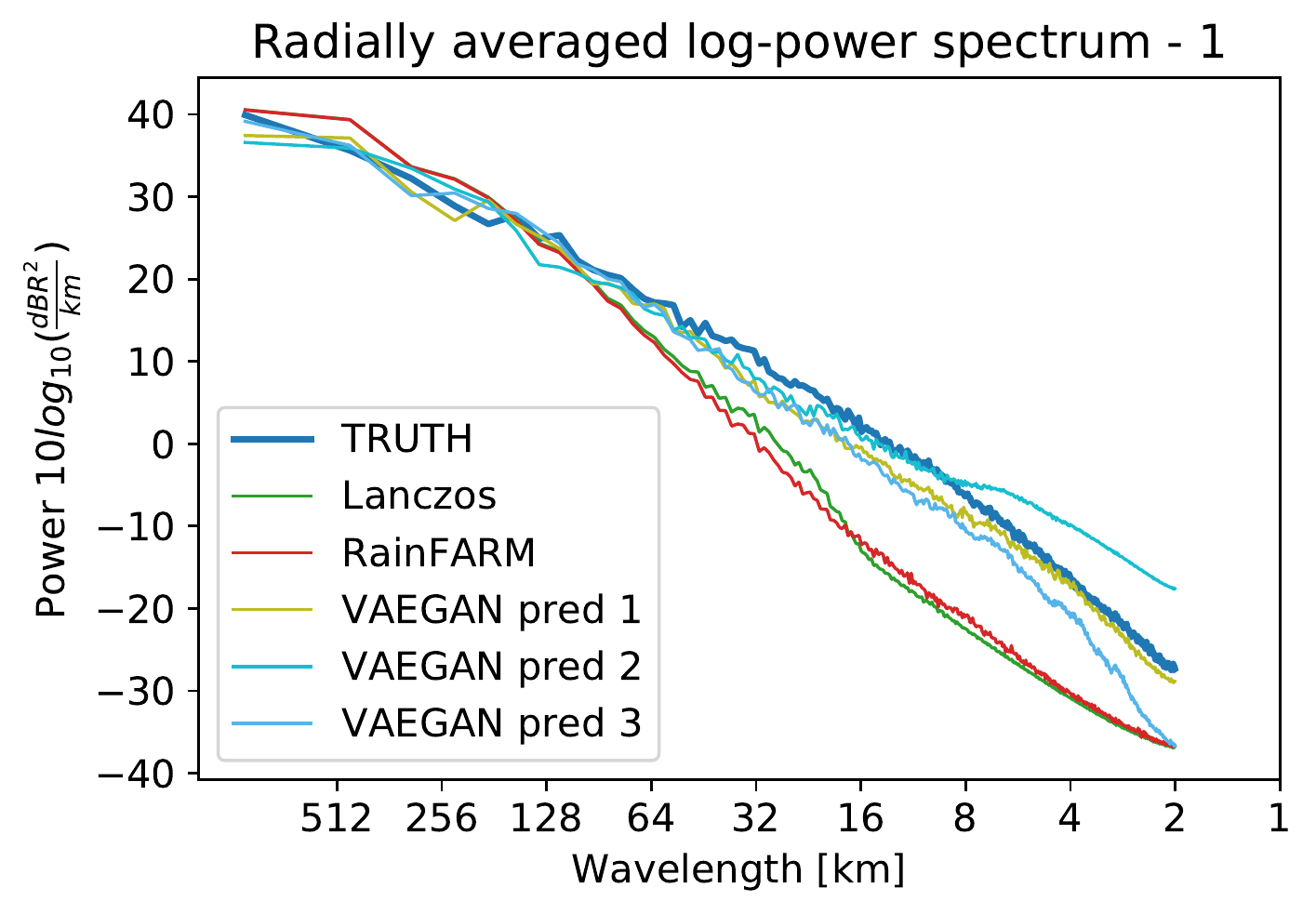}
         \caption{VAE-GAN model}
         \label{fig:VAEGAN-RAPSD}
     \end{subfigure}
     \caption{Plot displaying the radially-averaged power spectrum of images with decreasing scale produced by the GAN and VAE-GAN models, on the first example situation, compared to both Lanczos interpolation and the RainFARM method.}
     \label{fig:RAPSD-1}
\end{figure}

In addition to the scores displayed in Table~\ref{table:model-eval-results} for the RALSD, Figures~\ref{fig:GAN-RAPSD} and \ref{fig:VAEGAN-RAPSD} show radially averaged power spectral density (RAPSD) plots for the GAN and VAE-GAN models for the first of the four example situations, compared to the ground-truth NIMROD data, and the existing RainFARM and Lanczos models. Details of the RAPSD implementation are given in section~\ref{subsubsec:image-quality-metrics}. These plots show, firstly, that the RainFARM and Lanczos models are missing a lot of detail at lower scales (really, any grid scale under $\sim$100km), which is unsurprising as this is identifiable by eye in all of the example plots. The GAN and VAE-GAN are both much closer to the truth in terms of retaining energy in the image at much finer scales. Only one example is analysed here (example 1 in Figure~\ref{fig:GAN-comparison}); further examples are included in the supporting information document. Interestingly, in this particular example, the VAE-GAN \change{is closer to the truth}{contains more information} than the GAN at the \change{lowest}{smallest} scales. This is not always the case, as shown in Table~\ref{table:model-eval-results}. There is also typically more variation between members of the GAN ensemble than those of the VAE-GAN, unlike in this particular case.

\subsection{ROC curves}
The ROC curves for the GAN, VAE-GAN and ecPoint (partial-correlation and no-correlation) models \remove{shown here were generated by generating 100 member ensembles for 256 randomly chosen full-image examples. ROC curves} are shown here for the 0.5 mm/hr and 5 mm/hr thresholds, using pixel-wise analysis. Additional plots using spatial pooling, and for other precipitation thresholds, are included in the supporting information document. A perfect prediction would yield a point in the upper left corner of the ROC space, representing 100\% sensitivity (no false negatives) and 100\% specificity (no false positives). The dashed diagonal line represents random chance. Consequently, points far above the diagonal represent good classification results.

\begin{figure}
     \centering
     \begin{subfigure}[b]{0.495\textwidth}
         \centering
        \includegraphics[width=\textwidth]{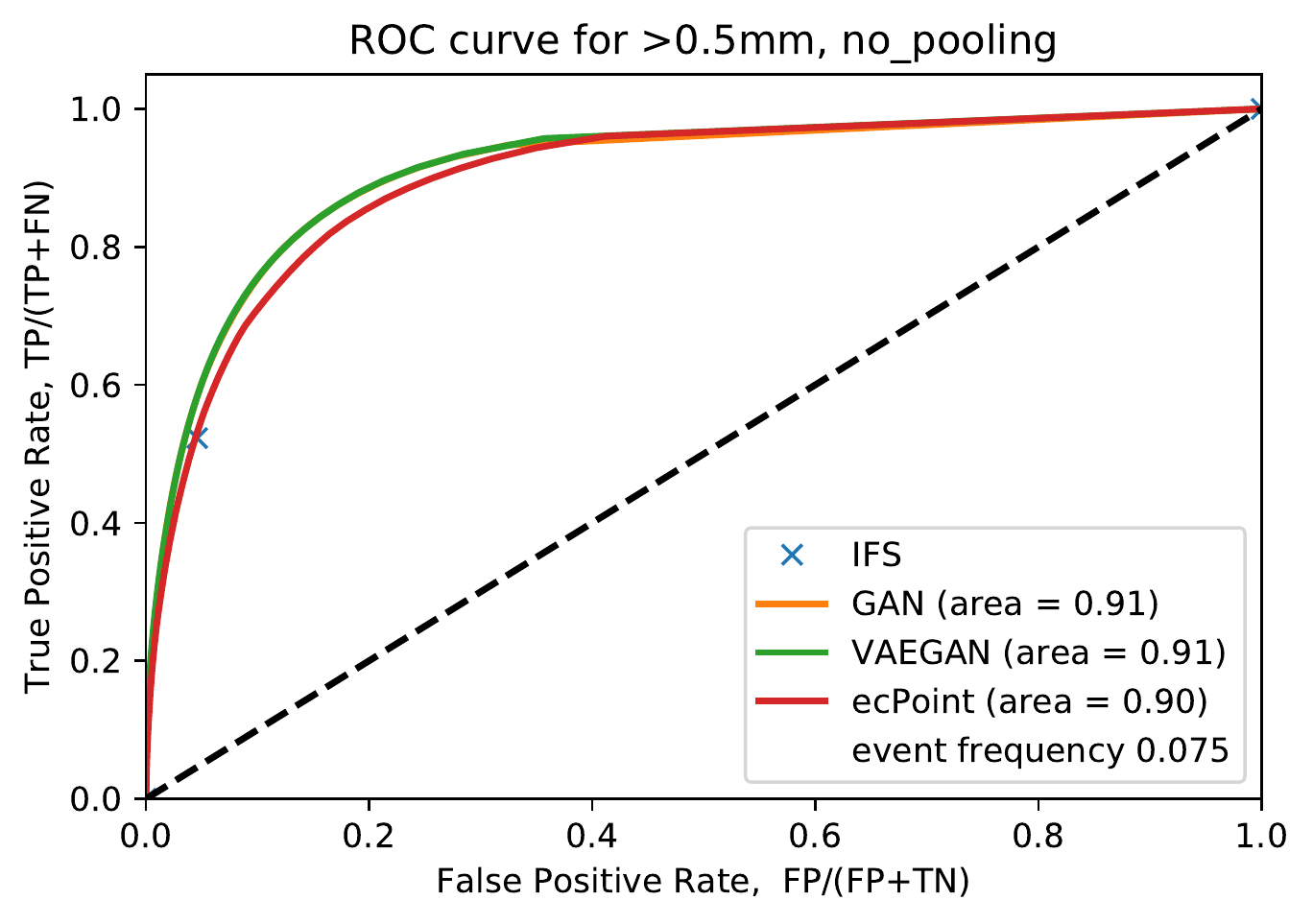}
         \caption{0.5 mm/hr precipitation threshold}
        \label{fig:ROC-0.5-none}
     \end{subfigure}
     \hfill
     \begin{subfigure}[b]{0.495\textwidth}
         \centering
        \includegraphics[width=\textwidth]{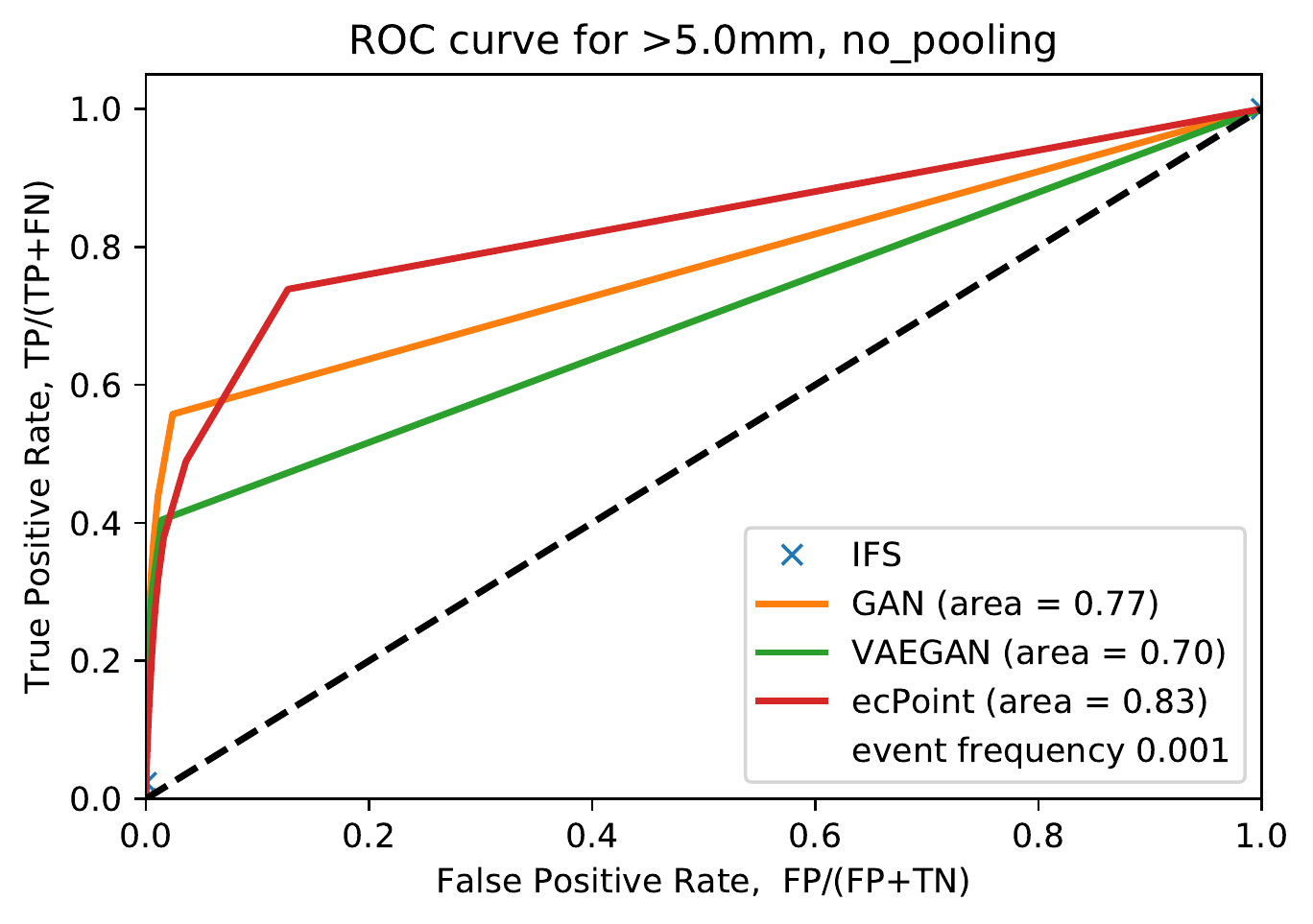}
         \caption{5.0 mm/hr precipitation threshold}
         \label{fig:ROC-5.0-none}
     \end{subfigure}
     \caption{ROC curves for the GAN, VAE-GAN and ecPoint models for 0.5 and 5.0 mm/hr precipitation thresholds. The upsampled input data, labelled `IFS', is represented by a cross, as it is a single prediction rather than an ensemble.}
     \label{fig:ROC-none}
\end{figure}

For the $0.5$mm/hr threshold, shown in Figure~\ref{fig:ROC-0.5-none}, \change{the GAN shows the best results and beats the VAE-GAN and the ecPoint approach. The GAN line is generally above and left of the other lines, and has the largest area under the curve.}{the GAN and VAE-GAN slightly outperform the ecPoint approach. The GAN and VAE-GAN lines are generally above and left of the ecPoint line, and have the largest areas under the curve.} For the $5$mm/hr threshold, shown in Figure~\ref{fig:ROC-5.0-none}, the results are harder to interpret. The curves are somewhat distorted due to the finite ensemble size and the rarity of the event (the event frequency is 0.001). \change{The VAE-GAN clearly under-performs the GAN and the ecPoint approach, which correlates with the results identified visually: the VAE-GAN is much less likely to correctly predict intense rainfall. Although ecPoint has a slightly higher area under the curve than the GAN, the GAN line is clearly to the left of the ecPoint line in the initial portion of the graph.}{Although the ecPoint line has the largest area under the curve, the GAN line is the furthest left in the initial portion of the graph, followed by the VAE-GAN.} This likely equates to better performance in the limiting case of unlimited ensemble members (see \citeA{Zied2022} for more on interpreting the area under ROC curves in the case of rare events).

\subsection{Fractions skill score}

\begin{figure}
     \centering
     \begin{subfigure}[b]{0.495\textwidth}
         \centering
         \includegraphics[width=\textwidth]{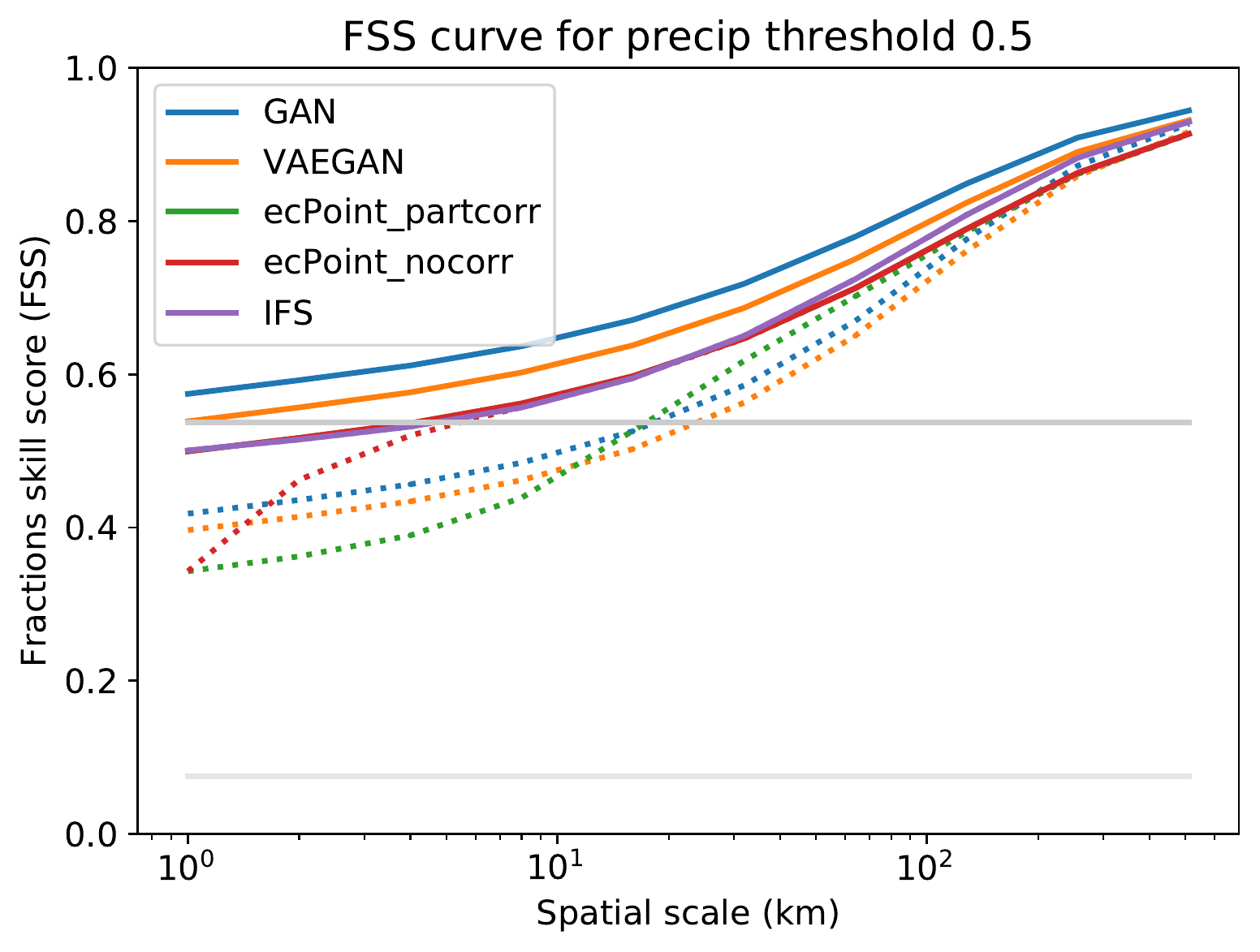}
         \caption{0.5 mm/hr precipitation threshold}
         \label{fig:FSS-0.5}
     \end{subfigure}
     \hfill
     \begin{subfigure}[b]{0.495\textwidth}
         \centering
         \includegraphics[width=\textwidth]{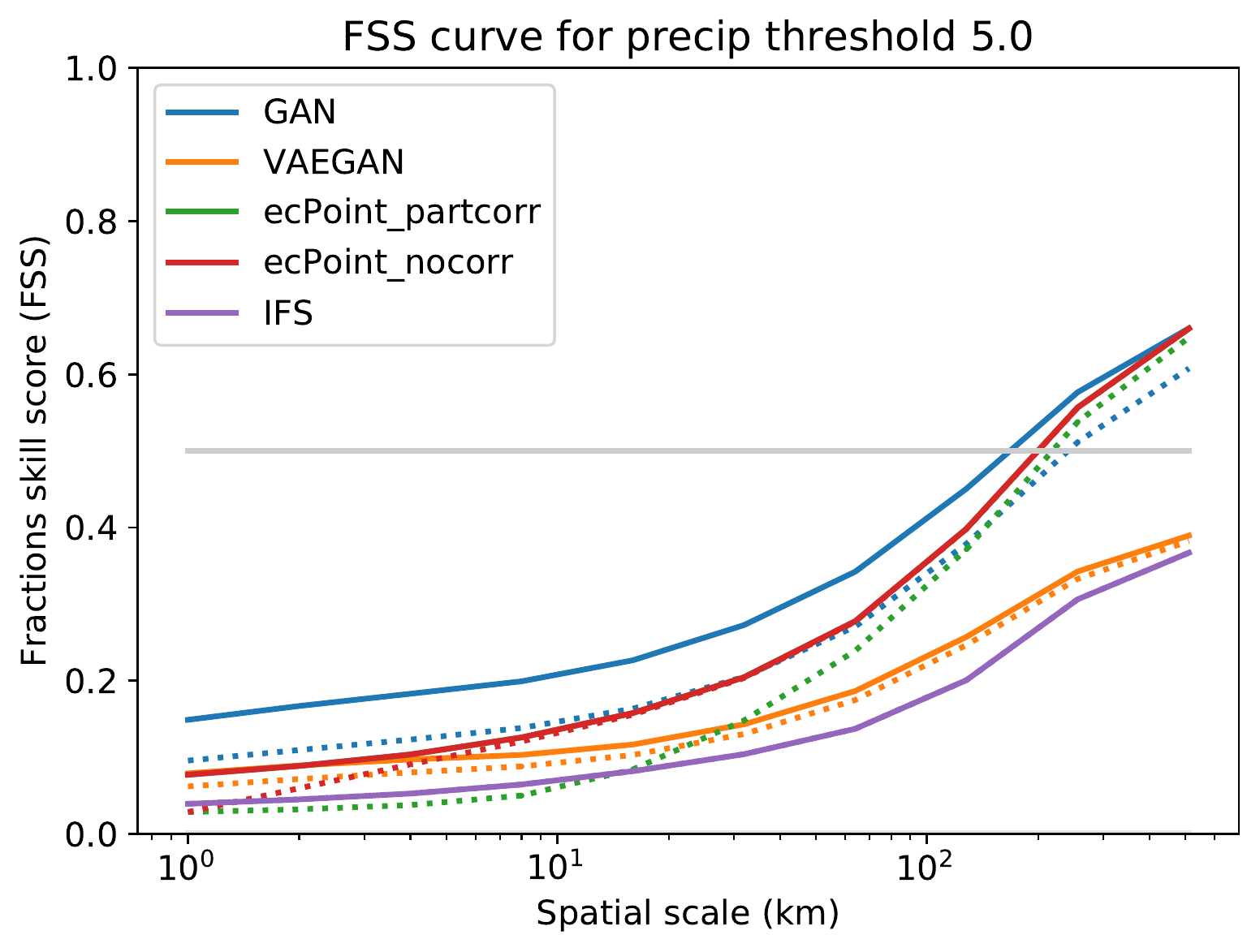}
         \caption{5.0 mm/hr precipitation threshold}
         \label{fig:FSS-5.0}
     \end{subfigure}
     \caption{FSS curves for the GAN, VAE-GAN and ecPoint models. The solid lines represent `ensemble FSS scores', as described in Section~\ref{subsubsec:fss}, while the dashed lines represent basic FSS scores applied to individual ensemble members. The grey lines represent commonly-used `no-skill' and `useful skill' thresholds, of $p$ and $\frac{1+p}{2}$, where $p$ is the event probability.}
     \label{fig:FSS}
\end{figure}

Fractions skill score (FSS) curves for the GAN and VAE-GAN models are plotted in Figure~\ref{fig:FSS}. \remove{These were generated by evaluating over 256 different images, with an ensemble size of 100.} FSS curves are shown here for the 0.5 mm/hr and 5 mm/hr thresholds only, with additional plots for 0.1 mm/hr and 2 mm/hr in the supporting information document.

For the light (0.5 mm/hr) rain threshold, both the GAN and VAE-GAN models produce a noticeably better ensemble FSS than the ecPoint variants, and have useful skill even at the pixel level. The FSS of individual \emph{ecPoint no-corr} members is particularly high at intermediate spatial scales, but we believe this is an artefact of the metric when applied to very `noisy' images, as discussed in Section~\ref{subsubsec:fss}, and not a sign of genuinely useful output. \remove{The individual VAE-GAN ensemble members have a higher FSS than those of the GAN, whilst the GAN ensemble FSS is better than that of the VAE-GAN. This perhaps reflects lower variability (and skill) in the VAE-GAN ensemble.} \add{The individual GAN members have a higher FSS than the individual VAE-GAN members, and the GAN ensemble FSS is better than that of the VAE-GAN.}

For the heavy (5.0 mm/hr) rain threshold, the GAN significantly outperforms the VAE-GAN for both ensemble and individual member FSS. \change{The lower variability of the VAE-GAN makes predicting extreme values less likely, so the GAN model is much more useful in these scenarios.}{The VAE-GAN struggles at producing the highest intensities of precipitation.} \change{The GAN ensemble outperforms the ecPoint ensemble at small and intermediate spatial scales, but loses out slightly on the largest spatial scales, suggesting that ecPoint predicts the overall extreme event frequency more accurately. ecPoint is a very well calibrated method so we expect it to score well on extreme events.}{The GAN ensemble clearly outperforms the ecPoint ensemble at small and intermediate spatial scales. At the largest spatial scales, the methods perform similarly, reflecting similar skill at predicting the overall extreme event frequency.}

\subsection{Model performance with increasing lead time}

\begin{figure}
     \centering
     \begin{subfigure}[b]{0.495\textwidth}
         \centering
    \includegraphics[width=\textwidth]{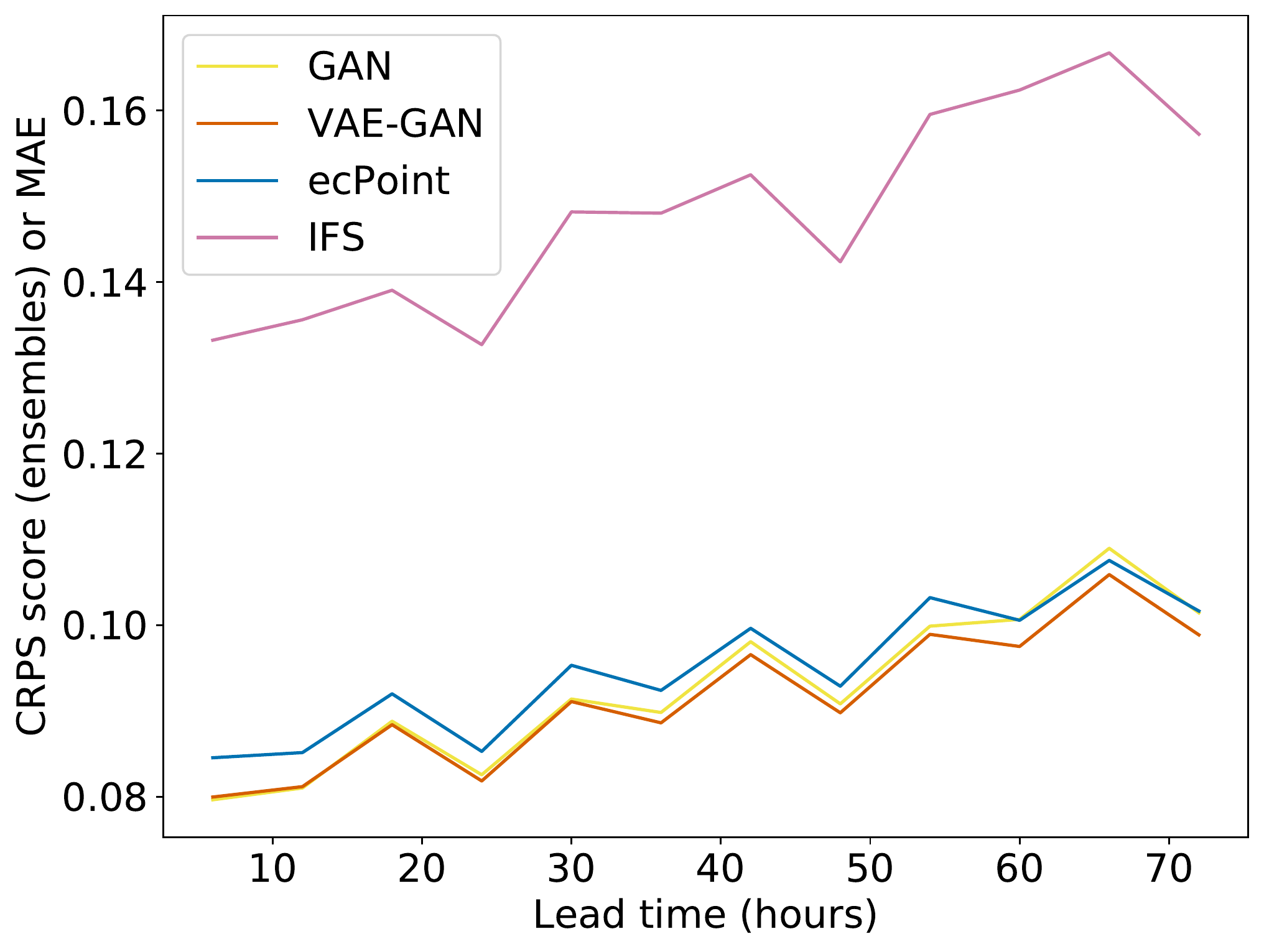}
    \caption{CRPS (lower = better)}
    \label{fig:lead-time-10}
    \end{subfigure}
     \hfill
     \begin{subfigure}[b]{0.495\textwidth}
         \centering
    \includegraphics[width=\textwidth]{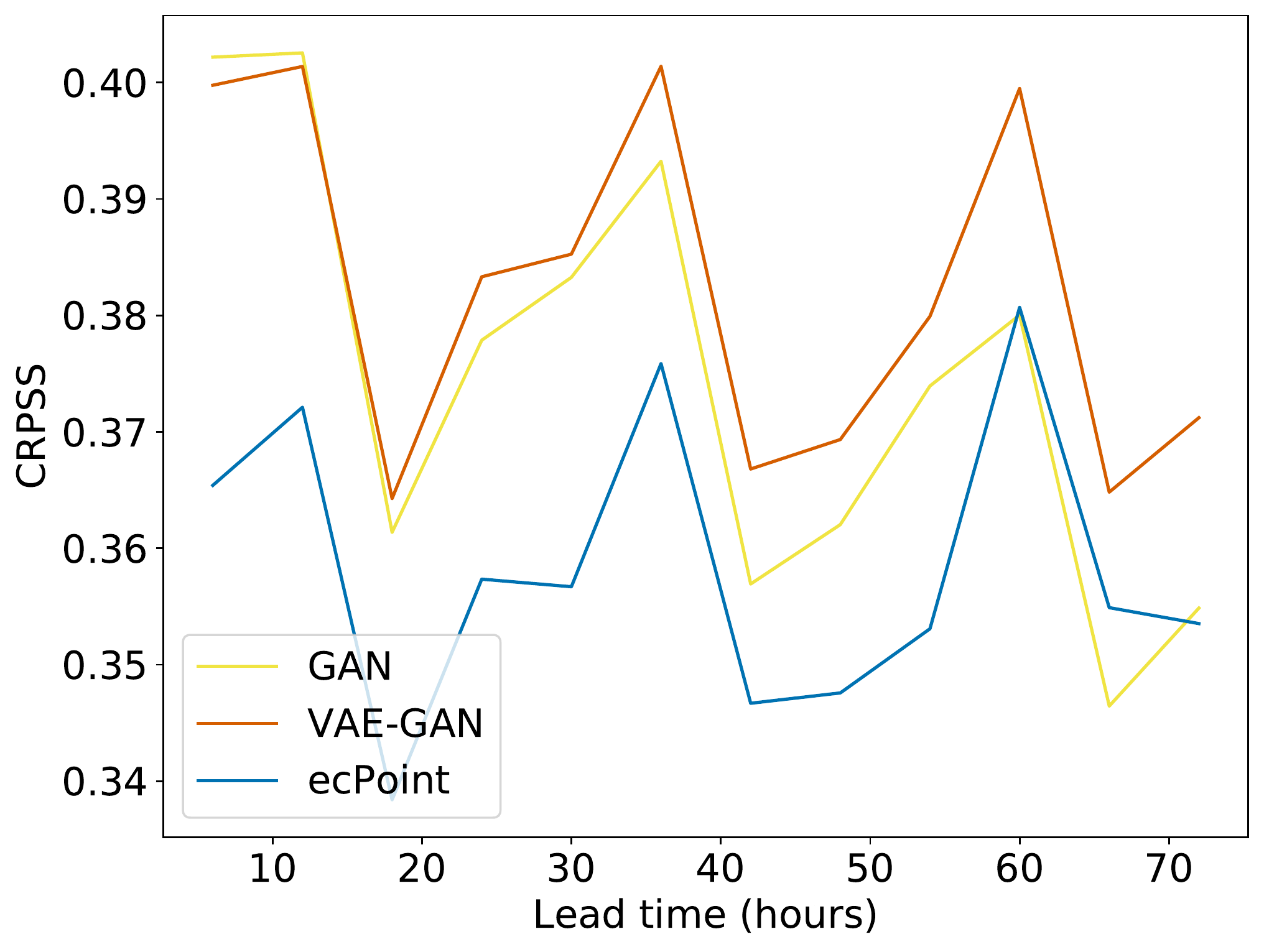}
    \caption{CRPSS (higher = better)}
    \label{fig:lead-time-100}
     \end{subfigure}
     \caption{Scores for the GAN, VAE-GAN and ecPoint models with increasing lead time, compared to the baseline case of the IFS forecast.}
     \label{fig:lead-time}
\end{figure}

To minimise the difference between our model and truth datasets during training we restricted training to lead times between 7--17 hours. Previous results were assessed solely on this time period. However, we are also interested in using our tool on shorter \& particularly on longer lead times. 
Figure~\ref{fig:lead-time} shows plots of the pixel-wise CRPS and CRPSS ($= 1 - \frac{\textrm{CRPS}}{\textrm{CRPS}_{\textrm{IFS}}}$) metrics for the GAN and VAE-GAN models, for increasing lead time, compared to the IFS forecast data and the ecPoint approach. These are obtained \emph{without} retraining the model with data from other lead times.

\change{The lead time investigation went through every image in the 2020 dataset, with an ensemble size of 100, for lead times from 6--72 hours, every 6 hours.}{The lead time investigation was carried out using every available 00Z forecast in our 2020 dataset, at lead times every 6\textsuperscript{th} hour from 6--72 hours, again with an ensemble size of $100$.} All models show generally increasing CRPS with lead time, with visible diurnal cycle effects. \add{The GAN and VAE-GAN show decreasing CRPSS as lead time increases, consistent with them being applied on longer lead times than they were trained on. However, the ecPoint approach shows CRPSS increasing with lead time, despite also being calibrated on 7--17 hour data.} The GAN, VAE-GAN and ecPoint approach all have \add{somewhat} similar CRPS scores, compared with the IFS input data. \change{The GAN slightly outperforms the VAE-GAN throughout. Interestingly, the GAN is better than the ecPoint approach at shorter lead times, but worse at longer lead times, despite both being trained on the same data. In an earlier version of this plot with only 10 ensemble members, the VAE-GAN outperformed the GAN, but this was reversed when the full 100 ensemble members was used. This may reflect the greater variability in the GAN predictions compared with those of the VAE-GAN.}{The GAN and VAE-GAN slightly out-perform the ecPoint approach, although the GAN is overtaken by ecPoint at the longest lead times.}

\subsection{Pure super-resolution tests}
Due to the different origins of our input data (IFS) and truth data (NIMROD), we are asking any machine learning solution to undertake two tasks: super-resolution, and bias/spread correction to account for forecast error. As a sanity check of our models, we also trained them to instead ingest coarsened NIMROD data (area-averaged to 0.1\degree{}) and full-resolution geographic fields, and predict the full-resolution 0.01\degree{} NIMROD field. The resulting problem is close to that tackled in \citeA{Leinonen2020}, without the temporal component, and is inherently easier than our full problem. The value of these experiments is to establish the limits of our ML models, and to understand whether the performance is limited by the super-resolution component or the forecast error correction component of the problem. The resulting models from this experiment are unlikely to perform well on the full downscaling problem since coarsened NIMROD radar data is not interchangeable with IFS forecast data. We have included results from this study in the appendix, section~\ref{sec:sr-section}.

\section{Discussion and conclusions}
\label{sec:conclusions}

We present two models: GAN-based and VAE-GAN-based, both capable of increasing the resolution of forecast data by a factor of 10 and calibrating the forecast. \remove{Both models demonstrably add skill to the forecast. Broadly, the VAE-GAN model is less likely to produce extreme values than the GAN, and produces less sharply varying spatial structure, with less variation between different ensemble members. However, the VAE-GAN still produces reasonable predictions, and the CRPS scores from the model evaluation detailed in Section} \ref{sec:model-evaluation} \remove{are comparable. The VAE-GAN model does improve on the IFS forecast for intense rainfall, most noticeable in example 3. However, the GAN model consistently outperforms the VAE-GAN model, particularly for intense rainfall.} \add{Both models demonstrably add skill to the forecast, and produce similar forecasts with similar spatial structure. The VAE-GAN produced slightly better CRPS scores than the GAN. However, the GAN is better-calibrated than the VAE-GAN, is more capable of producing intense rainfall than the VAE-GAN, and perhaps produces slightly more large-scale variation than the VAE-GAN.} \change{We}{Both models} produce much better results than simple alternative methods, and \remove{we} achieve similar or slightly improved scores compared to the precipitation downscaling state-of-the-art ecPoint method, whilst producing spatially coherent and visually realistic images, which are easier to interpret. Both the GAN and VAE-GAN models allow for an ensemble of predictions to be produced, providing an estimate of uncertainty quantification, which is essential in weather forecasting applications. From the perspective of the rank histograms, figure \ref{fig:GAN-histograms}, ecPoint produces a better calibrated ensemble than either GAN method. This is consistent with the design of ecPoint which was configured to produce a well-calibrated ensemble. It is interesting that both GANs approach the calibration of ecPoint, in this plot and the ROC plots, with no explicit training for calibration. \add{Furthermore, when restricted to extreme events, the neural network approaches are roughly as miscalibrated as ecPoint.} \remove{Future work could add the CPRS as an additional content loss term to further improve the calibration.}

To better understand the bounds of our success we also used the same GAN-based architecture to carry out a NIMROD to NIMROD mapping, similar to that of \citeA{Leinonen2020}, where the input field was an average-coarsened version of the NIMROD precipitation field (plus the static fields). For this changed problem we see a significant drop in CRPS (from 0.0856 mm/hr to 0.0230 mm/hr) and improvement in calibration. We interpret these auxiliary results, which are presented in the appendix, as a demonstration that the difference between the IFS and NIMROD datasets in the full problem (i.e., \emph{forecast error}) is the main factor limiting the success of the model. This is caused by the misalignment of fronts and other precipitation events between the two datasets. Future work could explore methods to limit these effects, e.g., pre-processing the dataset to include only well-aligned events.

We believe there are several avenues for future exploration with this work. Foremost is the application of our model to the downstream tasks of flood modelling, where the bias correction and higher resolution could help improve the accuracy of the flood forecast. We would like to investigate the potential of applying the model to post-process \emph{ensembles} of forecasts, which may guide the network further towards an accurate assessment of forecast error. \citeA{Price2022} incorporated this data in their approach, but did so by fixing the number of ensemble members ingested. Using a transformer architecture in the ensemble dimension could be an exciting, ensemble-size-agnostic, approach to try. There may be some advantages to a split approach, in which coarse-scale bias/spread correction is applied before the downscaling step. Further work with more significant computational power could re-visit the temporal aspect, to give temporally consistent downscaled results. We barely explored modifications to the network architecture, and gains could perhaps be made here. We expect our model could be extended to work more generally across different geographical regions. Publicly available datasets could be used to build a downscaling model that could be applied globally, although this would have challenges in areas of the world where reliable truth data is not presently available. More generally, further work will be required to produce an operational-ready product.

\section{Open Research}
The code for the GAN and VAE-GAN models used in this paper is available at \url{https://doi.org/10.5281/zenodo.6922291}. This was adapted from Jussi Leinonen's GAN model, available at \url{https://github.com/jleinonen/downscaling-rnn-gan}. \change{All experiments in this paper were performed within TensorFlow 2.2.0.}{All experiments in this paper were mostly performed within TensorFlow 2.7.0, except the deterministic CNN and the `natural' and `equal' training data ablation studies, which were performed within our older Tensorflow 2.2.0 environment. We did not find any scientific difference between models produced using the different Tensorflow versions.}

The ECMWF forecast archive can be obtained through MARS; more details are available at \url{https://www.ecmwf.int/en/forecasts/access-forecasts/access-archive-datasets}. \add{MARS accounts for academic use are available for free, subject to certain conditions; see} \url{https://www.ecmwf.int/en/forecasts/accessing-forecasts/licences-available}. The NIMROD radar dataset can be obtained through CEDA; more details are available at \url{https://catalogue.ceda.ac.uk/uuid/27dd6ffba67f667a18c62de5c3456350}. \add{A CEDA Archive account is required in order to access this data.}

\acknowledgments
We are grateful to Stephan Rasp, Jussi Leinonen, Suman Ravuri \& his colleagues at DeepMind, Peter Watson, Tim Hewson, Zied Ben Bouallegue, \add{Campbell Watson}, Hannah Christensen, Milan Klöwer, and Fenwick Cooper for many useful conversations and ideas. This project has received funding from the European Research Council (ERC) under the European Union’s Horizon 2020 research and innovation programme (\add{ITHACA}, grant agreement no 741112). Computing resources were provided by the European Weather Cloud, which is an ECMWF and EUMETSAT project. PD gratefully acknowledges funding from the Royal Society for his University Research Fellowship as well as the ESiWACE project funded under Horizon 2020 No. 823988. PD and MC gratefully acknowledge funding from the MAELSTROM EuroHPC-JU project (JU) under No 955513. The JU receives support from the European Union’s Horizon research and innovation programme and United Kingdom, Germany, Italy, Luxembourg, Switzerland, and Norway.

\clearpage

\bibliography{references}

\begin{thebibliography}{}

\bibitem [\protect \citeauthoryear {%
Adewoyin%
, Dueben%
, Watson%
, He%
\BCBL {}\ \BBA {} Dutta%
}{%
Adewoyin%
\ \protect \BOthers {.}}{%
{\protect \APACyear {2021}}%
}]{%
Adewoyin2021TRU-Net}
\APACinsertmetastar {%
Adewoyin2021TRU-Net}%
\begin{APACrefauthors}%
Adewoyin, R\BPBI A.%
, Dueben, P.%
, Watson, P.%
, He, Y.%
\BCBL {}\ \BBA {} Dutta, R.%
\end{APACrefauthors}%
\unskip\
\newblock
\APACrefYearMonthDay{2021}{}{}.
\newblock
{\BBOQ}\APACrefatitle {{TRU-NET: a deep learning approach to high resolution
  prediction of rainfall}} {{TRU-NET: a deep learning approach to high
  resolution prediction of rainfall}}.{\BBCQ}
\newblock
\APACjournalVolNumPages{Machine Learning}{110}{8}{2035--2062}.
\newblock
\begin{APACrefDOI} \doi{10.1007/s10994-021-06022-6} \end{APACrefDOI}
\PrintBackRefs{\CurrentBib}

\bibitem [\protect \citeauthoryear {%
Agrawal%
\ \protect \BOthers {.}}{%
Agrawal%
\ \protect \BOthers {.}}{%
{\protect \APACyear {2019}}%
}]{%
Agrawal2019}
\APACinsertmetastar {%
Agrawal2019}%
\begin{APACrefauthors}%
Agrawal, S.%
, Barrington, L.%
, Bromberg, C.%
, Burge, J.%
, Gazen, C.%
\BCBL {}\ \BBA {} Hickey, J.%
\end{APACrefauthors}%
\unskip\
\newblock
\APACrefYearMonthDay{2019}{}{}.
\newblock
\APACrefbtitle {{Machine Learning for Precipitation Nowcasting from Radar
  Images}.} {{Machine Learning for Precipitation Nowcasting from Radar
  Images}.}
\newblock
\begin{APACrefURL} \url{https://arxiv.org/abs/1912.12132} \end{APACrefURL}
\PrintBackRefs{\CurrentBib}

\bibitem [\protect \citeauthoryear {%
Applequist%
, Gahrs%
, Pfeffer%
\BCBL {}\ \BBA {} Niu%
}{%
Applequist%
\ \protect \BOthers {.}}{%
{\protect \APACyear {2002}}%
}]{%
Applequist2002}
\APACinsertmetastar {%
Applequist2002}%
\begin{APACrefauthors}%
Applequist, S.%
, Gahrs, G\BPBI E.%
, Pfeffer, R\BPBI L.%
\BCBL {}\ \BBA {} Niu, X\BHBI F.%
\end{APACrefauthors}%
\unskip\
\newblock
\APACrefYearMonthDay{2002}{}{}.
\newblock
{\BBOQ}\APACrefatitle {{Comparison of Methodologies for Probabilistic
  Quantitative Precipitation Forecasting}} {{Comparison of Methodologies for
  Probabilistic Quantitative Precipitation Forecasting}}.{\BBCQ}
\newblock
\APACjournalVolNumPages{Weather and Forecasting}{17}{4}{783-799}.
\newblock
\begin{APACrefDOI} \doi{10.1175/1520-0434(2002)017<0783:COMFPQ>2.0.CO;2}
  \end{APACrefDOI}
\PrintBackRefs{\CurrentBib}

\bibitem [\protect \citeauthoryear {%
Arjovsky%
, Chintala%
\BCBL {}\ \BBA {} Bottou%
}{%
Arjovsky%
\ \protect \BOthers {.}}{%
{\protect \APACyear {2017}}%
}]{%
Arjovsky2017}
\APACinsertmetastar {%
Arjovsky2017}%
\begin{APACrefauthors}%
Arjovsky, M.%
, Chintala, S.%
\BCBL {}\ \BBA {} Bottou, L.%
\end{APACrefauthors}%
\unskip\
\newblock
\APACrefYearMonthDay{2017}{}{}.
\newblock
\APACrefbtitle {{Wasserstein GAN}.} {{Wasserstein GAN}.}
\newblock
\begin{APACrefURL} \url{https://arxiv.org/abs/1701.07875} \end{APACrefURL}
\PrintBackRefs{\CurrentBib}

\bibitem [\protect \citeauthoryear {%
Ben~Bouallegue%
\ \BBA {} Richardson%
}{%
Ben~Bouallegue%
\ \BBA {} Richardson%
}{%
{\protect \APACyear {2022}}%
}]{%
Zied2022}
\APACinsertmetastar {%
Zied2022}%
\begin{APACrefauthors}%
Ben~Bouallegue, Z.%
\BCBT {}\ \BBA {} Richardson, D\BPBI S.%
\end{APACrefauthors}%
\unskip\
\newblock
\APACrefYearMonthDay{2022}{}{}.
\newblock
{\BBOQ}\APACrefatitle {{On the ROC Area of Ensemble Forecasts for Rare Events}}
  {{On the ROC Area of Ensemble Forecasts for Rare Events}}.{\BBCQ}
\newblock
\APACjournalVolNumPages{Weather and Forecasting}{To appear}{}{}.
\newblock
\APACrefnote{Available at
  \url{https://www.preprints.org/manuscript/202111.0535/v1}}
\PrintBackRefs{\CurrentBib}

\bibitem [\protect \citeauthoryear {%
Berrocal%
, Raftery%
\BCBL {}\ \BBA {} Gneiting%
}{%
Berrocal%
\ \protect \BOthers {.}}{%
{\protect \APACyear {2008}}%
}]{%
Berrocal2008}
\APACinsertmetastar {%
Berrocal2008}%
\begin{APACrefauthors}%
Berrocal, V\BPBI J.%
, Raftery, A\BPBI E.%
\BCBL {}\ \BBA {} Gneiting, T.%
\end{APACrefauthors}%
\unskip\
\newblock
\APACrefYearMonthDay{2008}{}{}.
\newblock
{\BBOQ}\APACrefatitle {{Probabilistic quantitative precipitation field
  forecasting using a two-stage spatial model}} {{Probabilistic quantitative
  precipitation field forecasting using a two-stage spatial model}}.{\BBCQ}
\newblock
\APACjournalVolNumPages{The Annals of Applied Statistics}{2}{4}{1170--1193}.
\newblock
\begin{APACrefDOI} \doi{10.1214/08-AOAS203} \end{APACrefDOI}
\PrintBackRefs{\CurrentBib}

\bibitem [\protect \citeauthoryear {%
Bihlo%
}{%
Bihlo%
}{%
{\protect \APACyear {2020}}%
}]{%
Bihlo2020}
\APACinsertmetastar {%
Bihlo2020}%
\begin{APACrefauthors}%
Bihlo, A.%
\end{APACrefauthors}%
\unskip\
\newblock
\APACrefYearMonthDay{2020}{}{}.
\newblock
\APACrefbtitle {{A generative adversarial network approach to (ensemble)
  weather prediction}.} {{A generative adversarial network approach to
  (ensemble) weather prediction}.}
\newblock
\begin{APACrefURL} \url{https://arxiv.org/abs/2006.07718} \end{APACrefURL}
\PrintBackRefs{\CurrentBib}

\bibitem [\protect \citeauthoryear {%
Dong%
, Loy%
, He%
\BCBL {}\ \BBA {} Tang%
}{%
Dong%
\ \protect \BOthers {.}}{%
{\protect \APACyear {2015}}%
}]{%
Dong2015}
\APACinsertmetastar {%
Dong2015}%
\begin{APACrefauthors}%
Dong, C.%
, Loy, C\BPBI C.%
, He, K.%
\BCBL {}\ \BBA {} Tang, X.%
\end{APACrefauthors}%
\unskip\
\newblock
\APACrefYearMonthDay{2015}{}{}.
\newblock
\APACrefbtitle {{Image Super-Resolution Using Deep Convolutional Networks}.}
  {{Image Super-Resolution Using Deep Convolutional Networks}.}
\newblock
\begin{APACrefURL} \url{https://arxiv.org/abs/1501.00092} \end{APACrefURL}
\PrintBackRefs{\CurrentBib}

\bibitem [\protect \citeauthoryear {%
D'Onofrio%
, Palazzi%
, von Hardenberg%
, Provenzale%
\BCBL {}\ \BBA {} Calmanti%
}{%
D'Onofrio%
\ \protect \BOthers {.}}{%
{\protect \APACyear {2014}}%
}]{%
DOnofrio2014}
\APACinsertmetastar {%
DOnofrio2014}%
\begin{APACrefauthors}%
D'Onofrio, D.%
, Palazzi, E.%
, von Hardenberg, J.%
, Provenzale, A.%
\BCBL {}\ \BBA {} Calmanti, S.%
\end{APACrefauthors}%
\unskip\
\newblock
\APACrefYearMonthDay{2014}{}{}.
\newblock
{\BBOQ}\APACrefatitle {{Stochastic Rainfall Downscaling of Climate Models}}
  {{Stochastic Rainfall Downscaling of Climate Models}}.{\BBCQ}
\newblock
\APACjournalVolNumPages{Journal of Hydrometeorology}{15}{2}{830--843}.
\newblock
\begin{APACrefDOI} \doi{10.1175/JHM-D-13-096.1} \end{APACrefDOI}
\PrintBackRefs{\CurrentBib}

\bibitem [\protect \citeauthoryear {%
Duc%
, Saito%
\BCBL {}\ \BBA {} Seko%
}{%
Duc%
\ \protect \BOthers {.}}{%
{\protect \APACyear {2013}}%
}]{%
Duc2013}
\APACinsertmetastar {%
Duc2013}%
\begin{APACrefauthors}%
Duc, L.%
, Saito, K.%
\BCBL {}\ \BBA {} Seko, H.%
\end{APACrefauthors}%
\unskip\
\newblock
\APACrefYearMonthDay{2013}{}{}.
\newblock
{\BBOQ}\APACrefatitle {{Spatial-temporal fractions verification for
  high-resolution ensemble forecasts}} {{Spatial-temporal fractions
  verification for high-resolution ensemble forecasts}}.{\BBCQ}
\newblock
\APACjournalVolNumPages{Tellus A: Dynamic Meteorology and
  Oceanography}{65}{18171}{1--22}.
\newblock
\begin{APACrefDOI} \doi{10.3402/tellusa.v65i0.18171} \end{APACrefDOI}
\PrintBackRefs{\CurrentBib}

\bibitem [\protect \citeauthoryear {%
Espeholt%
\ \protect \BOthers {.}}{%
Espeholt%
\ \protect \BOthers {.}}{%
{\protect \APACyear {2021}}%
}]{%
Espeholt2021}
\APACinsertmetastar {%
Espeholt2021}%
\begin{APACrefauthors}%
Espeholt, L.%
, Agrawal, S.%
, Sønderby, C.%
, Kumar, M.%
, Heek, J.%
, Bromberg, C.%
\BDBL {}Kalchbrenner, N.%
\end{APACrefauthors}%
\unskip\
\newblock
\APACrefYearMonthDay{2021}{}{}.
\newblock
\APACrefbtitle {{Skillful Twelve Hour Precipitation Forecasts using Large
  Context Neural Networks}.} {{Skillful Twelve Hour Precipitation Forecasts
  using Large Context Neural Networks}.}
\newblock
\begin{APACrefURL} \url{https://arxiv.org/abs/2111.07470} \end{APACrefURL}
\PrintBackRefs{\CurrentBib}

\bibitem [\protect \citeauthoryear {%
Feser%
, Rockel%
, von Storch%
, Winterfeldt%
\BCBL {}\ \BBA {} Zahn%
}{%
Feser%
\ \protect \BOthers {.}}{%
{\protect \APACyear {2011}}%
}]{%
Feser2011}
\APACinsertmetastar {%
Feser2011}%
\begin{APACrefauthors}%
Feser, F.%
, Rockel, B.%
, von Storch, H.%
, Winterfeldt, J.%
\BCBL {}\ \BBA {} Zahn, M.%
\end{APACrefauthors}%
\unskip\
\newblock
\APACrefYearMonthDay{2011}{}{}.
\newblock
{\BBOQ}\APACrefatitle {{Regional Climate Models Add Value to Global Model Data:
  A Review and Selected Examples}} {{Regional Climate Models Add Value to
  Global Model Data: A Review and Selected Examples}}.{\BBCQ}
\newblock
\APACjournalVolNumPages{Bulletin of the American Meteorological
  Society}{92}{9}{1181--1192}.
\newblock
\begin{APACrefDOI} \doi{10.1175/2011BAMS3061.1} \end{APACrefDOI}
\PrintBackRefs{\CurrentBib}

\bibitem [\protect \citeauthoryear {%
Gascón%
, Hewson%
\BCBL {}\ \BBA {} Haiden%
}{%
Gascón%
\ \protect \BOthers {.}}{%
{\protect \APACyear {2018}}%
}]{%
Gascon2018}
\APACinsertmetastar {%
Gascon2018}%
\begin{APACrefauthors}%
Gascón, E.%
, Hewson, T.%
\BCBL {}\ \BBA {} Haiden, T.%
\end{APACrefauthors}%
\unskip\
\newblock
\APACrefYearMonthDay{2018}{}{}.
\newblock
{\BBOQ}\APACrefatitle {{Improving Predictions of Precipitation Type at the
  Surface: Description and Verification of Two New Products from the ECMWF
  Ensemble}} {{Improving Predictions of Precipitation Type at the Surface:
  Description and Verification of Two New Products from the ECMWF
  Ensemble}}.{\BBCQ}
\newblock
\APACjournalVolNumPages{Weather and Forecasting}{33}{1}{89--108}.
\newblock
\begin{APACrefDOI} \doi{10.1175/WAF-D-17-0114.1} \end{APACrefDOI}
\PrintBackRefs{\CurrentBib}

\bibitem [\protect \citeauthoryear {%
Gneiting%
\ \BBA {} Raftery%
}{%
Gneiting%
\ \BBA {} Raftery%
}{%
{\protect \APACyear {2007}}%
}]{%
Gneiting2007}
\APACinsertmetastar {%
Gneiting2007}%
\begin{APACrefauthors}%
Gneiting, T.%
\BCBT {}\ \BBA {} Raftery, A\BPBI E.%
\end{APACrefauthors}%
\unskip\
\newblock
\APACrefYearMonthDay{2007}{}{}.
\newblock
{\BBOQ}\APACrefatitle {{Strictly Proper Scoring Rules, Prediction, and
  Estimation}} {{Strictly Proper Scoring Rules, Prediction, and
  Estimation}}.{\BBCQ}
\newblock
\APACjournalVolNumPages{Journal of the American Statistical
  Association}{102}{477}{359--378}.
\newblock
\begin{APACrefDOI} \doi{10.1198/016214506000001437} \end{APACrefDOI}
\PrintBackRefs{\CurrentBib}

\bibitem [\protect \citeauthoryear {%
Gong%
\ \protect \BOthers {.}}{%
Gong%
\ \protect \BOthers {.}}{%
{\protect \APACyear {2022}}%
}]{%
Gong2022}
\APACinsertmetastar {%
Gong2022}%
\begin{APACrefauthors}%
Gong, B.%
, Langguth, M.%
, Ji, Y.%
, Mozaffari, A.%
, Stadtler, S.%
, Mache, K.%
\BCBL {}\ \BBA {} Schultz, M\BPBI G.%
\end{APACrefauthors}%
\unskip\
\newblock
\APACrefYearMonthDay{2022}{}{}.
\newblock
{\BBOQ}\APACrefatitle {{Temperature forecasting by deep learning methods}}
  {{Temperature forecasting by deep learning methods}}.{\BBCQ}
\newblock
\APACjournalVolNumPages{Geoscientific Model Development
  Discussions}{2022}{}{1--35}.
\newblock
\begin{APACrefDOI} \doi{10.5194/gmd-2021-430} \end{APACrefDOI}
\PrintBackRefs{\CurrentBib}

\bibitem [\protect \citeauthoryear {%
Goodfellow%
\ \protect \BOthers {.}}{%
Goodfellow%
\ \protect \BOthers {.}}{%
{\protect \APACyear {2014}}%
}]{%
Goodfellow2014}
\APACinsertmetastar {%
Goodfellow2014}%
\begin{APACrefauthors}%
Goodfellow, I\BPBI J.%
, Pouget-Abadie, J.%
, Mirza, M.%
, Xu, B.%
, Warde-Farley, D.%
, Ozair, S.%
\BDBL {}Bengio, Y.%
\end{APACrefauthors}%
\unskip\
\newblock
\APACrefYearMonthDay{2014}{}{}.
\newblock
\APACrefbtitle {{Generative Adversarial Networks}.} {{Generative Adversarial
  Networks}.}
\newblock
\begin{APACrefURL} \url{https://arxiv.org/abs/1406.2661} \end{APACrefURL}
\PrintBackRefs{\CurrentBib}

\bibitem [\protect \citeauthoryear {%
Gulrajani%
, Ahmed%
, Arjovsky%
, Dumoulin%
\BCBL {}\ \BBA {} Courville%
}{%
Gulrajani%
\ \protect \BOthers {.}}{%
{\protect \APACyear {2017}}%
}]{%
Gulrajani2017}
\APACinsertmetastar {%
Gulrajani2017}%
\begin{APACrefauthors}%
Gulrajani, I.%
, Ahmed, F.%
, Arjovsky, M.%
, Dumoulin, V.%
\BCBL {}\ \BBA {} Courville, A.%
\end{APACrefauthors}%
\unskip\
\newblock
\APACrefYearMonthDay{2017}{}{}.
\newblock
\APACrefbtitle {{Improved Training of Wasserstein GANs}.} {{Improved Training
  of Wasserstein GANs}.}
\newblock
\begin{APACrefURL} \url{https://arxiv.org/abs/1704.00028} \end{APACrefURL}
\PrintBackRefs{\CurrentBib}

\bibitem [\protect \citeauthoryear {%
Hamill%
}{%
Hamill%
}{%
{\protect \APACyear {2000}}%
}]{%
Hamill2000}
\APACinsertmetastar {%
Hamill2000}%
\begin{APACrefauthors}%
Hamill, T.%
\end{APACrefauthors}%
\unskip\
\newblock
\APACrefYearMonthDay{2000}{}{}.
\newblock
{\BBOQ}\APACrefatitle {{Interpretation of Rank Histograms for Verifying
  Ensemble Forecasts}} {{Interpretation of Rank Histograms for Verifying
  Ensemble Forecasts}}.{\BBCQ}
\newblock
\APACjournalVolNumPages{Monthly Weather Review}{129}{3}{550--560}.
\newblock
\begin{APACrefDOI} \doi{10.1175/1520-0493(2001)129<0550:IORHFV>2.0.CO;2}
  \end{APACrefDOI}
\PrintBackRefs{\CurrentBib}

\bibitem [\protect \citeauthoryear {%
He%
, Zhang%
, Ren%
\BCBL {}\ \BBA {} Sun%
}{%
He%
\ \protect \BOthers {.}}{%
{\protect \APACyear {2015}}%
}]{%
He2015}
\APACinsertmetastar {%
He2015}%
\begin{APACrefauthors}%
He, K.%
, Zhang, X.%
, Ren, S.%
\BCBL {}\ \BBA {} Sun, J.%
\end{APACrefauthors}%
\unskip\
\newblock
\APACrefYearMonthDay{2015}{}{}.
\newblock
\APACrefbtitle {{Deep Residual Learning for Image Recognition}.} {{Deep
  Residual Learning for Image Recognition}.}
\newblock
\begin{APACrefURL} \url{https://arxiv.org/abs/1512.03385} \end{APACrefURL}
\PrintBackRefs{\CurrentBib}

\bibitem [\protect \citeauthoryear {%
Hersbach%
}{%
Hersbach%
}{%
{\protect \APACyear {2000}}%
}]{%
Hersbach2000}
\APACinsertmetastar {%
Hersbach2000}%
\begin{APACrefauthors}%
Hersbach, H.%
\end{APACrefauthors}%
\unskip\
\newblock
\APACrefYearMonthDay{2000}{10}{}.
\newblock
{\BBOQ}\APACrefatitle {{Decomposition of the Continuous Ranked Probability
  Score for Ensemble Prediction Systems}} {{Decomposition of the Continuous
  Ranked Probability Score for Ensemble Prediction Systems}}.{\BBCQ}
\newblock
\APACjournalVolNumPages{Weather and Forecasting}{15}{}{559-570}.
\newblock
\begin{APACrefDOI} \doi{10.1175/1520-0434(2000)} \end{APACrefDOI}
\PrintBackRefs{\CurrentBib}

\bibitem [\protect \citeauthoryear {%
Hess%
\ \BBA {} Boers%
}{%
Hess%
\ \BBA {} Boers%
}{%
{\protect \APACyear {2022}}%
}]{%
Hess2022}
\APACinsertmetastar {%
Hess2022}%
\begin{APACrefauthors}%
Hess, P.%
\BCBT {}\ \BBA {} Boers, N.%
\end{APACrefauthors}%
\unskip\
\newblock
\APACrefYearMonthDay{2022}{}{}.
\newblock
{\BBOQ}\APACrefatitle {{Deep Learning for Improving Numerical Weather
  Prediction of Heavy Rainfall}} {{Deep Learning for Improving Numerical
  Weather Prediction of Heavy Rainfall}}.{\BBCQ}
\newblock
\APACjournalVolNumPages{Journal of Advances in Modeling Earth
  Systems}{14}{3}{e2021MS002765}.
\newblock
\begin{APACrefDOI} \doi{10.1029/2021MS002765} \end{APACrefDOI}
\PrintBackRefs{\CurrentBib}

\bibitem [\protect \citeauthoryear {%
Hewson%
\ \BBA {} Pillosu%
}{%
Hewson%
\ \BBA {} Pillosu%
}{%
{\protect \APACyear {2021}}%
}]{%
ecPoint}
\APACinsertmetastar {%
ecPoint}%
\begin{APACrefauthors}%
Hewson, T\BPBI D.%
\BCBT {}\ \BBA {} Pillosu, F\BPBI M.%
\end{APACrefauthors}%
\unskip\
\newblock
\APACrefYearMonthDay{2021}{}{}.
\newblock
{\BBOQ}\APACrefatitle {{A low-cost post-processing technique improves weather
  forecasts around the world}} {{A low-cost post-processing technique improves
  weather forecasts around the world}}.{\BBCQ}
\newblock
\APACjournalVolNumPages{Communications Earth \& Environment}{2}{132}{1--10}.
\newblock
\begin{APACrefDOI} \doi{10.1038/s43247-021-00185-9} \end{APACrefDOI}
\PrintBackRefs{\CurrentBib}

\bibitem [\protect \citeauthoryear {%
Holden%
, Abatzoglou%
, Luce%
\BCBL {}\ \BBA {} Baggett%
}{%
Holden%
\ \protect \BOthers {.}}{%
{\protect \APACyear {2011}}%
}]{%
Holden2011}
\APACinsertmetastar {%
Holden2011}%
\begin{APACrefauthors}%
Holden, Z\BPBI A.%
, Abatzoglou, J\BPBI T.%
, Luce, C\BPBI H.%
\BCBL {}\ \BBA {} Baggett, L\BPBI S.%
\end{APACrefauthors}%
\unskip\
\newblock
\APACrefYearMonthDay{2011}{}{}.
\newblock
{\BBOQ}\APACrefatitle {{Empirical downscaling of daily minimum air temperature
  at very fine resolutions in complex terrain}} {{Empirical downscaling of
  daily minimum air temperature at very fine resolutions in complex
  terrain}}.{\BBCQ}
\newblock
\APACjournalVolNumPages{Agricultural and Forest
  Meteorology}{151}{8}{1066-1073}.
\newblock
\begin{APACrefDOI} \doi{10.1016/j.agrformet.2011.03.011} \end{APACrefDOI}
\PrintBackRefs{\CurrentBib}

\bibitem [\protect \citeauthoryear {%
Huang%
}{%
Huang%
}{%
{\protect \APACyear {2020}}%
}]{%
Huang2020}
\APACinsertmetastar {%
Huang2020}%
\begin{APACrefauthors}%
Huang, X.%
\end{APACrefauthors}%
\unskip\
\newblock
\APACrefYearMonthDay{2020}{}{}.
\newblock
{\BBOQ}\APACrefatitle {{Deep-learning based climate downscaling using the
  super-resolution method: a case study over the western US}} {{Deep-learning
  based climate downscaling using the super-resolution method: a case study
  over the western US}}.{\BBCQ}
\newblock
\APACjournalVolNumPages{Geoscientific Model Development
  Discussions}{}{}{1--18}.
\newblock
\begin{APACrefDOI} \doi{10.5194/gmd-2020-214} \end{APACrefDOI}
\PrintBackRefs{\CurrentBib}

\bibitem [\protect \citeauthoryear {%
Kingma%
\ \BBA {} Ba%
}{%
Kingma%
\ \BBA {} Ba%
}{%
{\protect \APACyear {2014}}%
}]{%
Kingma2014}
\APACinsertmetastar {%
Kingma2014}%
\begin{APACrefauthors}%
Kingma, D\BPBI P.%
\BCBT {}\ \BBA {} Ba, J.%
\end{APACrefauthors}%
\unskip\
\newblock
\APACrefYearMonthDay{2014}{}{}.
\newblock
\APACrefbtitle {{Adam: A Method for Stochastic Optimization}.} {{Adam: A Method
  for Stochastic Optimization}.}
\newblock
\begin{APACrefURL} \url{https://arxiv.org/abs/1412.6980} \end{APACrefURL}
\PrintBackRefs{\CurrentBib}

\bibitem [\protect \citeauthoryear {%
Klocek%
\ \protect \BOthers {.}}{%
Klocek%
\ \protect \BOthers {.}}{%
{\protect \APACyear {2021}}%
}]{%
Klocek2021}
\APACinsertmetastar {%
Klocek2021}%
\begin{APACrefauthors}%
Klocek, S.%
, Dong, H.%
, Dixon, M.%
, Kanengoni, P.%
, Kazmi, N.%
, Luferenko, P.%
\BDBL {}Xiang, S.%
\end{APACrefauthors}%
\unskip\
\newblock
\APACrefYearMonthDay{2021}{}{}.
\newblock
{\BBOQ}\APACrefatitle {{MS-nowcasting: Operational Precipitation Nowcasting
  with Convolutional LSTMs at Microsoft Weather}} {{MS-nowcasting: Operational
  Precipitation Nowcasting with Convolutional LSTMs at Microsoft
  Weather}}.{\BBCQ}
\newblock
\BIn{} \APACrefbtitle {{NeurIPS 2021 Workshop on Tackling Climate Change with
  Machine Learning}.} {{NeurIPS 2021 Workshop on Tackling Climate Change with
  Machine Learning}.}
\PrintBackRefs{\CurrentBib}

\bibitem [\protect \citeauthoryear {%
Kumar%
\ \protect \BOthers {.}}{%
Kumar%
\ \protect \BOthers {.}}{%
{\protect \APACyear {2021}}%
}]{%
Kumar2021}
\APACinsertmetastar {%
Kumar2021}%
\begin{APACrefauthors}%
Kumar, B.%
, Chattopadhyay, R.%
, Singh, M.%
, Chaudhari, N.%
, Kodari, K.%
\BCBL {}\ \BBA {} Barve, A.%
\end{APACrefauthors}%
\unskip\
\newblock
\APACrefYearMonthDay{2021}{}{}.
\newblock
{\BBOQ}\APACrefatitle {{Deep learning–based downscaling of summer monsoon
  rainfall data over Indian region}} {{Deep learning–based downscaling of
  summer monsoon rainfall data over Indian region}}.{\BBCQ}
\newblock
\APACjournalVolNumPages{Theoretical and Applied
  Climatology}{143}{3}{1145--1156}.
\newblock
\begin{APACrefDOI} \doi{10.1007/s00704-020-03489-6} \end{APACrefDOI}
\PrintBackRefs{\CurrentBib}

\bibitem [\protect \citeauthoryear {%
Kurach%
, Lucic%
, Zhai%
, Michalski%
\BCBL {}\ \BBA {} Gelly%
}{%
Kurach%
\ \protect \BOthers {.}}{%
{\protect \APACyear {2018}}%
}]{%
Kurach2018}
\APACinsertmetastar {%
Kurach2018}%
\begin{APACrefauthors}%
Kurach, K.%
, Lucic, M.%
, Zhai, X.%
, Michalski, M.%
\BCBL {}\ \BBA {} Gelly, S.%
\end{APACrefauthors}%
\unskip\
\newblock
\APACrefYearMonthDay{2018}{}{}.
\newblock
\APACrefbtitle {{A Large-Scale Study on Regularization and Normalization in
  GANs}.} {{A Large-Scale Study on Regularization and Normalization in GANs}.}
\newblock
\begin{APACrefURL} \url{https://arxiv.org/abs/1807.04720} \end{APACrefURL}
\PrintBackRefs{\CurrentBib}

\bibitem [\protect \citeauthoryear {%
Ledig%
\ \protect \BOthers {.}}{%
Ledig%
\ \protect \BOthers {.}}{%
{\protect \APACyear {2017}}%
}]{%
Ledig2017}
\APACinsertmetastar {%
Ledig2017}%
\begin{APACrefauthors}%
Ledig, C.%
, Theis, L.%
, Huszár, F.%
, Caballero, J.%
, Cunningham, A.%
, Acosta, A.%
\BDBL {}Shi, W.%
\end{APACrefauthors}%
\unskip\
\newblock
\APACrefYearMonthDay{2017}{}{}.
\newblock
{\BBOQ}\APACrefatitle {{Photo-Realistic Single Image Super-Resolution Using a
  Generative Adversarial Network}} {{Photo-Realistic Single Image
  Super-Resolution Using a Generative Adversarial Network}}.{\BBCQ}
\newblock
\BIn{} \APACrefbtitle {{2017 IEEE Conference on Computer Vision and Pattern
  Recognition (CVPR)}} {{2017 IEEE Conference on Computer Vision and Pattern
  Recognition (CVPR)}}\ (\BPGS\ 105--114).
\newblock
\begin{APACrefDOI} \doi{10.1109/CVPR.2017.19} \end{APACrefDOI}
\PrintBackRefs{\CurrentBib}

\bibitem [\protect \citeauthoryear {%
Leinonen%
, Nerini%
\BCBL {}\ \BBA {} Berne%
}{%
Leinonen%
\ \protect \BOthers {.}}{%
{\protect \APACyear {2020}}%
}]{%
Leinonen2020}
\APACinsertmetastar {%
Leinonen2020}%
\begin{APACrefauthors}%
Leinonen, J.%
, Nerini, D.%
\BCBL {}\ \BBA {} Berne, A.%
\end{APACrefauthors}%
\unskip\
\newblock
\APACrefYearMonthDay{2020}{}{}.
\newblock
{\BBOQ}\APACrefatitle {{Stochastic Super-Resolution for Downscaling
  Time-Evolving Atmospheric Fields With a Generative Adversarial Network}}
  {{Stochastic Super-Resolution for Downscaling Time-Evolving Atmospheric
  Fields With a Generative Adversarial Network}}.{\BBCQ}
\newblock
\APACjournalVolNumPages{IEEE Transactions on Geoscience and Remote
  Sensing}{59}{9}{7211--7223}.
\newblock
\begin{APACrefDOI} \doi{10.1109/TGRS.2020.3032790} \end{APACrefDOI}
\PrintBackRefs{\CurrentBib}

\bibitem [\protect \citeauthoryear {%
Lin%
, Wu%
, Huang%
, Qiu%
\BCBL {}\ \BBA {} Chen%
}{%
Lin%
\ \protect \BOthers {.}}{%
{\protect \APACyear {2017}}%
}]{%
Lin2017-DCSR}
\APACinsertmetastar {%
Lin2017-DCSR}%
\begin{APACrefauthors}%
Lin, G.%
, Wu, Q.%
, Huang, X.%
, Qiu, L.%
\BCBL {}\ \BBA {} Chen, X.%
\end{APACrefauthors}%
\unskip\
\newblock
\APACrefYearMonthDay{2017}{}{}.
\newblock
{\BBOQ}\APACrefatitle {Deep Convolutional Networks-Based Image
  Super-Resolution} {Deep convolutional networks-based image
  super-resolution}.{\BBCQ}
\newblock
\BIn{} D\BHBI S.~Huang, V.~Bevilacqua, P.~Premaratne\BCBL {}\ \BBA {} P.~Gupta\
  (\BEDS), \APACrefbtitle {{Intelligent Computing Theories and Application}}
  {{Intelligent Computing Theories and Application}}\ (\BPGS\ 338--344).
\newblock
\APACaddressPublisher{Cham}{Springer International Publishing}.
\newblock
\begin{APACrefDOI} \doi{10.1007/978-3-319-63309-1_31} \end{APACrefDOI}
\PrintBackRefs{\CurrentBib}

\bibitem [\protect \citeauthoryear {%
Maas%
, Hannun%
\BCBL {}\ \BBA {} Ng%
}{%
Maas%
\ \protect \BOthers {.}}{%
{\protect \APACyear {2013}}%
}]{%
Maas13}
\APACinsertmetastar {%
Maas13}%
\begin{APACrefauthors}%
Maas, A\BPBI L.%
, Hannun, A\BPBI Y.%
\BCBL {}\ \BBA {} Ng, A\BPBI Y.%
\end{APACrefauthors}%
\unskip\
\newblock
\APACrefYearMonthDay{2013}{}{}.
\newblock
{\BBOQ}\APACrefatitle {{Rectifier Nonlinearities Improve Neural Network
  Acoustic Models}} {{Rectifier Nonlinearities Improve Neural Network Acoustic
  Models}}.{\BBCQ}
\newblock
\BIn{} \APACrefbtitle {{ICML Workshop on Deep Learning for Audio, Speech and
  Language Processing}.} {{ICML Workshop on Deep Learning for Audio, Speech and
  Language Processing}.}
\PrintBackRefs{\CurrentBib}

\bibitem [\protect \citeauthoryear {%
Matheson%
\ \BBA {} Winkler%
}{%
Matheson%
\ \BBA {} Winkler%
}{%
{\protect \APACyear {1976}}%
}]{%
MathesonWinkler1976}
\APACinsertmetastar {%
MathesonWinkler1976}%
\begin{APACrefauthors}%
Matheson, J\BPBI E.%
\BCBT {}\ \BBA {} Winkler, R\BPBI L.%
\end{APACrefauthors}%
\unskip\
\newblock
\APACrefYearMonthDay{1976}{}{}.
\newblock
{\BBOQ}\APACrefatitle {{Scoring Rules for Continuous Probability
  Distributions}} {{Scoring Rules for Continuous Probability
  Distributions}}.{\BBCQ}
\newblock
\APACjournalVolNumPages{Management Science}{22}{10}{1087-1096}.
\PrintBackRefs{\CurrentBib}

\bibitem [\protect \citeauthoryear {%
{Met Office}%
}{%
{Met Office}%
}{%
{\protect \APACyear {2003}}%
}]{%
NIMROD}
\APACinsertmetastar {%
NIMROD}%
\begin{APACrefauthors}%
{Met Office}.%
\end{APACrefauthors}%
\unskip\
\newblock
\APACrefYearMonthDay{2003}{}{}.
\newblock
\APACrefbtitle {{1 km Resolution UK Composite Rainfall Data from the Met Office
  Nimrod System. NCAS British Atmospheric Data Centre}.} {{1 km Resolution UK
  Composite Rainfall Data from the Met Office Nimrod System. NCAS British
  Atmospheric Data Centre}.}
\newblock
\begin{APACrefURL}
  \url{https://catalogue.ceda.ac.uk/uuid/27dd6ffba67f667a18c62de5c3456350}
  \end{APACrefURL}
\newblock
\APACrefnote{Accessed: 2022-03-31}
\PrintBackRefs{\CurrentBib}

\bibitem [\protect \citeauthoryear {%
Mirza%
\ \BBA {} Osindero%
}{%
Mirza%
\ \BBA {} Osindero%
}{%
{\protect \APACyear {2014}}%
}]{%
mirza2014conditional}
\APACinsertmetastar {%
mirza2014conditional}%
\begin{APACrefauthors}%
Mirza, M.%
\BCBT {}\ \BBA {} Osindero, S.%
\end{APACrefauthors}%
\unskip\
\newblock
\APACrefYearMonthDay{2014}{}{}.
\newblock
\APACrefbtitle {{Conditional Generative Adversarial Nets}.} {{Conditional
  Generative Adversarial Nets}.}
\newblock
\begin{APACrefURL} \url{https://arxiv.org/abs/1411.1784} \end{APACrefURL}
\PrintBackRefs{\CurrentBib}

\bibitem [\protect \citeauthoryear {%
Palmer%
}{%
Palmer%
}{%
{\protect \APACyear {2020}}%
{\protect \APACexlab {{\protect \BCnt {1}}}}}]{%
PalmerarXiv}
\APACinsertmetastar {%
PalmerarXiv}%
\begin{APACrefauthors}%
Palmer, T.%
\end{APACrefauthors}%
\unskip\
\newblock
\APACrefYearMonthDay{2020{\protect \BCnt {1}}}{}{}.
\newblock
\APACrefbtitle {{A Vision for Numerical Weather Prediction in 2030}.} {{A
  Vision for Numerical Weather Prediction in 2030}.}
\newblock
\begin{APACrefURL} \url{https://arxiv.org/abs/2007.04830} \end{APACrefURL}
\newblock
\begin{APACrefDOI} \doi{10.48550/ARXIV.2007.04830} \end{APACrefDOI}
\PrintBackRefs{\CurrentBib}

\bibitem [\protect \citeauthoryear {%
Palmer%
}{%
Palmer%
}{%
{\protect \APACyear {2020}}%
{\protect \APACexlab {{\protect \BCnt {2}}}}}]{%
PalmerWhitePaper}
\APACinsertmetastar {%
PalmerWhitePaper}%
\begin{APACrefauthors}%
Palmer, T.%
\end{APACrefauthors}%
\unskip\
\newblock
\APACrefYearMonthDay{2020{\protect \BCnt {2}}}{}{}.
\newblock
\APACrefbtitle {{White Paper One Contributor: Tim Palmer}.} {{White Paper One
  Contributor: Tim Palmer}.}
\newblock
\APAChowpublished {\url{https://ppe-openplatform.wmo.int/en/WP1TP}}.
\newblock
\APACrefnote{Accessed: 2022-07-28}
\PrintBackRefs{\CurrentBib}

\bibitem [\protect \citeauthoryear {%
Price%
\ \BBA {} Rasp%
}{%
Price%
\ \BBA {} Rasp%
}{%
{\protect \APACyear {2022}}%
}]{%
Price2022}
\APACinsertmetastar {%
Price2022}%
\begin{APACrefauthors}%
Price, I.%
\BCBT {}\ \BBA {} Rasp, S.%
\end{APACrefauthors}%
\unskip\
\newblock
\APACrefYearMonthDay{2022}{}{}.
\newblock
\APACrefbtitle {{Increasing the accuracy and resolution of precipitation
  forecasts using deep generative models}.} {{Increasing the accuracy and
  resolution of precipitation forecasts using deep generative models}.}
\newblock
\begin{APACrefURL} \url{https://arxiv.org/abs/2203.12297} \end{APACrefURL}
\PrintBackRefs{\CurrentBib}

\bibitem [\protect \citeauthoryear {%
Pulkkinen%
\ \protect \BOthers {.}}{%
Pulkkinen%
\ \protect \BOthers {.}}{%
{\protect \APACyear {2019}}%
}]{%
pysteps}
\APACinsertmetastar {%
pysteps}%
\begin{APACrefauthors}%
Pulkkinen, S.%
, Nerini, D.%
, Pérez~Hortal, A\BPBI A.%
, Velasco-Forero, C.%
, Seed, A.%
, Germann, U.%
\BCBL {}\ \BBA {} Foresti, L.%
\end{APACrefauthors}%
\unskip\
\newblock
\APACrefYearMonthDay{2019}{}{}.
\newblock
{\BBOQ}\APACrefatitle {{Pysteps: an open-source Python library for
  probabilistic precipitation nowcasting (v1.0)}} {{Pysteps: an open-source
  Python library for probabilistic precipitation nowcasting (v1.0)}}.{\BBCQ}
\newblock
\APACjournalVolNumPages{Geoscientific Model Development}{12}{10}{4185--4219}.
\newblock
\begin{APACrefDOI} \doi{10.5194/gmd-12-4185-2019} \end{APACrefDOI}
\PrintBackRefs{\CurrentBib}

\bibitem [\protect \citeauthoryear {%
Ravuri%
\ \protect \BOthers {.}}{%
Ravuri%
\ \protect \BOthers {.}}{%
{\protect \APACyear {2021}}%
}]{%
deepmind_2021}
\APACinsertmetastar {%
deepmind_2021}%
\begin{APACrefauthors}%
Ravuri, S.%
, Lenc, K.%
, Willson, M.%
, Kangin, D.%
, Lam, R.%
, Mirowski, P.%
\BDBL {}Mohamed, S.%
\end{APACrefauthors}%
\unskip\
\newblock
\APACrefYearMonthDay{2021}{}{}.
\newblock
{\BBOQ}\APACrefatitle {{Skilful precipitation nowcasting using deep generative
  models of radar}} {{Skilful precipitation nowcasting using deep generative
  models of radar}}.{\BBCQ}
\newblock
\APACjournalVolNumPages{Nature}{597}{}{672--677}.
\newblock
\begin{APACrefDOI} \doi{10.1038/s41586-021-03854-z} \end{APACrefDOI}
\PrintBackRefs{\CurrentBib}

\bibitem [\protect \citeauthoryear {%
Rebora%
, Ferraris%
, von Hardenberg%
\BCBL {}\ \BBA {} Provenzale%
}{%
Rebora%
\ \protect \BOthers {.}}{%
{\protect \APACyear {2006}}%
}]{%
rebora2006rainfarm}
\APACinsertmetastar {%
rebora2006rainfarm}%
\begin{APACrefauthors}%
Rebora, N.%
, Ferraris, L.%
, von Hardenberg, J.%
\BCBL {}\ \BBA {} Provenzale, A.%
\end{APACrefauthors}%
\unskip\
\newblock
\APACrefYearMonthDay{2006}{}{}.
\newblock
{\BBOQ}\APACrefatitle {{RainFARM: Rainfall Downscaling by a Filtered
  Autoregressive Model}} {{RainFARM: Rainfall Downscaling by a Filtered
  Autoregressive Model}}.{\BBCQ}
\newblock
\APACjournalVolNumPages{Journal of Hydrometeorology}{7}{4}{724--738}.
\newblock
\begin{APACrefDOI} \doi{10.1175/JHM517.1} \end{APACrefDOI}
\PrintBackRefs{\CurrentBib}

\bibitem [\protect \citeauthoryear {%
N.~Roberts%
}{%
N.~Roberts%
}{%
{\protect \APACyear {2008}}%
}]{%
Roberts2008}
\APACinsertmetastar {%
Roberts2008}%
\begin{APACrefauthors}%
Roberts, N.%
\end{APACrefauthors}%
\unskip\
\newblock
\APACrefYearMonthDay{2008}{}{}.
\newblock
{\BBOQ}\APACrefatitle {{Assessing the spatial and temporal variation in the
  skill of precipitation forecasts from an NWP model}} {{Assessing the spatial
  and temporal variation in the skill of precipitation forecasts from an NWP
  model}}.{\BBCQ}
\newblock
\APACjournalVolNumPages{Meteorological Applications}{15}{1}{163--169}.
\newblock
\begin{APACrefDOI} \doi{10.1002/met.57} \end{APACrefDOI}
\PrintBackRefs{\CurrentBib}

\bibitem [\protect \citeauthoryear {%
N\BPBI M.~Roberts%
\ \BBA {} Lean%
}{%
N\BPBI M.~Roberts%
\ \BBA {} Lean%
}{%
{\protect \APACyear {2008}}%
}]{%
RobertsLean2008}
\APACinsertmetastar {%
RobertsLean2008}%
\begin{APACrefauthors}%
Roberts, N\BPBI M.%
\BCBT {}\ \BBA {} Lean, H\BPBI W.%
\end{APACrefauthors}%
\unskip\
\newblock
\APACrefYearMonthDay{2008}{}{}.
\newblock
{\BBOQ}\APACrefatitle {{Scale-Selective Verification of Rainfall Accumulations
  from High-Resolution Forecasts of Convective Events}} {{Scale-Selective
  Verification of Rainfall Accumulations from High-Resolution Forecasts of
  Convective Events}}.{\BBCQ}
\newblock
\APACjournalVolNumPages{Monthly Weather Review}{136}{1}{78--97}.
\newblock
\begin{APACrefDOI} \doi{10.1175/2007MWR2123.1} \end{APACrefDOI}
\PrintBackRefs{\CurrentBib}

\bibitem [\protect \citeauthoryear {%
Rossa%
, Nurmi%
\BCBL {}\ \BBA {} Ebert%
}{%
Rossa%
\ \protect \BOthers {.}}{%
{\protect \APACyear {2008}}%
}]{%
double_penalty}
\APACinsertmetastar {%
double_penalty}%
\begin{APACrefauthors}%
Rossa, A.%
, Nurmi, P.%
\BCBL {}\ \BBA {} Ebert, E.%
\end{APACrefauthors}%
\unskip\
\newblock
\APACrefYearMonthDay{2008}{}{}.
\newblock
{\BBOQ}\APACrefatitle {{Overview of methods for the verification of
  quantitative precipitation forecasts}} {{Overview of methods for the
  verification of quantitative precipitation forecasts}}.{\BBCQ}
\newblock
\BIn{} S.~Michaelides\ (\BED), \APACrefbtitle {{Precipitation: Advances in
  Measurement, Estimation and Prediction}} {{Precipitation: Advances in
  Measurement, Estimation and Prediction}}\ (\BPGS\ 419--452).
\newblock
\APACaddressPublisher{Berlin, Heidelberg}{Springer Berlin Heidelberg}.
\newblock
\begin{APACrefDOI} \doi{10.1007/978-3-540-77655-0_16} \end{APACrefDOI}
\PrintBackRefs{\CurrentBib}

\bibitem [\protect \citeauthoryear {%
Ruzanski%
\ \BBA {} Chandrasekar%
}{%
Ruzanski%
\ \BBA {} Chandrasekar%
}{%
{\protect \APACyear {2011}}%
}]{%
ruzanski_scale_2011}
\APACinsertmetastar {%
ruzanski_scale_2011}%
\begin{APACrefauthors}%
Ruzanski, E.%
\BCBT {}\ \BBA {} Chandrasekar, V.%
\end{APACrefauthors}%
\unskip\
\newblock
\APACrefYearMonthDay{2011}{}{}.
\newblock
{\BBOQ}\APACrefatitle {{Scale Filtering for Improved Nowcasting Performance in
  a High-Resolution X-Band Radar Network}} {{Scale Filtering for Improved
  Nowcasting Performance in a High-Resolution X-Band Radar Network}}.{\BBCQ}
\newblock
\APACjournalVolNumPages{IEEE Transactions on Geoscience and Remote
  Sensing}{49}{6}{2296--2307}.
\newblock
\begin{APACrefDOI} \doi{10.1109/TGRS.2010.2103946} \end{APACrefDOI}
\PrintBackRefs{\CurrentBib}

\bibitem [\protect \citeauthoryear {%
Saito%
\ \BBA {} Rehmsmeier%
}{%
Saito%
\ \BBA {} Rehmsmeier%
}{%
{\protect \APACyear {2015}}%
}]{%
saito2015precision}
\APACinsertmetastar {%
saito2015precision}%
\begin{APACrefauthors}%
Saito, T.%
\BCBT {}\ \BBA {} Rehmsmeier, M.%
\end{APACrefauthors}%
\unskip\
\newblock
\APACrefYearMonthDay{2015}{}{}.
\newblock
{\BBOQ}\APACrefatitle {{The Precision-Recall Plot Is More Informative than the
  ROC Plot When Evaluating Binary Classifiers on Imbalanced Datasets}} {{The
  Precision-Recall Plot Is More Informative than the ROC Plot When Evaluating
  Binary Classifiers on Imbalanced Datasets}}.{\BBCQ}
\newblock
\APACjournalVolNumPages{PLOS ONE}{10}{3}{1--21}.
\newblock
\begin{APACrefDOI} \doi{10.1371/journal.pone.0118432} \end{APACrefDOI}
\PrintBackRefs{\CurrentBib}

\bibitem [\protect \citeauthoryear {%
Sha%
, Gagne~II%
, West%
\BCBL {}\ \BBA {} Stull%
}{%
Sha%
\ \protect \BOthers {.}}{%
{\protect \APACyear {2020}}%
}]{%
Sha2020}
\APACinsertmetastar {%
Sha2020}%
\begin{APACrefauthors}%
Sha, Y.%
, Gagne~II, D\BPBI J.%
, West, G.%
\BCBL {}\ \BBA {} Stull, R.%
\end{APACrefauthors}%
\unskip\
\newblock
\APACrefYearMonthDay{2020}{}{}.
\newblock
{\BBOQ}\APACrefatitle {{Deep-Learning-Based Gridded Downscaling of Surface
  Meteorological Variables in Complex Terrain. Part II: Daily Precipitation}}
  {{Deep-Learning-Based Gridded Downscaling of Surface Meteorological Variables
  in Complex Terrain. Part II: Daily Precipitation}}.{\BBCQ}
\newblock
\APACjournalVolNumPages{Journal of Applied Meteorology and
  Climatology}{59}{12}{2075--2092}.
\newblock
\begin{APACrefDOI} \doi{10.1175/JAMC-D-20-0058.1} \end{APACrefDOI}
\PrintBackRefs{\CurrentBib}

\bibitem [\protect \citeauthoryear {%
Shi%
\ \protect \BOthers {.}}{%
Shi%
\ \protect \BOthers {.}}{%
{\protect \APACyear {2015}}%
}]{%
ShiChen2015}
\APACinsertmetastar {%
ShiChen2015}%
\begin{APACrefauthors}%
Shi, X.%
, Chen, Z.%
, Wang, H.%
, Yeung, D\BHBI Y.%
, Wong, W\BHBI k.%
\BCBL {}\ \BBA {} Woo, W\BHBI c.%
\end{APACrefauthors}%
\unskip\
\newblock
\APACrefYearMonthDay{2015}{}{}.
\newblock
{\BBOQ}\APACrefatitle {{Convolutional LSTM Network: A Machine Learning Approach
  for Precipitation Nowcasting}} {{Convolutional LSTM Network: A Machine
  Learning Approach for Precipitation Nowcasting}}.{\BBCQ}
\newblock
\BIn{} \APACrefbtitle {{Proceedings of the 28th International Conference on
  Neural Information Processing Systems - Volume 1}} {{Proceedings of the 28th
  International Conference on Neural Information Processing Systems - Volume
  1}}\ (\BPGS\ 802--810).
\newblock
\APACaddressPublisher{Cambridge, MA, USA}{MIT Press}.
\PrintBackRefs{\CurrentBib}

\bibitem [\protect \citeauthoryear {%
Sønderby%
\ \protect \BOthers {.}}{%
Sønderby%
\ \protect \BOthers {.}}{%
{\protect \APACyear {2020}}%
}]{%
Sonderby2020}
\APACinsertmetastar {%
Sonderby2020}%
\begin{APACrefauthors}%
Sønderby, C\BPBI K.%
, Espeholt, L.%
, Heek, J.%
, Dehghani, M.%
, Oliver, A.%
, Salimans, T.%
\BDBL {}Kalchbrenner, N.%
\end{APACrefauthors}%
\unskip\
\newblock
\APACrefYearMonthDay{2020}{}{}.
\newblock
\APACrefbtitle {{MetNet: A Neural Weather Model for Precipitation
  Forecasting}.} {{MetNet: A Neural Weather Model for Precipitation
  Forecasting}.}
\newblock
\begin{APACrefURL} \url{https://arxiv.org/abs/2003.12140} \end{APACrefURL}
\PrintBackRefs{\CurrentBib}

\bibitem [\protect \citeauthoryear {%
Turkowski%
}{%
Turkowski%
}{%
{\protect \APACyear {1990}}%
}]{%
Turkowski1990}
\APACinsertmetastar {%
Turkowski1990}%
\begin{APACrefauthors}%
Turkowski, K.%
\end{APACrefauthors}%
\unskip\
\newblock
\APACrefYearMonthDay{1990}{}{}.
\newblock
{\BBOQ}\APACrefatitle {{Filters for Common Resampling Tasks}} {{Filters for
  Common Resampling Tasks}}.{\BBCQ}
\newblock
\BIn{} A\BPBI S.~Glassner\ (\BED), \APACrefbtitle {{Graphics Gems}} {{Graphics
  Gems}}\ (\BPG~147-165).
\newblock
\APACaddressPublisher{San Diego}{Morgan Kaufmann}.
\newblock
\begin{APACrefDOI} \doi{10.1016/B978-0-08-050753-8.50042-5} \end{APACrefDOI}
\PrintBackRefs{\CurrentBib}

\bibitem [\protect \citeauthoryear {%
Vaughan%
, Tebbutt%
, Hosking%
\BCBL {}\ \BBA {} Turner%
}{%
Vaughan%
\ \protect \BOthers {.}}{%
{\protect \APACyear {2022}}%
}]{%
vaughan2022convolutional}
\APACinsertmetastar {%
vaughan2022convolutional}%
\begin{APACrefauthors}%
Vaughan, A.%
, Tebbutt, W.%
, Hosking, J\BPBI S.%
\BCBL {}\ \BBA {} Turner, R\BPBI E.%
\end{APACrefauthors}%
\unskip\
\newblock
\APACrefYearMonthDay{2022}{}{}.
\newblock
{\BBOQ}\APACrefatitle {{Convolutional conditional neural processes for local
  climate downscaling}} {{Convolutional conditional neural processes for local
  climate downscaling}}.{\BBCQ}
\newblock
\APACjournalVolNumPages{Geoscientific Model Development}{15}{1}{251--268}.
\newblock
\begin{APACrefDOI} \doi{10.5194/gmd-15-251-2022} \end{APACrefDOI}
\PrintBackRefs{\CurrentBib}

\bibitem [\protect \citeauthoryear {%
F.~Wang%
, Tian%
, Lowe%
, Kalin%
\BCBL {}\ \BBA {} Lehrter%
}{%
F.~Wang%
\ \protect \BOthers {.}}{%
{\protect \APACyear {2021}}%
}]{%
Wang2021}
\APACinsertmetastar {%
Wang2021}%
\begin{APACrefauthors}%
Wang, F.%
, Tian, D.%
, Lowe, L.%
, Kalin, L.%
\BCBL {}\ \BBA {} Lehrter, J.%
\end{APACrefauthors}%
\unskip\
\newblock
\APACrefYearMonthDay{2021}{}{}.
\newblock
{\BBOQ}\APACrefatitle {{Deep Learning for Daily Precipitation and Temperature
  Downscaling}} {{Deep Learning for Daily Precipitation and Temperature
  Downscaling}}.{\BBCQ}
\newblock
\APACjournalVolNumPages{Water Resources Research}{57}{4}{e2020WR029308}.
\newblock
\begin{APACrefDOI} \doi{10.1029/2020WR029308} \end{APACrefDOI}
\PrintBackRefs{\CurrentBib}

\bibitem [\protect \citeauthoryear {%
Z.~Wang%
, Simoncelli%
\BCBL {}\ \BBA {} Bovik%
}{%
Z.~Wang%
\ \protect \BOthers {.}}{%
{\protect \APACyear {2003}}%
}]{%
Wang2003MSSSIM}
\APACinsertmetastar {%
Wang2003MSSSIM}%
\begin{APACrefauthors}%
Wang, Z.%
, Simoncelli, E\BPBI P.%
\BCBL {}\ \BBA {} Bovik, A\BPBI C.%
\end{APACrefauthors}%
\unskip\
\newblock
\APACrefYearMonthDay{2003}{}{}.
\newblock
{\BBOQ}\APACrefatitle {{Multiscale structural similarity for image quality
  assessment}} {{Multiscale structural similarity for image quality
  assessment}}.{\BBCQ}
\newblock
\BIn{} \APACrefbtitle {{The Thirty-Seventh Asilomar Conference on Signals,
  Systems \& Computers, 2003}} {{The Thirty-Seventh Asilomar Conference on
  Signals, Systems \& Computers, 2003}}\ (\BVOL~2, \BPGS\ 1398--1402 Vol.2).
\newblock
\begin{APACrefDOI} \doi{10.1109/ACSSC.2003.1292216} \end{APACrefDOI}
\PrintBackRefs{\CurrentBib}

\bibitem [\protect \citeauthoryear {%
Watson%
, Wang%
, Lynar%
\BCBL {}\ \BBA {} Weldemariam%
}{%
Watson%
\ \protect \BOthers {.}}{%
{\protect \APACyear {2020}}%
}]{%
Watson2020}
\APACinsertmetastar {%
Watson2020}%
\begin{APACrefauthors}%
Watson, C\BPBI D.%
, Wang, C.%
, Lynar, T.%
\BCBL {}\ \BBA {} Weldemariam, K.%
\end{APACrefauthors}%
\unskip\
\newblock
\APACrefYearMonthDay{2020}{}{}.
\newblock
\APACrefbtitle {{Investigating two super-resolution methods for downscaling
  precipitation: ESRGAN and CAR}.} {{Investigating two super-resolution methods
  for downscaling precipitation: ESRGAN and CAR}.}
\newblock
\begin{APACrefURL} \url{https://arxiv.org/abs/2012.01233} \end{APACrefURL}
\PrintBackRefs{\CurrentBib}

\bibitem [\protect \citeauthoryear {%
Yu%
, Nakakita%
, Kim%
\BCBL {}\ \BBA {} Yamaguchi%
}{%
Yu%
\ \protect \BOthers {.}}{%
{\protect \APACyear {2016}}%
}]{%
Yu2016}
\APACinsertmetastar {%
Yu2016}%
\begin{APACrefauthors}%
Yu, W.%
, Nakakita, E.%
, Kim, S.%
\BCBL {}\ \BBA {} Yamaguchi, K.%
\end{APACrefauthors}%
\unskip\
\newblock
\APACrefYearMonthDay{2016}{}{}.
\newblock
{\BBOQ}\APACrefatitle {{Impact Assessment of Uncertainty Propagation of
  Ensemble NWP Rainfall to Flood Forecasting with Catchment Scale}} {{Impact
  Assessment of Uncertainty Propagation of Ensemble NWP Rainfall to Flood
  Forecasting with Catchment Scale}}.{\BBCQ}
\newblock
\APACjournalVolNumPages{Advances in Meteorology}{2016}{}{1--17}.
\newblock
\begin{APACrefDOI} \doi{10.1155/2016/1384302} \end{APACrefDOI}
\PrintBackRefs{\CurrentBib}

\end{thebibliography}

\clearpage

\appendix
\section{Ablation studies}

\begin{threeparttable}[t]
    \centering
    \caption{\add{Table showing evaluation results for our final GAN, and various ablated versions.}}
    \label{table:ablation}
    \begin{tabular}{ccccccccc}
        \hline
        \multirow{2}{*}{Variant} & \multicolumn{8}{c}{Evaluation Metric}\\
        \cline{2-9}
         & \multicolumn{5}{c}{CRPS (mm/hr)} & \multirow{2}{*}{RALSD (dB)} & \multicolumn{2}{c}{RMSE (mm/hr)}\\
         & pixelwise & avg 4 & max 4 & avg 16 & max 16 & & ens-mean & individual\\
         \hline
         Main GAN   & \textbf{0.0856} & \textbf{0.0844} & \textbf{0.1151} & 0.0806 & \textbf{0.2117} & 4.88 & \textbf{0.404} & 0.528 \\
         `Natural'\tnote{1} & 0.0877 & 0.0866 & 0.1185 & 0.0832 & 0.2192 & 5.10 & 0.416 & 0.514 \\
         `Equal'\tnote{1} & 0.0912 & 0.0903 & 0.1226 & 0.0871 & 0.2272 & \textbf{4.33} & 0.417 & 0.533 \\
         No CL\tnote{2} & 0.0901 & 0.0890 & 0.1215 & 0.0857 & 0.2247 & 5.64 & 0.419 & \textbf{0.502} \\
         No geog\tnote{3} & 0.0857 & 0.0845 & 0.1159 & \textbf{0.0805} & 0.2155 & 4.43 & 0.407 & 0.541 \\
         \hline
    \end{tabular}
    \begin{tablenotes}
    \item[1] \add{Varying the distribution of data used during training.}
    \item[2] \add{No content loss term used during generator training, only discriminator loss.}
    \item[3] \add{No geographic fields used, i.e., the high-resolution orography and land-sea mask.}
    \end{tablenotes}
\end{threeparttable}

\subsection{Varying training data distribution}
\label{sec:data-weighting}

\begin{figure*}
\centering
    \includegraphics[width=0.9\textwidth]{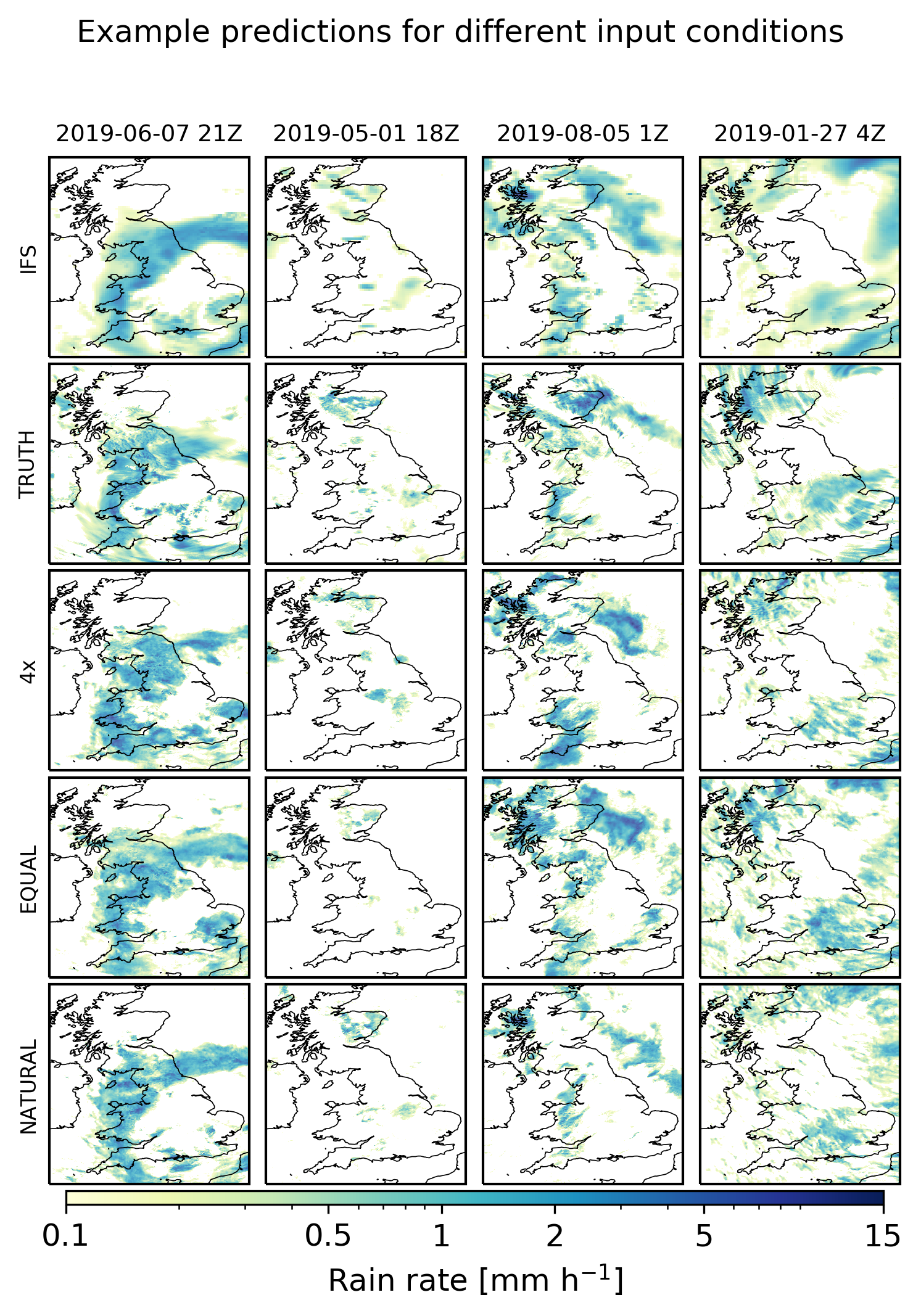}
    \caption{Example GAN model output predictions with different input sample weighting, compared to the input low-res information (IFS) and hi-res ground-truth data (NIMROD)}
\label{fig:input-data}
\end{figure*}

As discussed in Section~\ref{sec:datasubset}, the training data was pre-processed into sub-images and sorted into bins according to the proportion of pixels containing rain within the sub-image. We then trained the network on different frequency distributions of these bins. We anticipated that the model would otherwise under-predict precipitation if trained on images that overwhelmingly do not contain precipitation. We treated this training data distribution as a hyperparameter to be optimised, and explored a few different distributions.

The initial selection of sampling equally from these bins caused the GAN to over-predict rainfall, shown in Figure~\ref{fig:input-data}. We also investigated training the network on the natural distribution of images - sampling in the proportion that the data would naturally fall. This corresponds to 41$\times$ as many images from the least-rainy bin as the most-rainy bin. Sampling the sub-images in this way produced marginally better CRPS scores (see Table~\ref{table:ablation}). However, as anticipated, the network clearly under-predicts rainfall.

Finally, we tried showing the network $k$ times as many images from the ``least-rainy'' bin as the ``most-rainy'' bin, with $k$ varying between 2 and 12. The intermediate bin weights were interpolated linearly between these. Lower $k$-values tended to produce better rank histogram plots and RALSD scores, and higher $k$-values produced better results for CRPS. A $k$-value of 4 was determined to offer the best results overall. Example predictions for our best model are shown in Figure~\ref{fig:input-data}. Although the network still has a tendency to over-predict light rainfall, it retains the predictive power at the extremes of rainfall, so this was determined to be the best compromise.

The plots and evaluation numbers shown here are for the final, best version of the model. The differences here are subtle. The initial assessment of the data training weights was carried out on a more preliminary version, and the differences were much more stark. Improvements on the 4x-trained model also significantly improved models trained with other input data distributions, however, we still consider the balance of training data an important factor to be considered. Choosing an optimal input data distribution should also help to accelerate training, even if the respective models eventually converge to similar minima.

\subsection{Removal of content loss term}

\add{In Section} \ref{sec:training}, \add{we described that the generator is not just trained on discriminator loss, but an ensemble-mean-MSE content loss term is also added to the loss function. As shown in Table} \ref{table:ablation}, \add{this content loss term improves the resulting network considerably. The CRPS improves noticeably, and the RALSD and ensemble-mean RMSE also improve.}

\subsection{Removal of geographic fields}

\add{Our main models used not just low-resolution IFS forecast fields as input, but also high-resolution orographic and land-sea mask fields, since these are expected to affect precipitation locally due to physical principles. From Table} \ref{table:ablation}, \add{it would appear that a network without these high-resolution inputs has roughly equivalent skill. However, we found that removing geographic fields made the network perform noticeably worse on the 2019 validation dataset when averaged across a larger number of candidate model checkpoints, and earlier versions of the model showed a reduction in skill from removing geographic fields. We therefore believe that the geographic fields should still be included, even if the improvement in skill was not apparent in the final model training runs used in this paper.}

\section{Pure super-resolution problem}
\label{sec:sr-section}

We also assessed the performance of the model architecture on the pure super-resolution problem of increasing spatial resolution without accounting for forecast error. The input data is now NIMROD radar data that has been coarsened, using averaging, by a factor of 10. The output is compared with the original, high-resolution NIMROD truth. We again pass in high-resolution orography and land-sea masks to the model. The precise information flow and network architecture is detailed in Figure~\ref{fig:architecture}, with the ``9 IFS fields'' input replaced by a single ``coarsened NIMROD'' input.

\begin{figure*}
    \centering
    \includegraphics[width=0.75\textwidth]{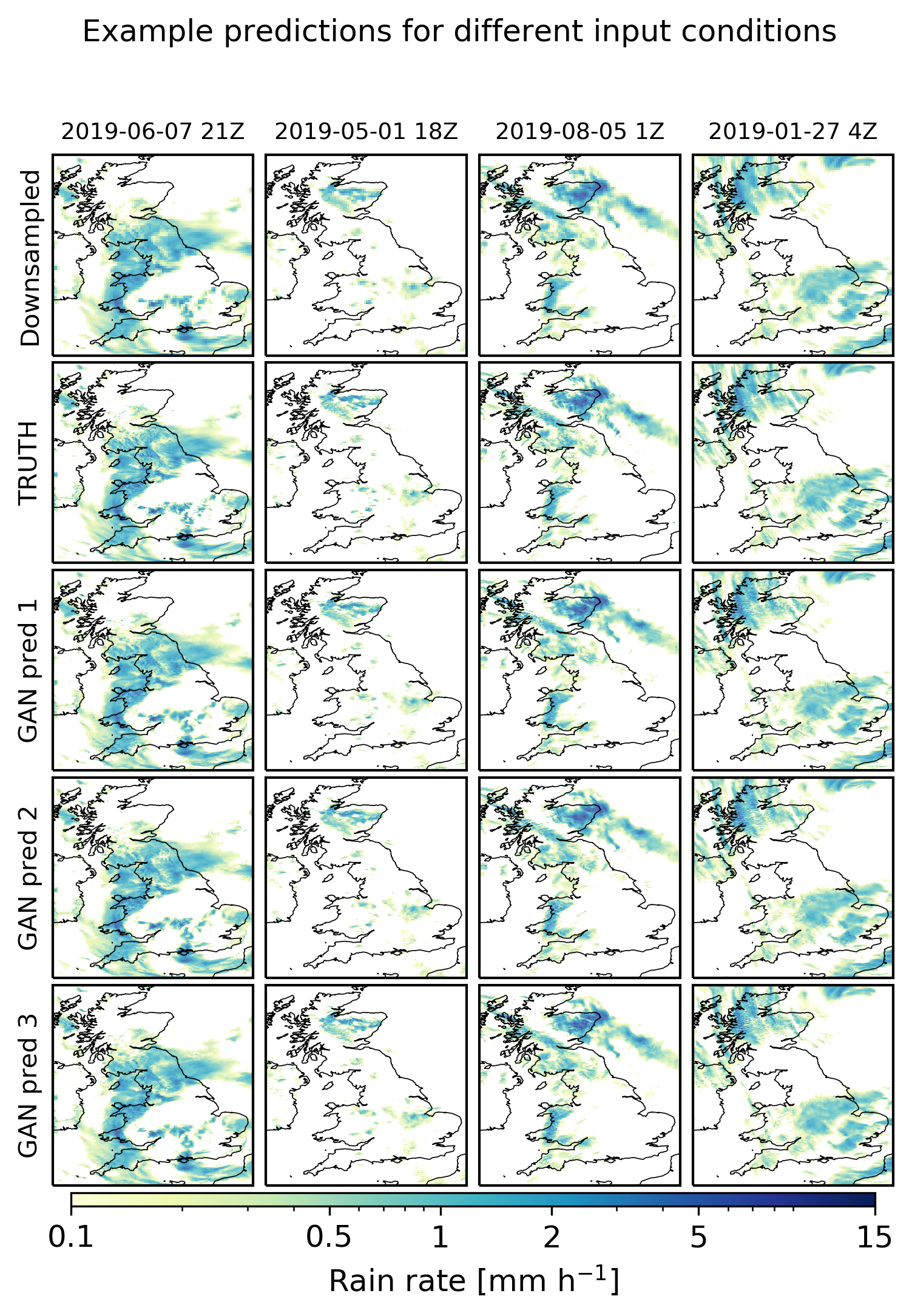}
    \caption{GAN model output predictions for the pure super-resolution problem: mapping coarsened NIMROD data to full-resolution NIMROD data}
    \label{fig:sr-prediction}
\end{figure*}

Example predictions for a model trained to map from coarsened to full-resolution NIMROD data are shown in Figure~\ref{fig:sr-prediction}. It is immediately clear that the model performs very well on this task. Comparing these results to Figures \ref{fig:GAN-comparison} and \ref{fig:predictions} indicates how much more challenging the full downscaling problem is, in the presence of forecast error. Quantitatively, the model obtains a CRPS of $0.0230$ mm/hr on this problem, less than one-third of the $0.0856$ mm/hr CRPS obtained on the full problem. Increasing the resolution of the forecast appears to be a less challenging problem than accounting for forecast error in a spatially-coherent manner, and there is potential for an approach that performs these two steps separately.

\section{Reduced numerical precision}
\add{The models were originally trained with 32-bit floating point numbers on an NVIDIA V100. We later gained access to an NVIDIA A100, on which TensorFlow automatically employed the ``TensorFloat-32'' (TF-32) format for many internal calculations. This number format has the range of 32-bit numbers but the precision of 16-bit numbers. The resulting trained models appeared to be equal in quality to the original models trained in 32-bit, although it is hard to be completely sure due to the random variations between different training runs.}

\add{We also tried using the TF-32 format only for inference, with models trained at full 32-bit. This was completely successful, giving practically identical metrics to 32-bit inference with approximately a 1/3 reduction in run time. We further tried training the model explicitly using 16-bit numbers. However, this quickly led to overflow during training and was unsuccessful.}
\end{document}


%
%


\title{Supporting Information for ``A Generative Deep Learning Approach to Stochastic Downscaling of Precipitation Forecasts"}
%

%
%


\authors{Lucy Harris\affil{1}, Andrew T. T. McRae\affil{1}, Matthew Chantry\affil{2}, Peter D. Dueben\affil{2}, Tim N. Palmer\affil{1}}

\affiliation{1}{Department of Physics, University of Oxford}
\affiliation{2}{European Centre for Medium-Range Weather Forecasts}

%
%

%

\begin{article}

%
%

\noindent\textbf{Contents of this file}

\begin{figure}
\centering
\noindent\includegraphics[width=0.75\textwidth]{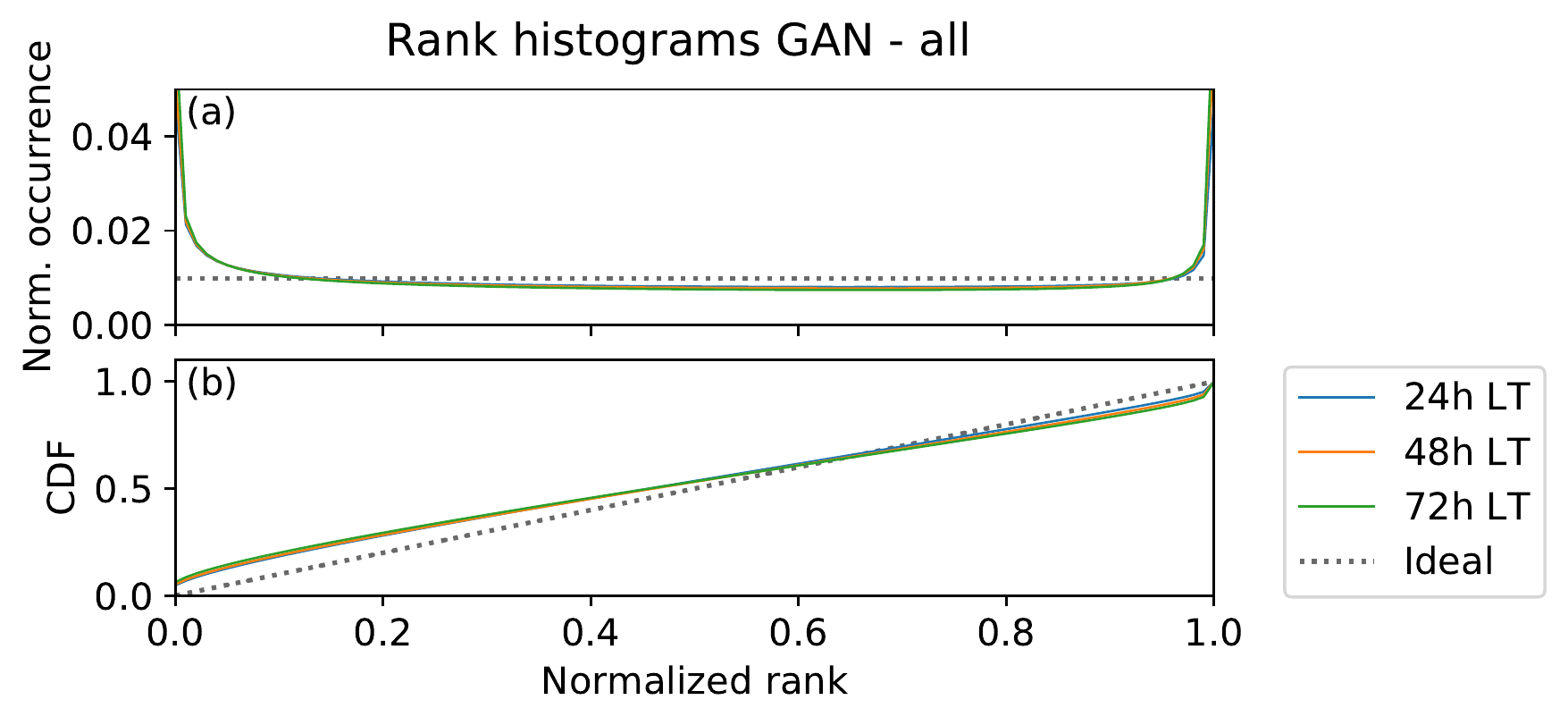}
    \caption{Calibration plot for the GAN model at increasing lead times: (a) shows the frequency of per-pixel normalised ranks for the trained GAN model evaluated on the hold-out dataset (2020). The dotted grey line shows the ideal distribution for comparison. (b) shows the same as panel a, except displaying the CDFs of the distributions.}
    \label{fig:ranks-GAN-lead-time}
\end{figure}

\begin{figure}
\centering
\noindent\includegraphics[width=0.75\textwidth]{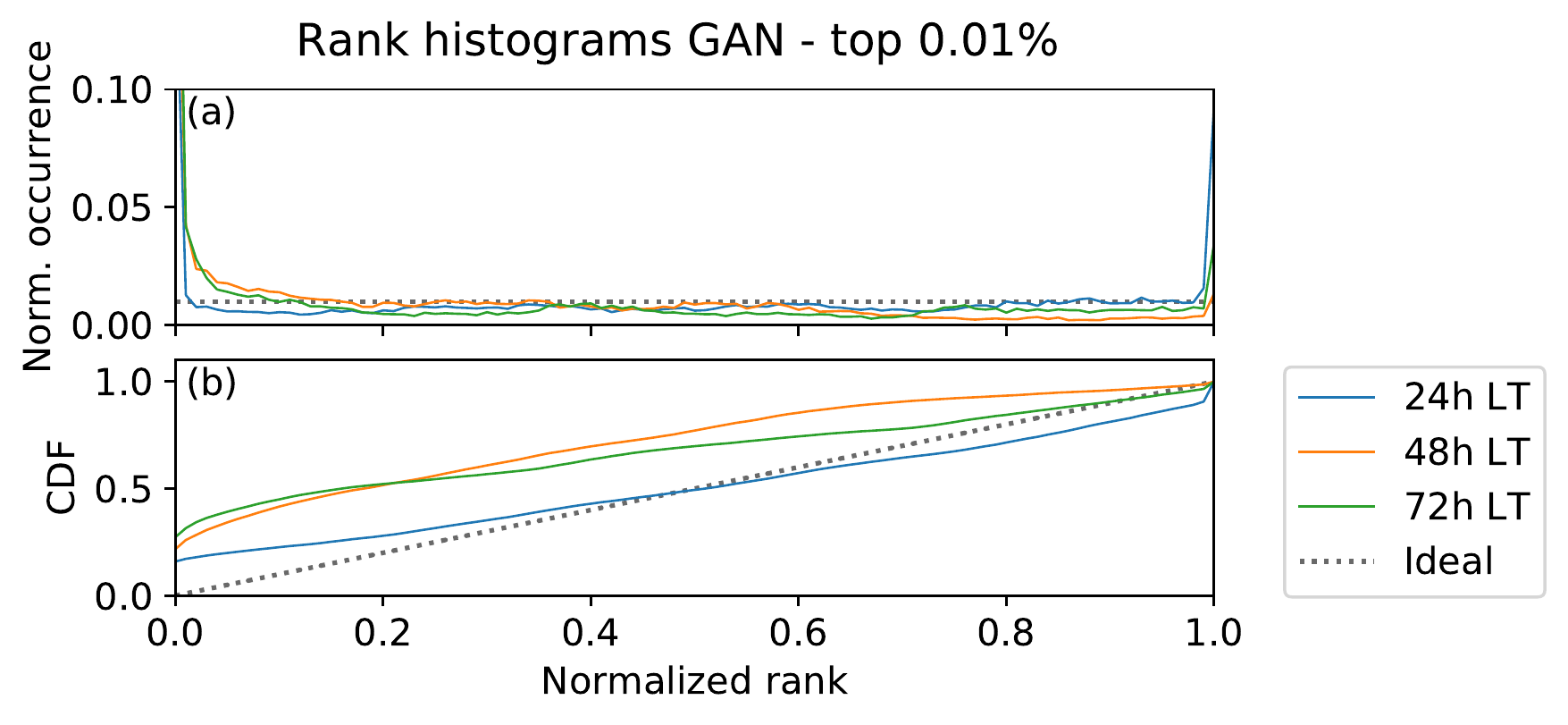}
    \caption{Thresholded calibration plot for the GAN model at increasing lead times: (a) shows the frequency of per-pixel normalised ranks over the 0.01\% threshold for the trained GAN model evaluated on the hold-out dataset (2020). The dotted grey line shows the ideal distribution for comparison. (b) shows the same as panel a, except displaying the CDFs of the distributions.}
    \label{fig:ranks-GAN-lead-time-thresh}
\end{figure}

\begin{figure}
\centering
\noindent\includegraphics[width=0.75\textwidth]{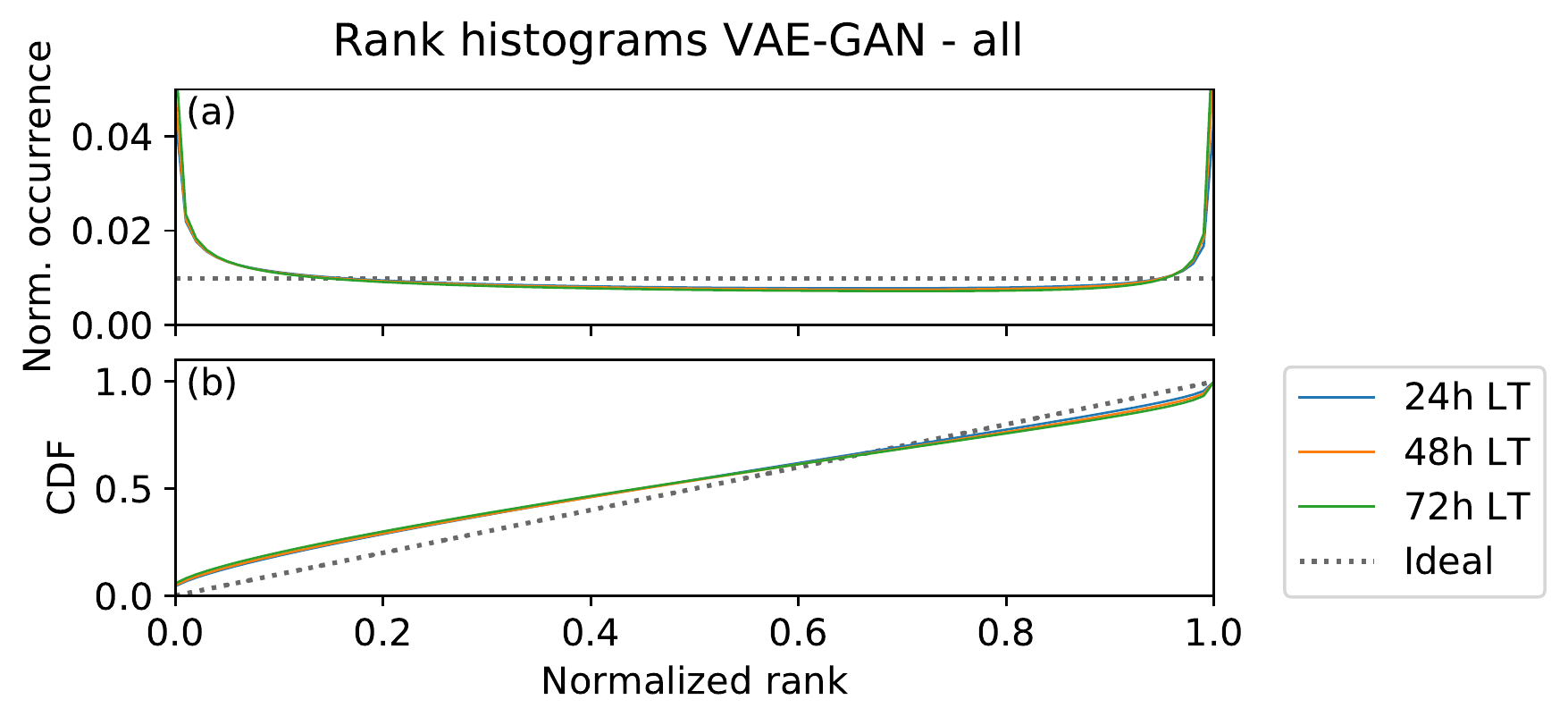}
    \caption{Calibration plot for the VAE-GAN model at increasing lead times: (a) shows the frequency of per-pixel normalised ranks for the trained VAE-GAN model evaluated on the hold-out dataset (2020). The dotted grey line shows the ideal distribution for comparison. (b) shows the same as panel a, except displaying the CDFs of the distributions.}
    \label{fig:ranks-VAEGAN-lead-time}
\end{figure}

\begin{figure}
\centering
\noindent\includegraphics[width=0.75\textwidth]{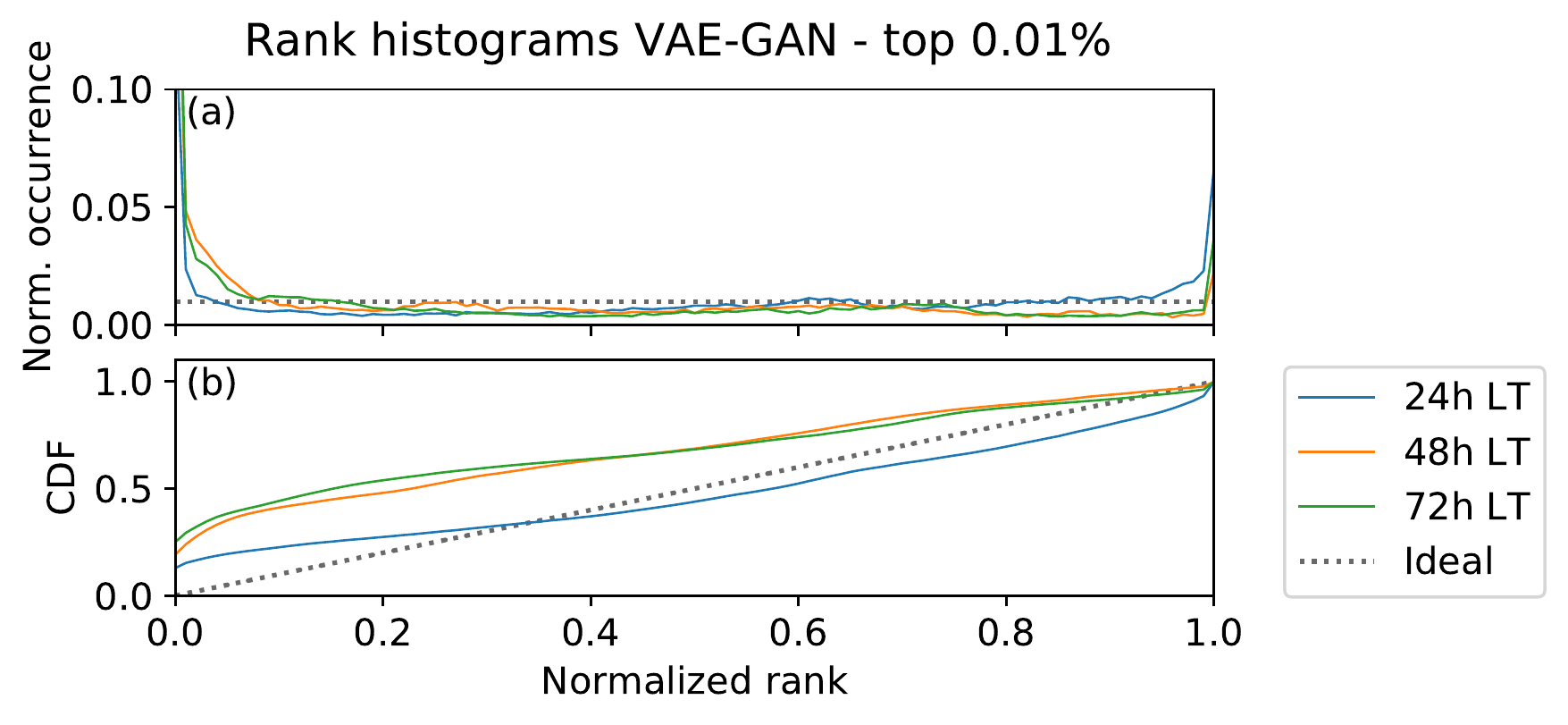}
    \caption{Thresholded calibration plot for the VAE-GAN model at increasing lead times: (a) shows the frequency of per-pixel normalised ranks over the 0.01\% threshold for the trained VAE-GAN model evaluated on the hold-out dataset (2020). The dotted grey line shows the ideal distribution for comparison. (b) shows the same as panel a, except displaying the CDFs of the distributions.}
    \label{fig:ranks-VAEGAN-lead-time-thresh}
\end{figure}

\begin{figure}
\centering
\noindent\includegraphics[width=0.75\textwidth]{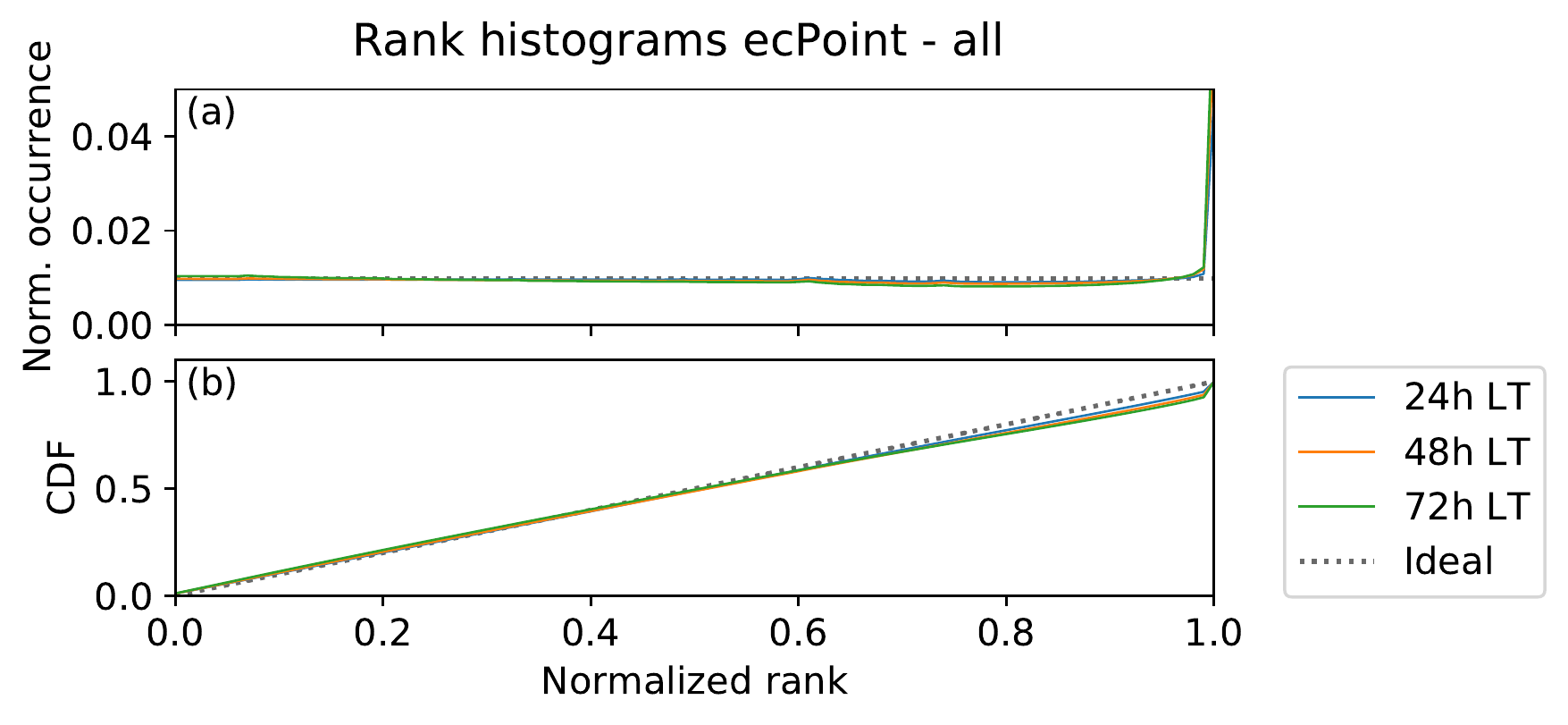}
    \caption{Calibration plot for the ecPoint model at increasing lead times: (a) shows the frequency of per-pixel normalised ranks for the ecPoint model evaluated on the hold-out dataset (2020). The dotted grey line shows the ideal distribution for comparison. (b) shows the same as panel a, except displaying the CDFs of the distributions.}
    \label{fig:ranks-ecpoint-lead-time}
\end{figure}

\begin{figure}
\centering
\noindent\includegraphics[width=0.75\textwidth]{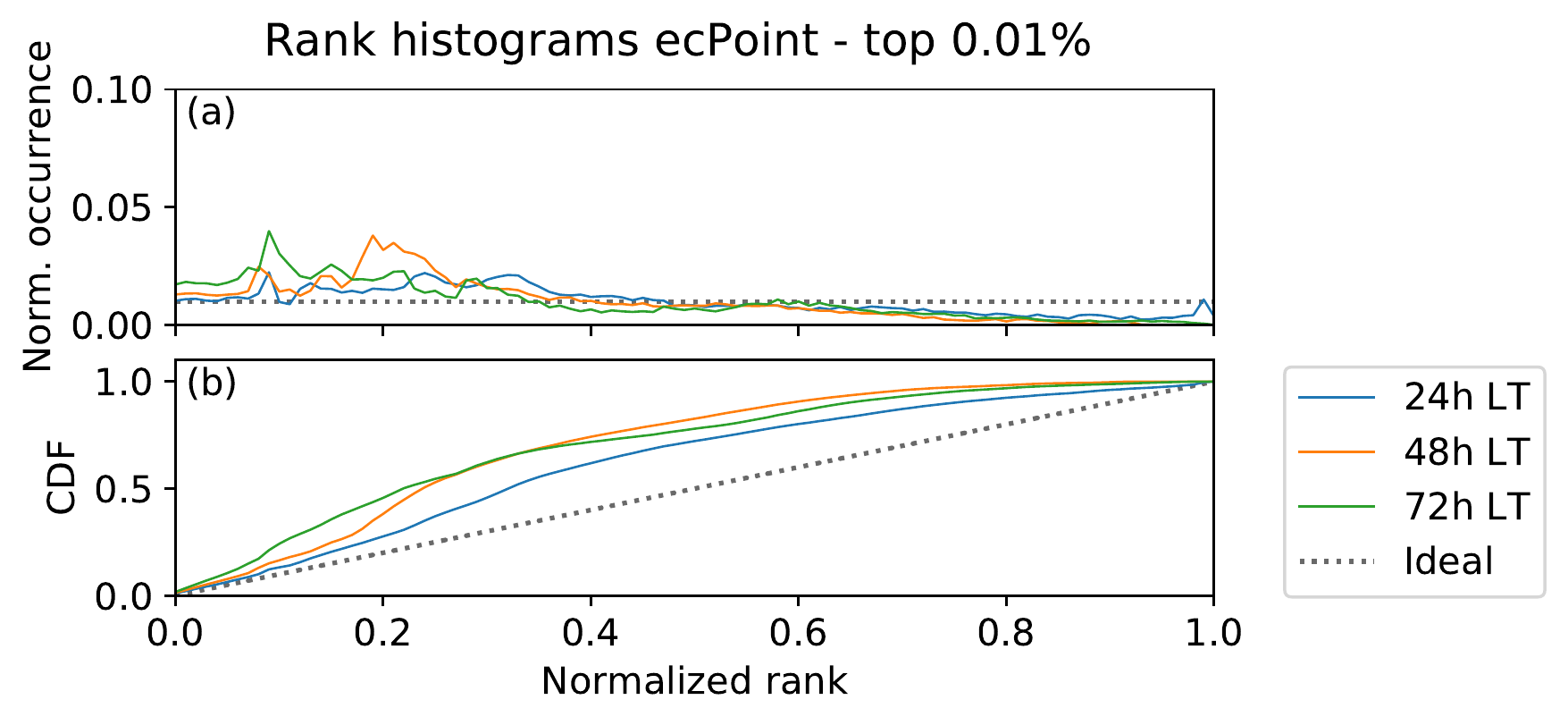}
    \caption{Thresholded calibration plot for the ecPoint model at increasing lead times: (a) shows the frequency of per-pixel normalised ranks over the 0.01\% threshold for the trained ecPoint model evaluated on the hold-out dataset (2020). The dotted grey line shows the ideal distribution for comparison. (b) shows the same as panel a, except displaying the CDFs of the distributions.}
    \label{fig:ranks-ecpoint-lead-time-thresh}
\end{figure}

\begin{figure}
\centering
\noindent\includegraphics[width=0.75\textwidth]{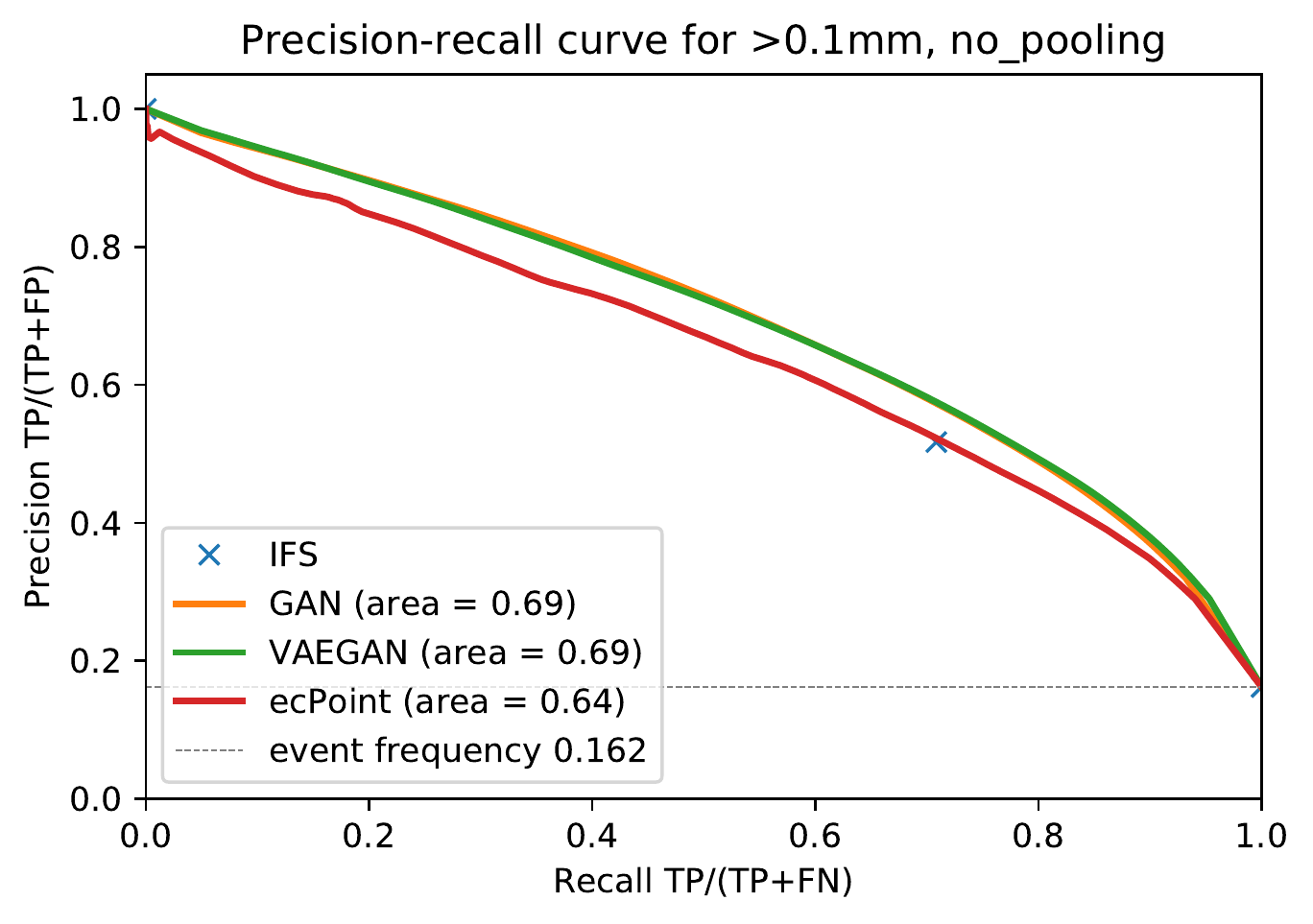}
    \caption{PR curves for the GAN, VAE-GAN and ecPoint models for a 0.1 mm/hr precipitation threshold, pixel-wise}
    \label{fig:PRC-0.1-none}
\end{figure}

\begin{figure}
\centering
\noindent\includegraphics[width=0.75\textwidth]{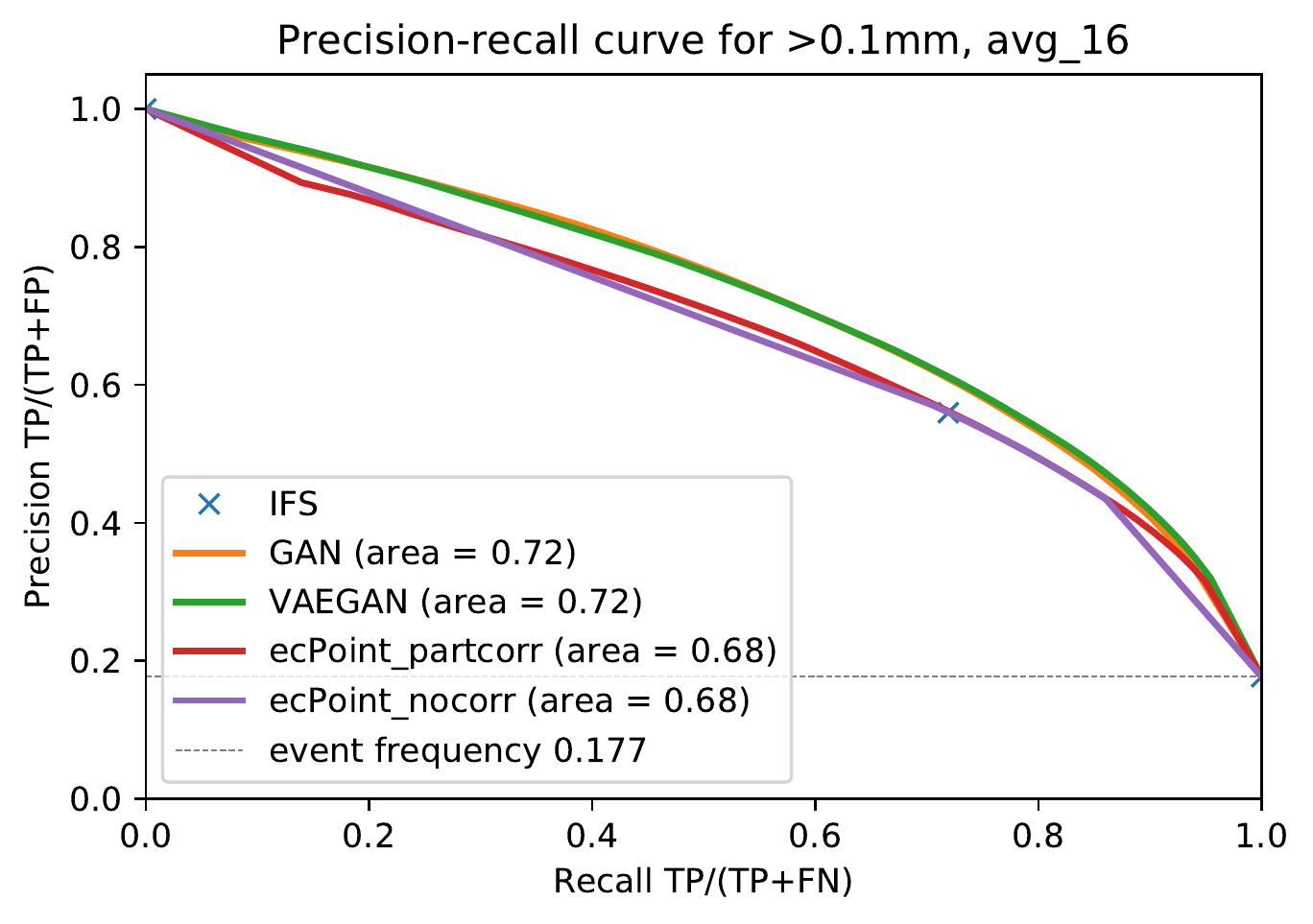}
    \caption{PR curves for the GAN, VAE-GAN and ecPoint models for a 0.1 mm/hr precipitation threshold, average pooling over 16$\times$16 pixels}
    \label{fig:PRC-0.1-avg}
\end{figure}

\begin{figure}
\centering
\noindent\includegraphics[width=0.75\textwidth]{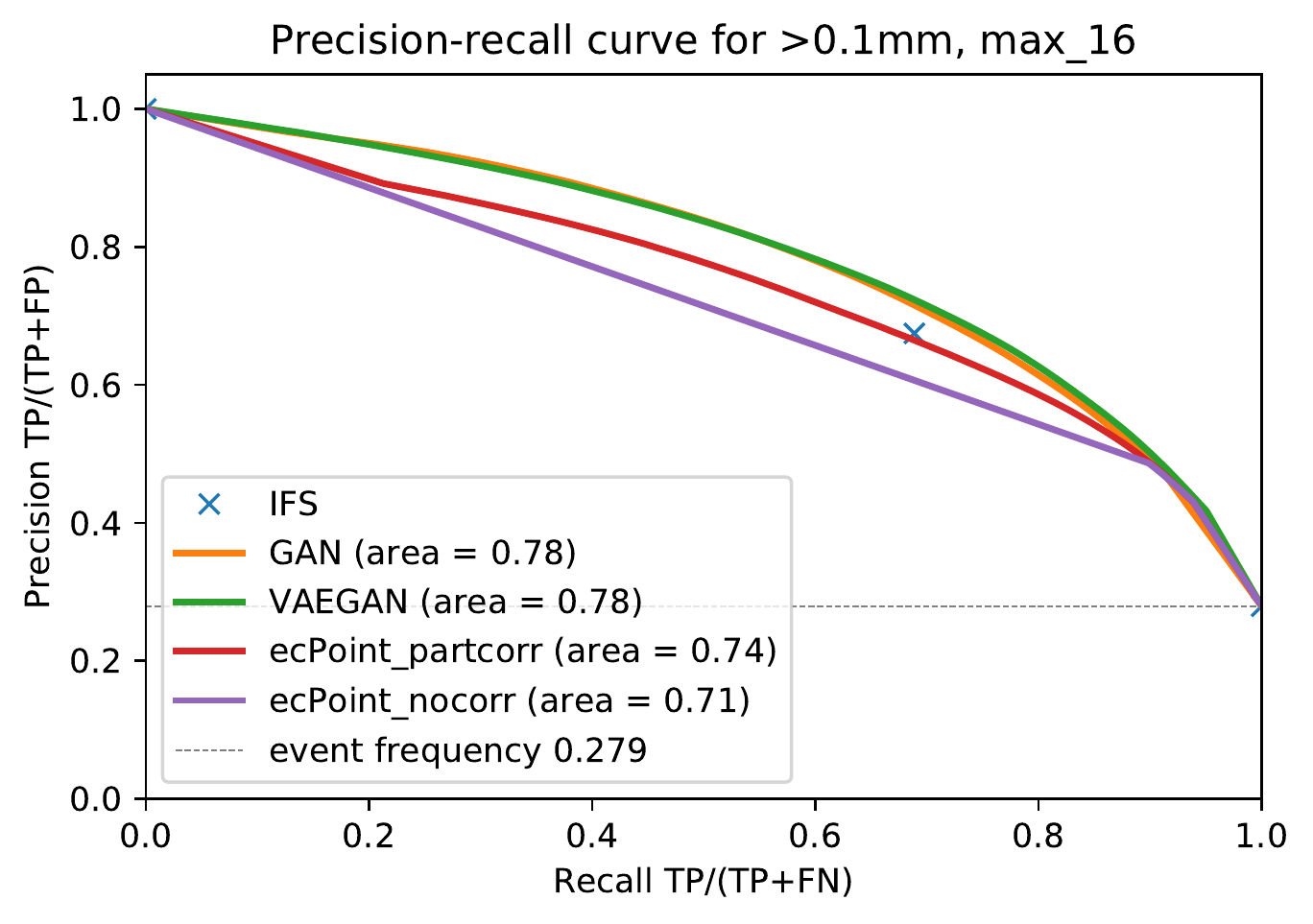}
    \caption{PR curves for the GAN, VAE-GAN and ecPoint models for a 0.1 mm/hr precipitation threshold, max pooling over 16$\times$16 pixels}
    \label{fig:PRC-0.1-max}
\end{figure}

\begin{figure}
\centering
\noindent\includegraphics[width=0.75\textwidth]{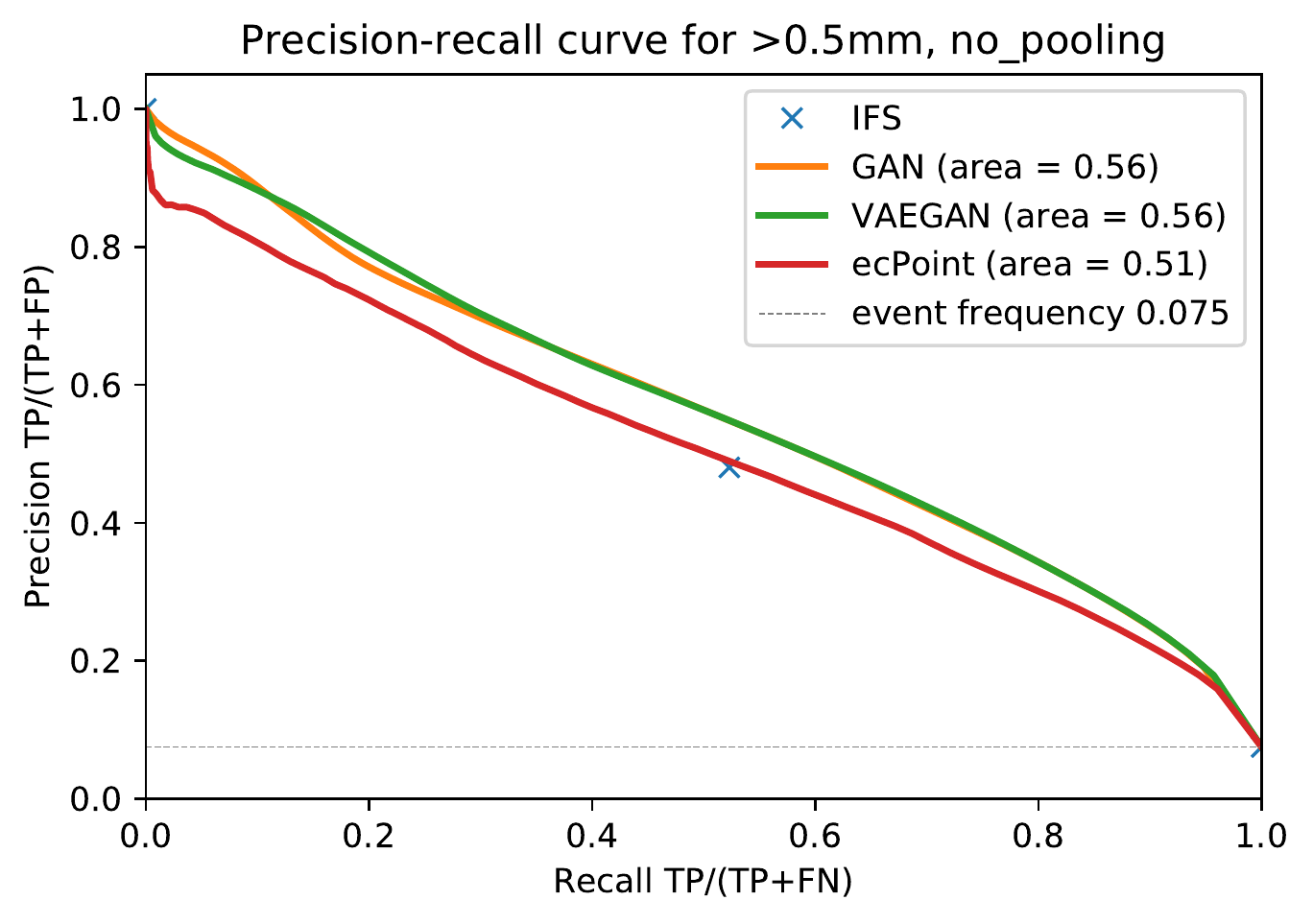}
    \caption{PR curves for the GAN, VAE-GAN and ecPoint models for a 0.5 mm/hr precipitation threshold, pixel-wise}
    \label{fig:PRC-0.5-none}
\end{figure}

\begin{figure}
\centering
\noindent\includegraphics[width=0.75\textwidth]{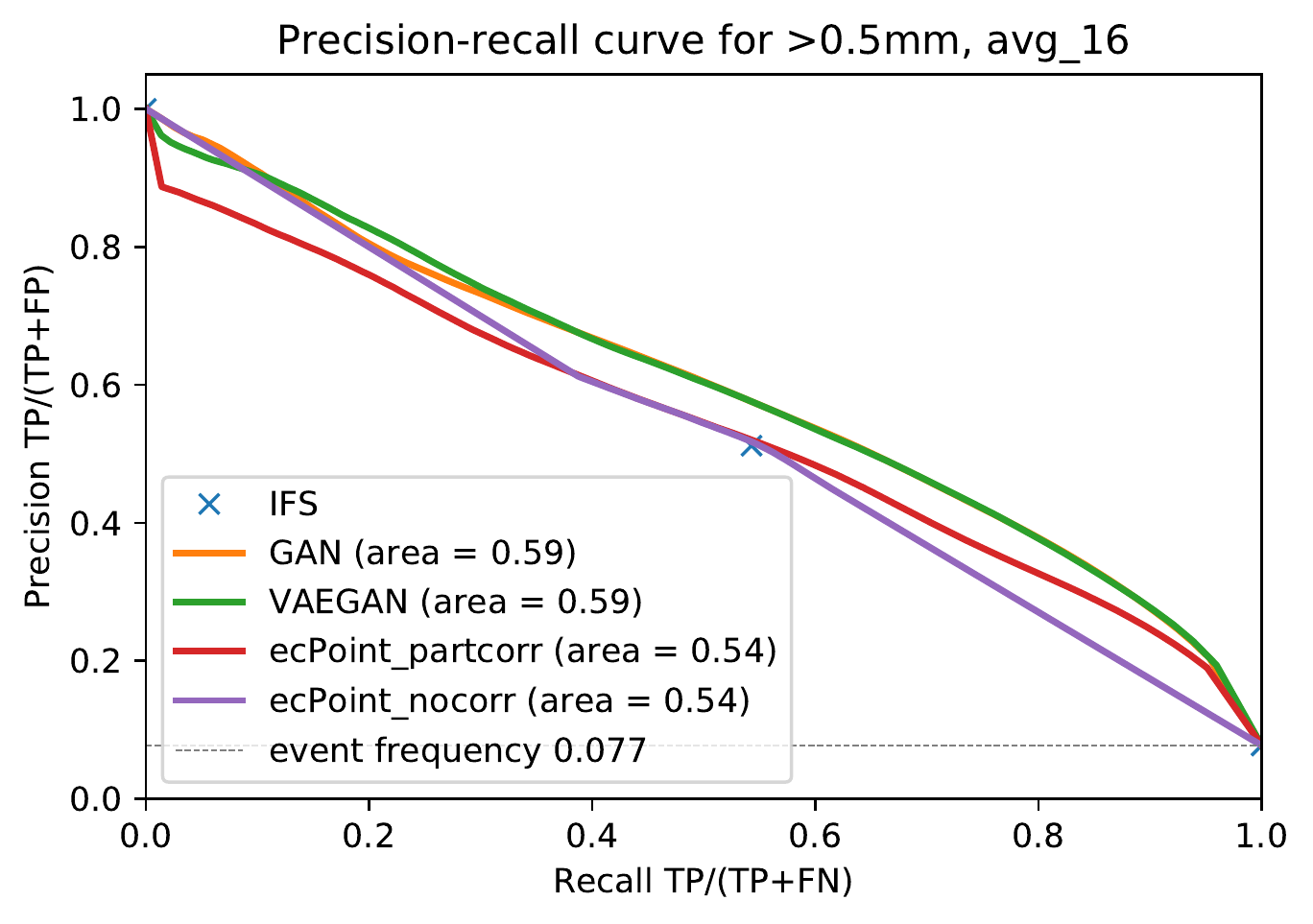}
    \caption{PR curves for the GAN, VAE-GAN and ecPoint models for a 0.5 mm/hr precipitation threshold, average pooling over 16$\times$16 pixels}
    \label{fig:PRC-0.5-avg}
\end{figure}

\begin{figure}
\centering
\noindent\includegraphics[width=0.75\textwidth]{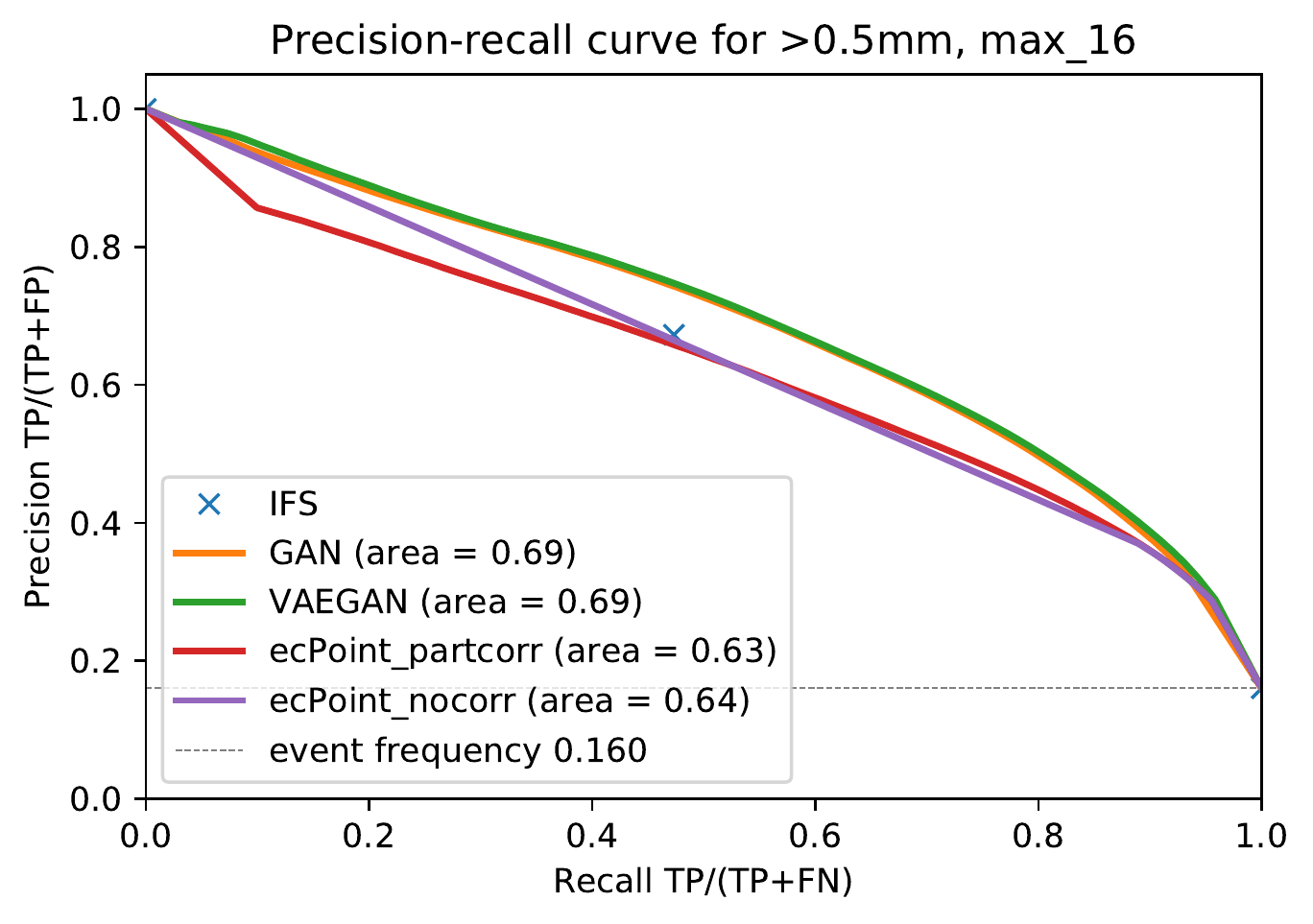}
    \caption{PR curves for the GAN, VAE-GAN and ecPoint models for a 0.5 mm/hr precipitation threshold, max pooling over 16$\times$16 pixels}
    \label{fig:PRC-0.5-max}
\end{figure}

\begin{figure}
\centering
\noindent\includegraphics[width=0.75\textwidth]{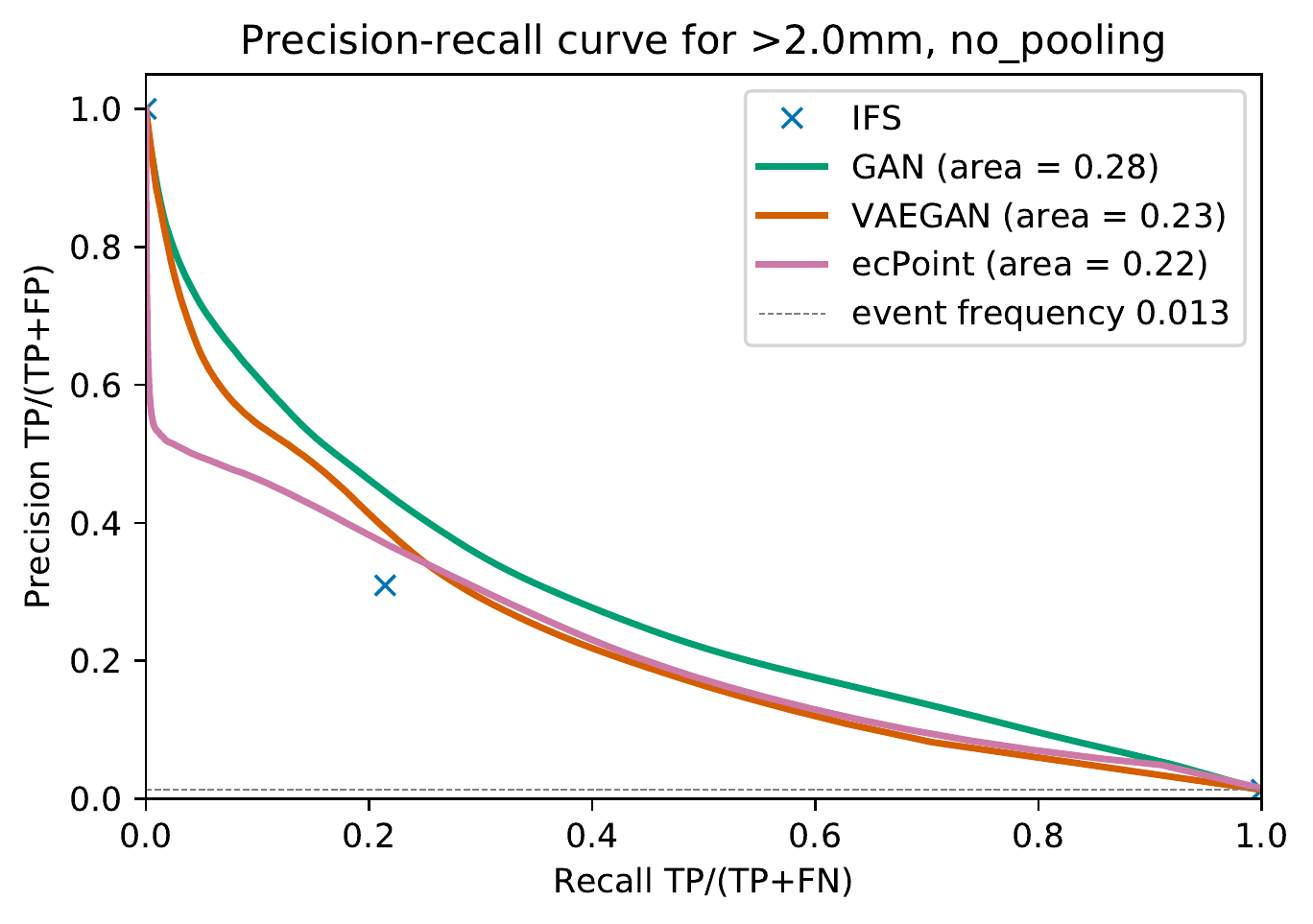}
    \caption{PR curves for the GAN, VAE-GAN and ecPoint models for a 2.0 mm/hr precipitation threshold, pixel-wise}
    \label{fig:PRC-2.0-none}
\end{figure}

\begin{figure}
\centering
\noindent\includegraphics[width=0.75\textwidth]{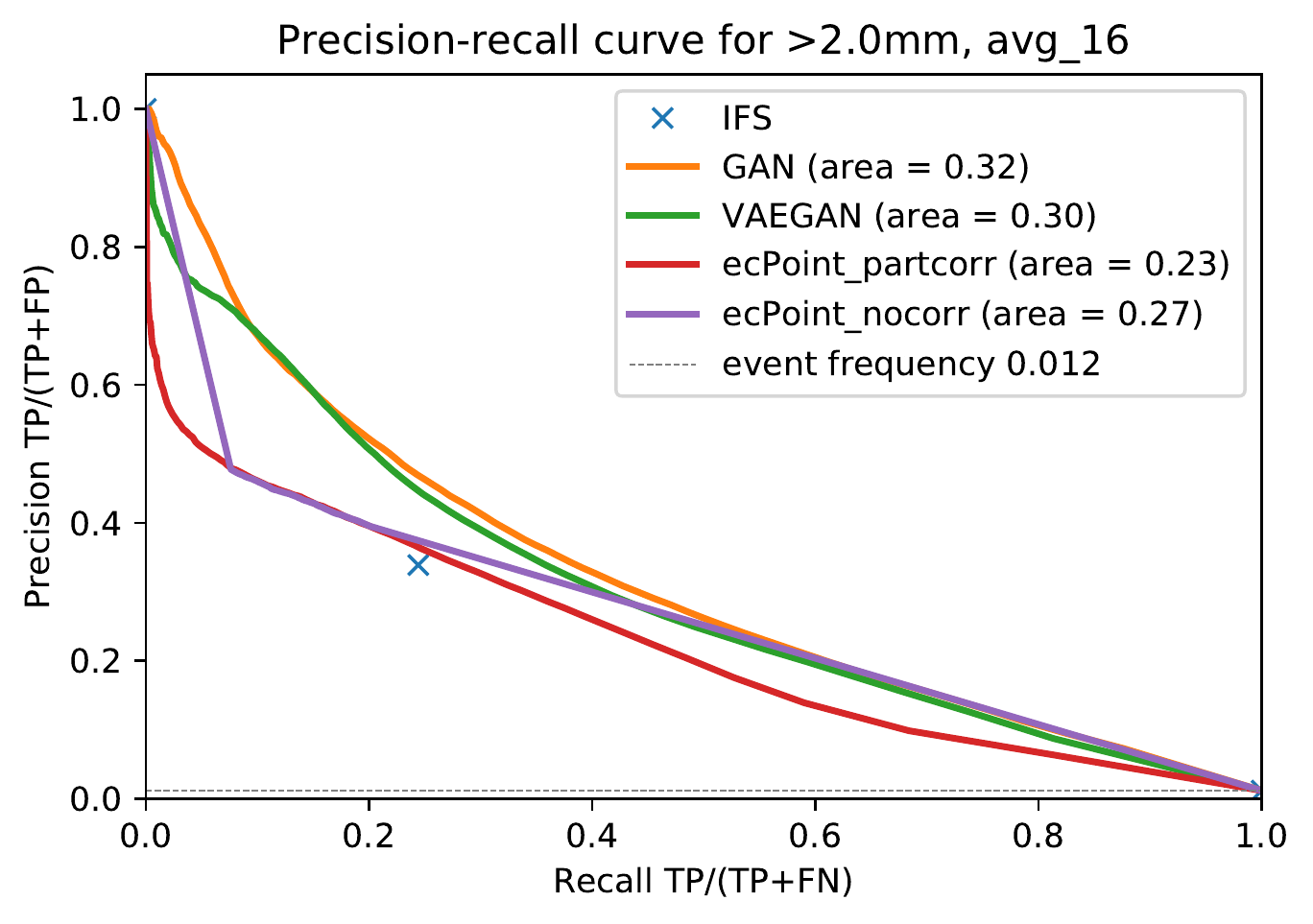}
    \caption{PR curves for the GAN, VAE-GAN and ecPoint models for a 2.0 mm/hr precipitation threshold, average pooling over 16$\times$16 pixels}
    \label{fig:PRC-2.0-avg}
\end{figure}

\begin{figure}
\centering
\noindent\includegraphics[width=0.75\textwidth]{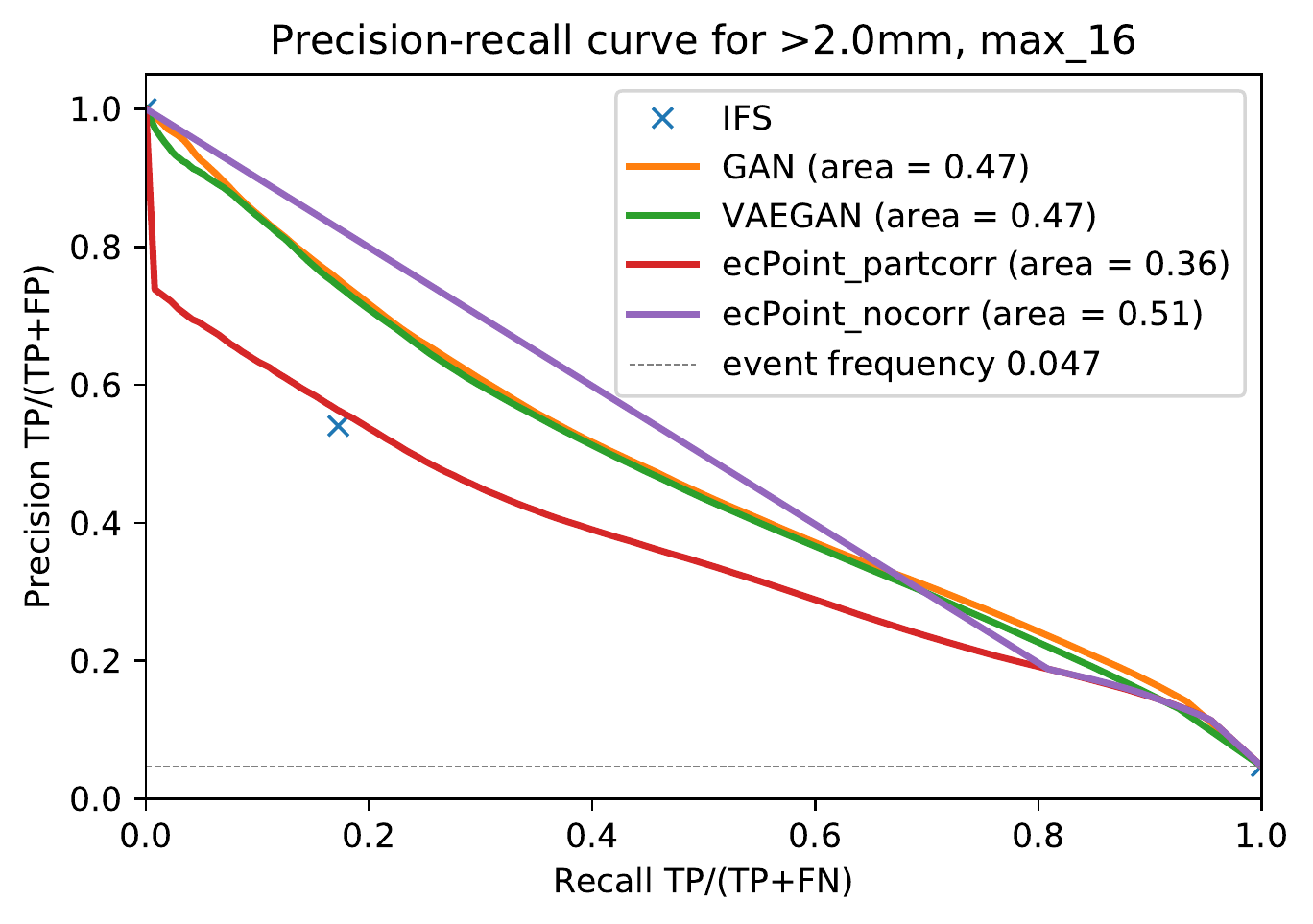}
    \caption{PR curves for the GAN, VAE-GAN and ecPoint models for a 2.0 mm/hr precipitation threshold, max pooling over 16$\times$16 pixels}
    \label{fig:PRC-2.0-max}
\end{figure}

\begin{figure}
\centering
\noindent\includegraphics[width=0.75\textwidth]{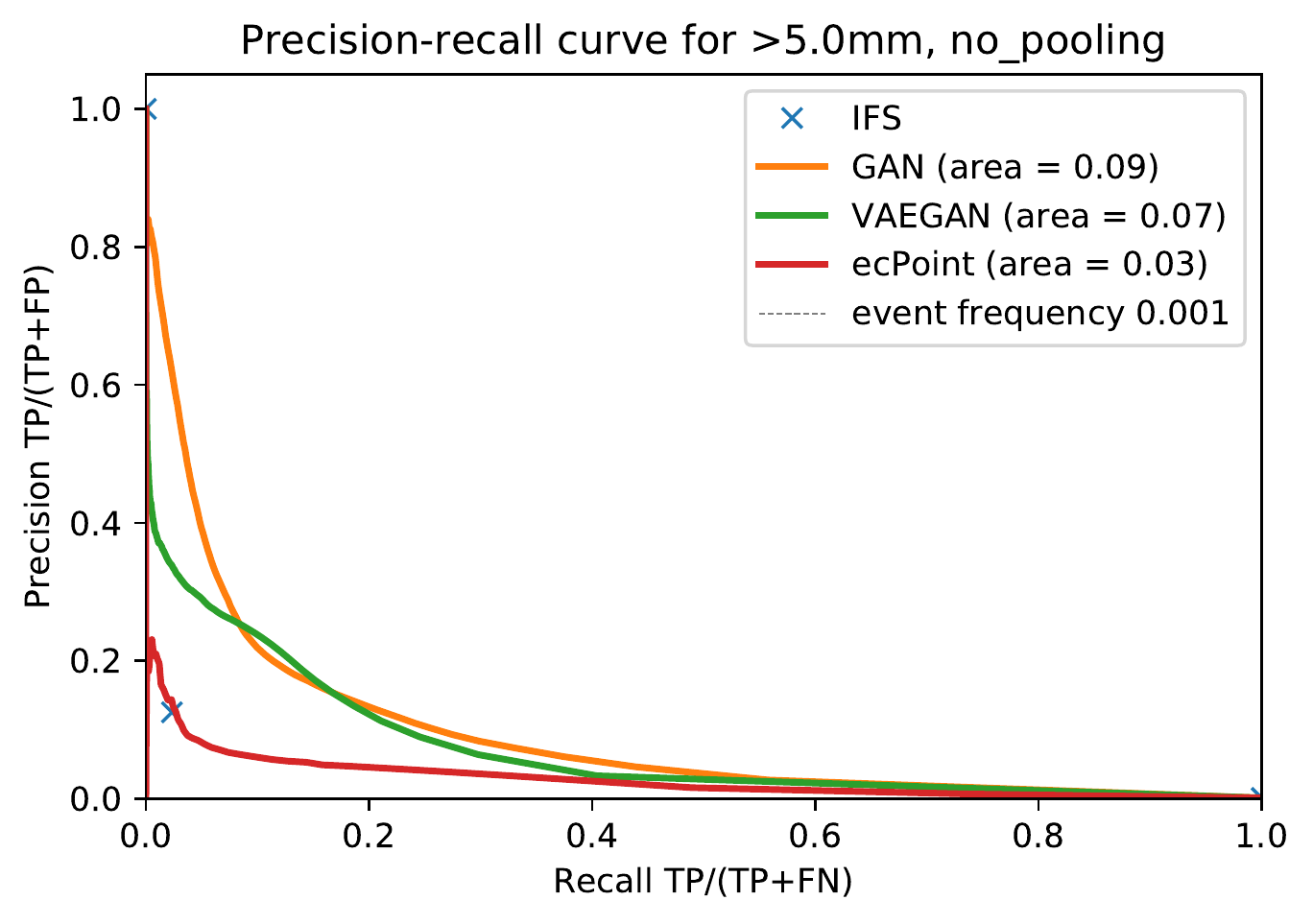}
    \caption{PR curves for the GAN, VAE-GAN and ecPoint models for a 5.0 mm/hr precipitation threshold, pixel-wise}
    \label{fig:PRC-5.0-none}
\end{figure}

\begin{figure}
\centering
\noindent\includegraphics[width=0.75\textwidth]{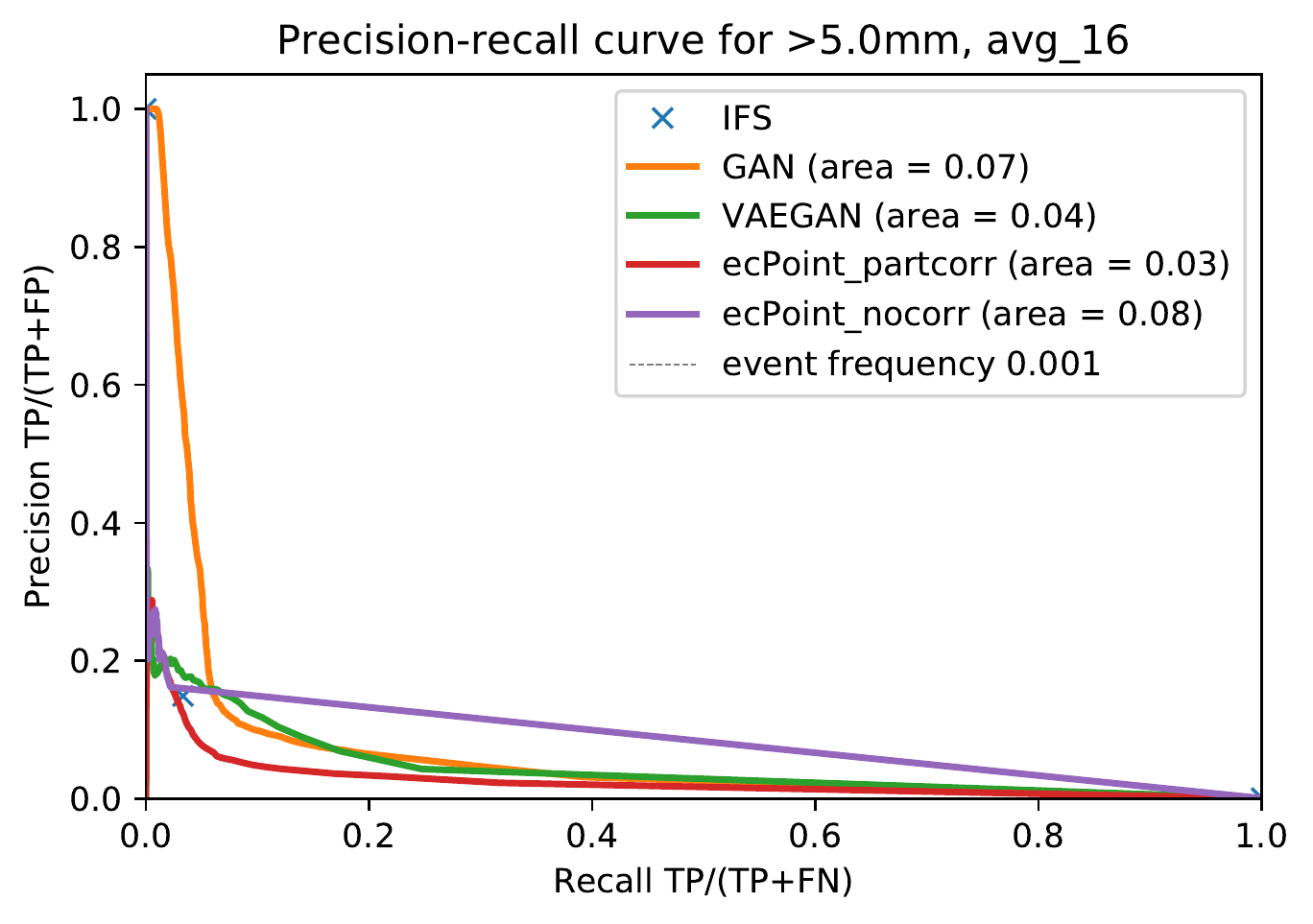}
    \caption{PR curves for the GAN, VAE-GAN and ecPoint models for a 5.0 mm/hr precipitation threshold, average pooling over 16$\times$16 pixels}
    \label{fig:PRC-5.0-avg}
\end{figure}

\begin{figure}
\centering
\noindent\includegraphics[width=0.75\textwidth]{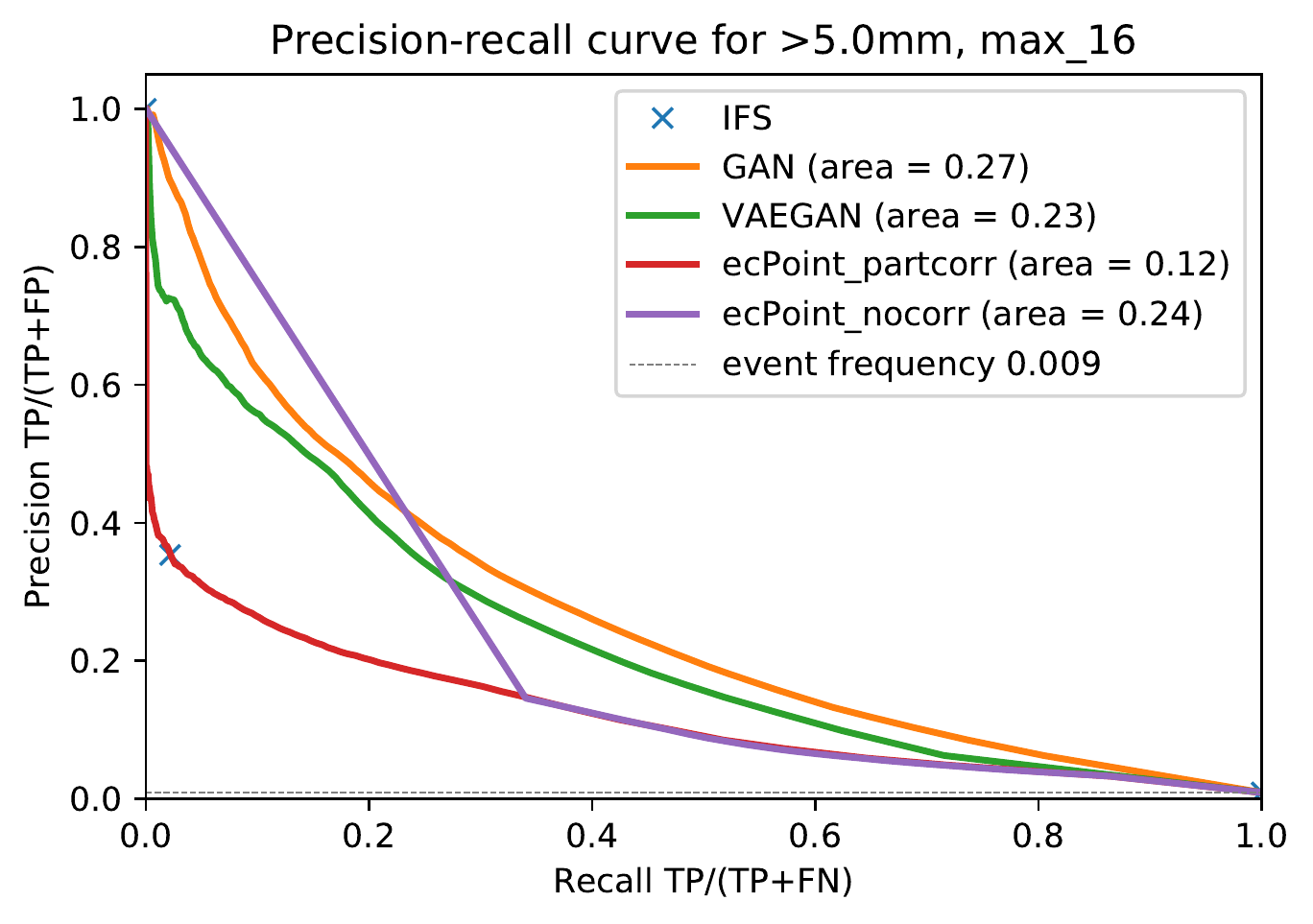}
    \caption{PR curves for the GAN, VAE-GAN and ecPoint models for a 5.0 mm/hr precipitation threshold, max pooling over 16$\times$16 pixels}
    \label{fig:PRC-5.0-max}
\end{figure}

\begin{figure}
\centering
\noindent\includegraphics[width=0.75\textwidth]{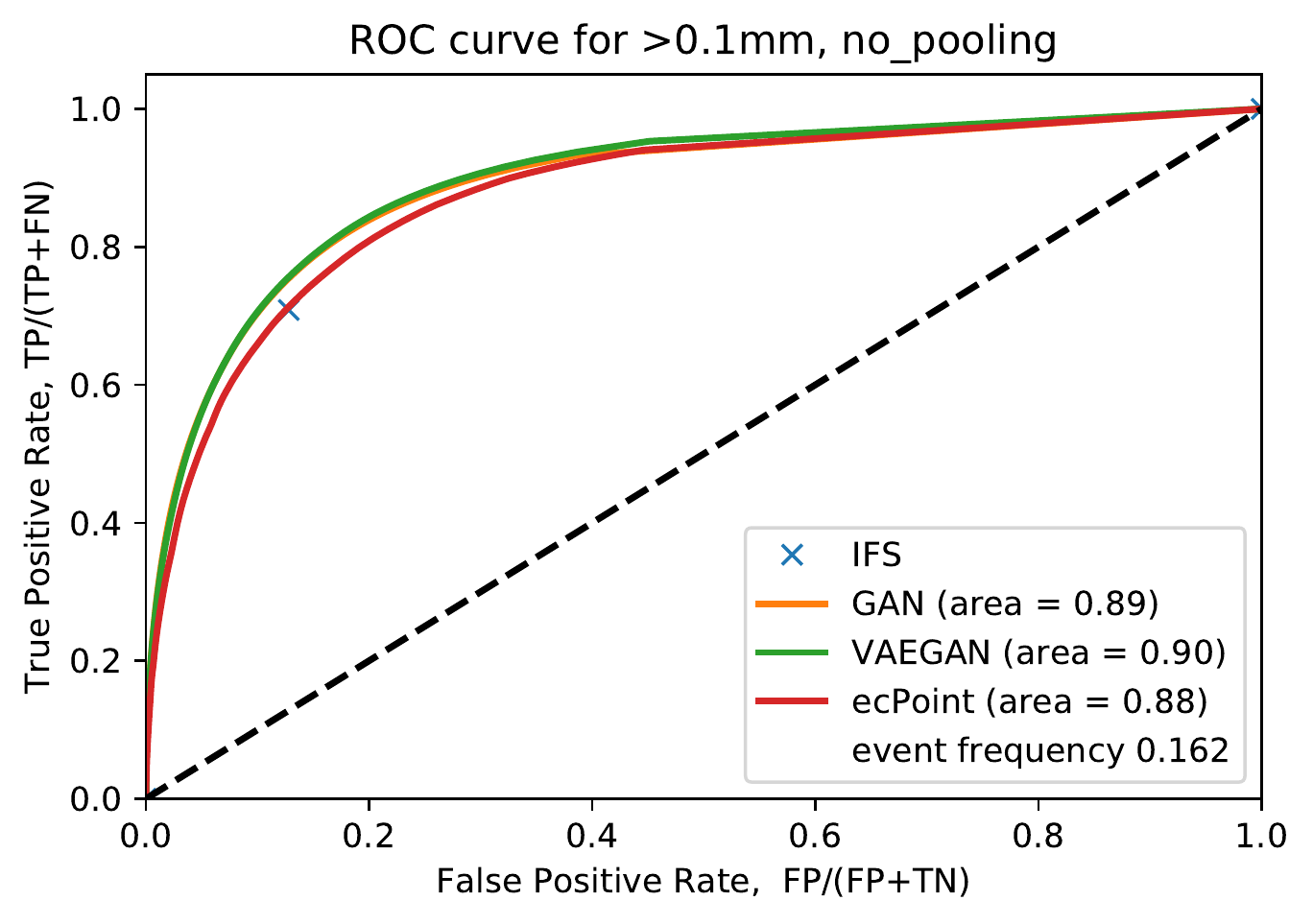}
\caption{ROC curves for the GAN, VAE-GAN and ecPoint models for a 0.1 mm/hr precipitation threshold, pixel-wise}
\label{fig:ROC-0.1-none}
\end{figure}

\begin{figure}
\centering
\noindent\includegraphics[width=0.75\textwidth]{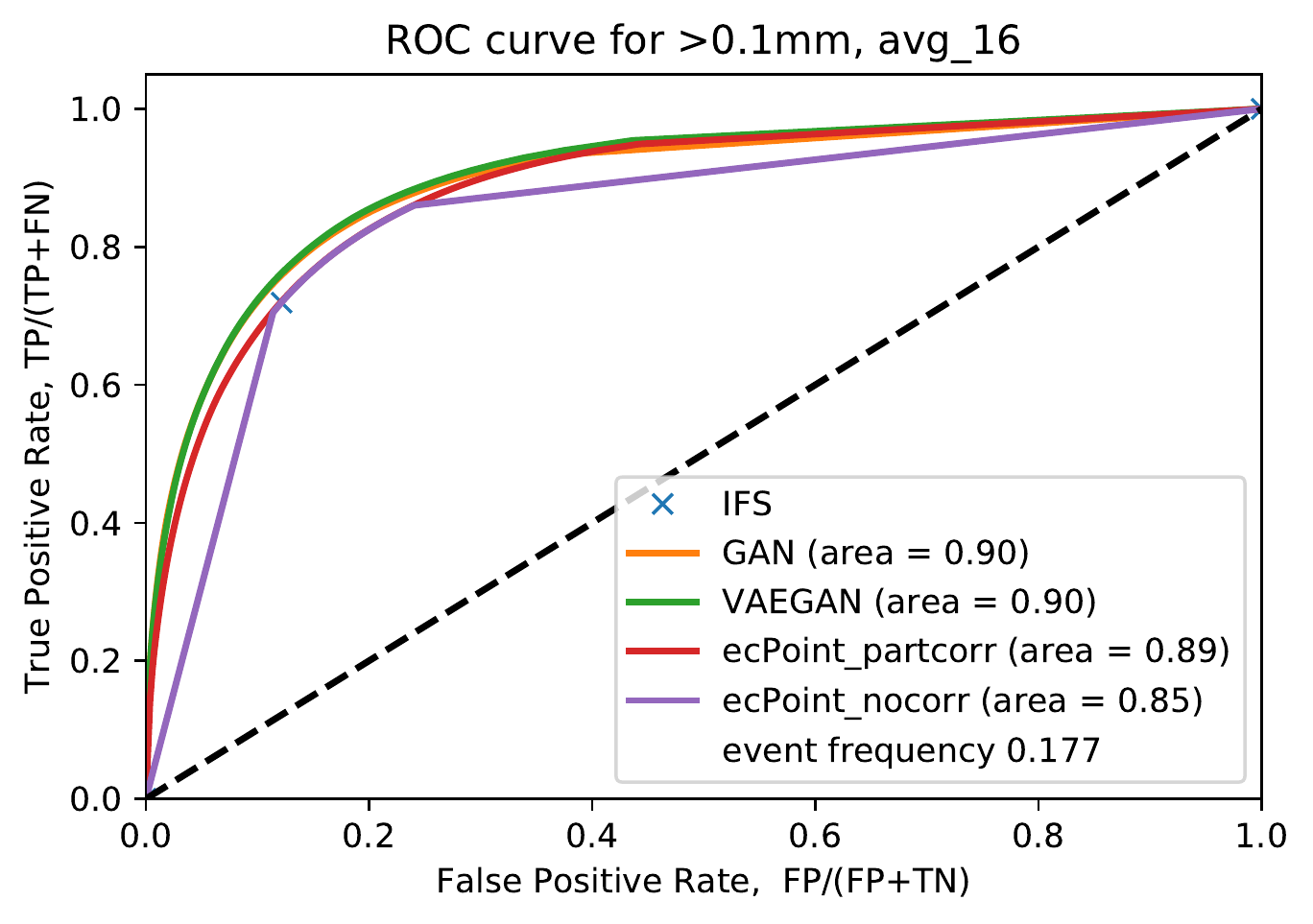}
\caption{ROC curves for the GAN, VAE-GAN and ecPoint models for a 0.1 mm/hr precipitation threshold, average pooling (16$\times$16 pixels)}
\label{fig:ROC-0.1-avg}
\end{figure}

\begin{figure}
\centering
\noindent\includegraphics[width=0.75\textwidth]{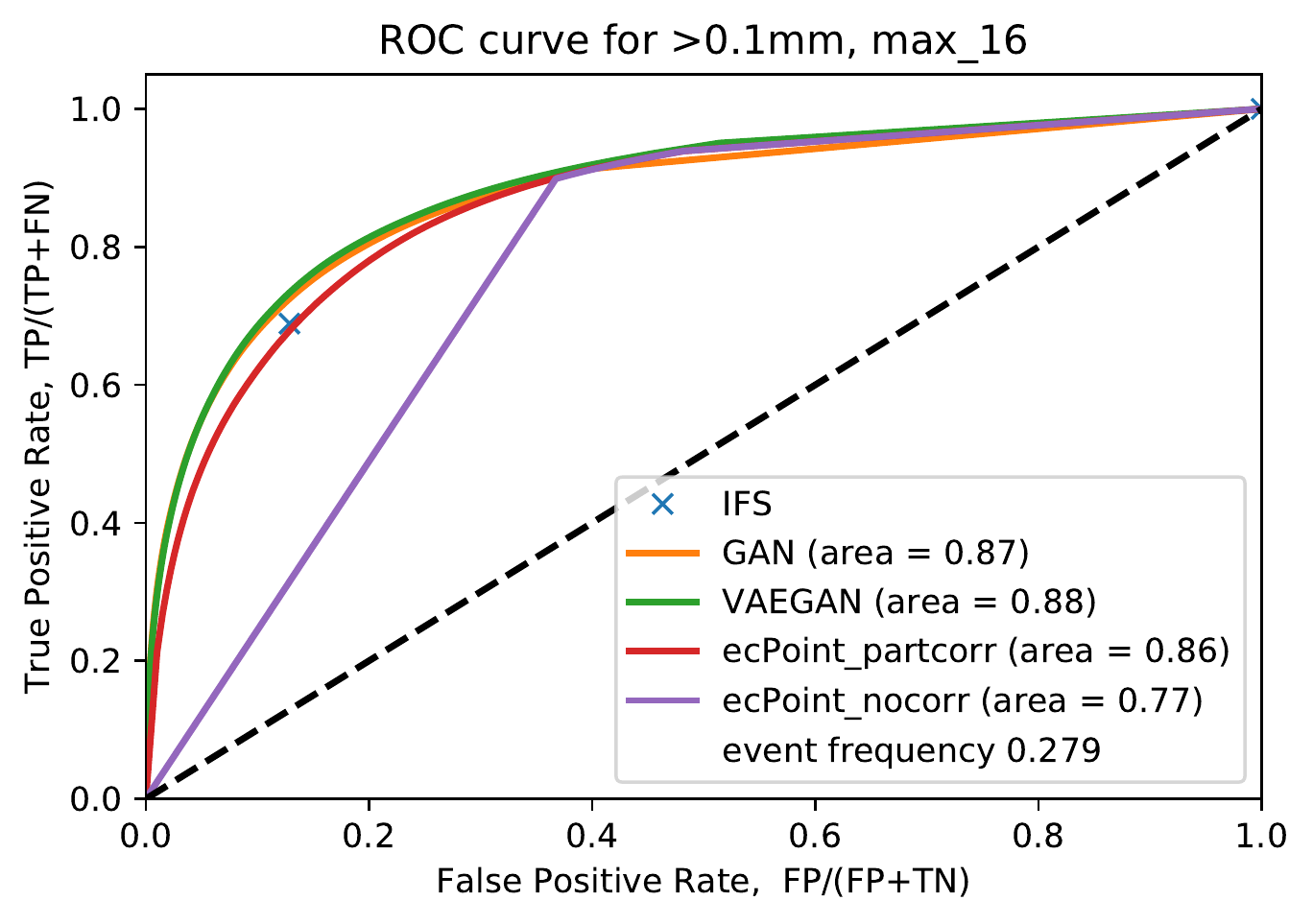}
\caption{ROC curves for the GAN, VAE-GAN and ecPoint models for a 0.1 mm/hr precipitation threshold, max pooling (16$\times$16 pixels)}
\label{fig:ROC-0.1-max}
\end{figure}

\begin{figure}
\centering
\noindent\includegraphics[width=0.75\textwidth]{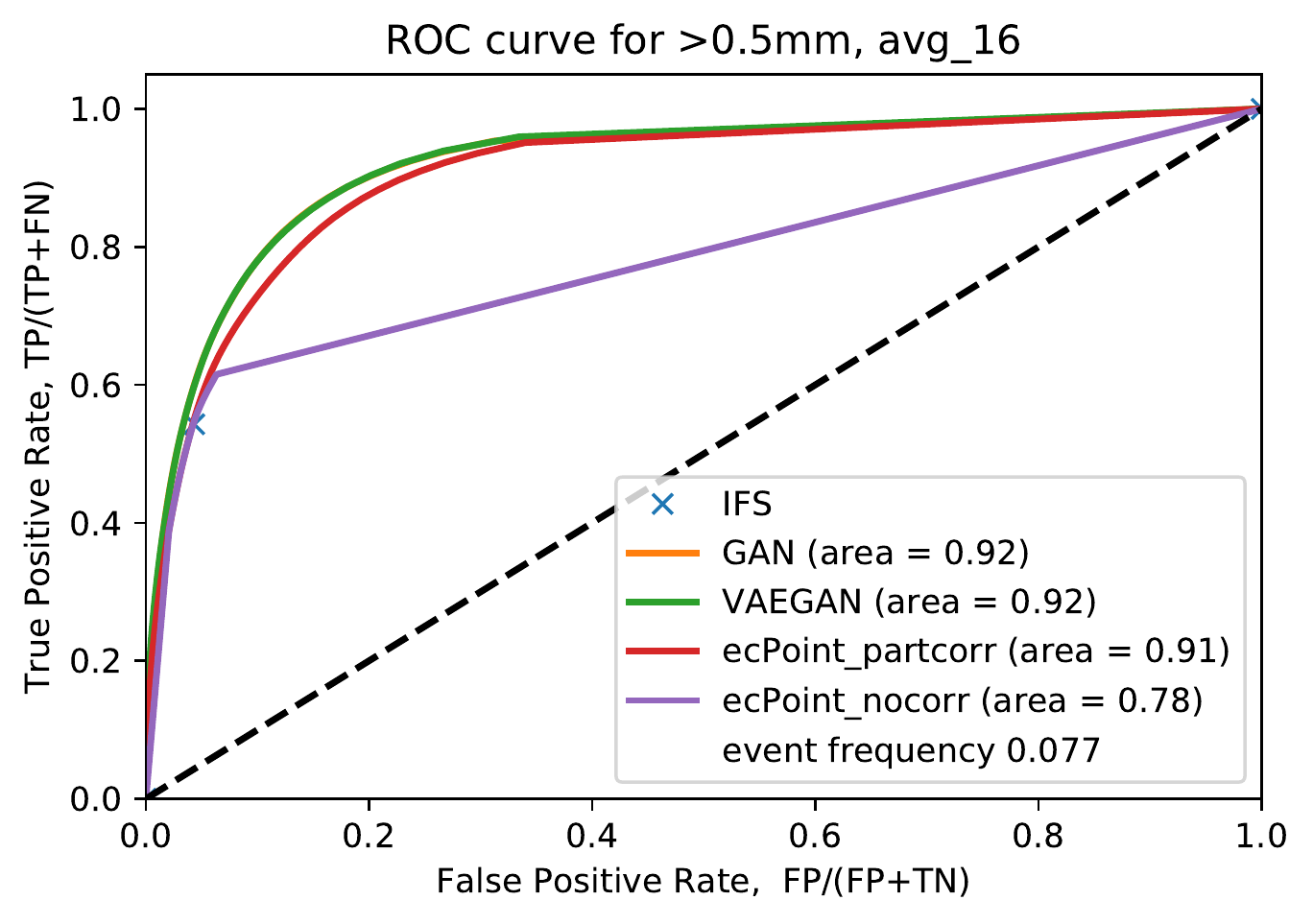}
\caption{ROC curves for the GAN, VAE-GAN and ecPoint models for a 0.5 mm/hr precipitation threshold, average pooling (16$\times$16 pixels)}
\label{fig:ROC-0.5-avg}
\end{figure}

\begin{figure}
\centering
\noindent\includegraphics[width=0.75\textwidth]{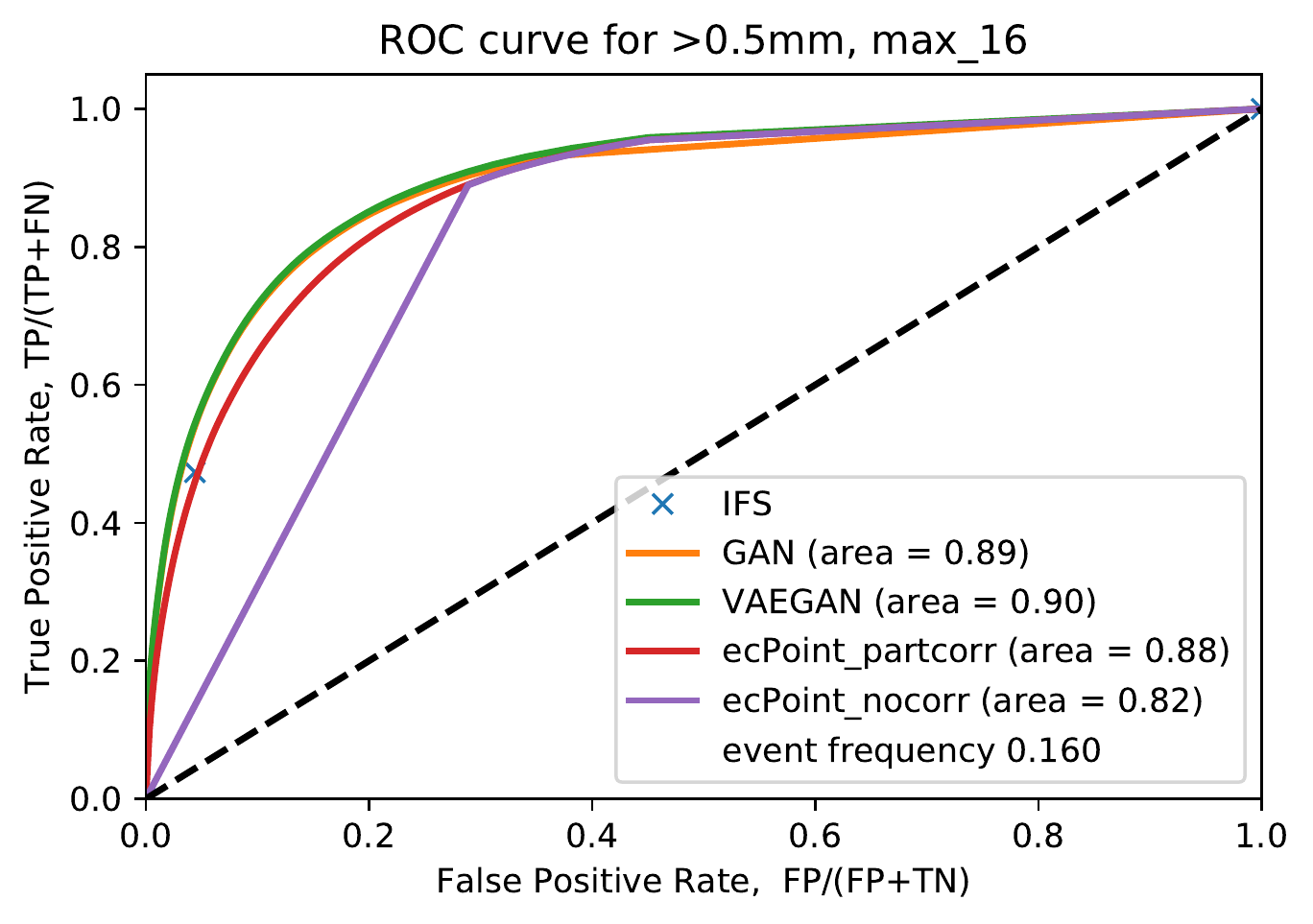}
\caption{ROC curves for the GAN, VAE-GAN and ecPoint models for a 0.5 mm/hr precipitation threshold, max pooling (16$\times$16 pixels)}
\label{fig:ROC-0.5-max}
\end{figure}

\begin{figure}
\centering
\noindent\includegraphics[width=0.75\textwidth]{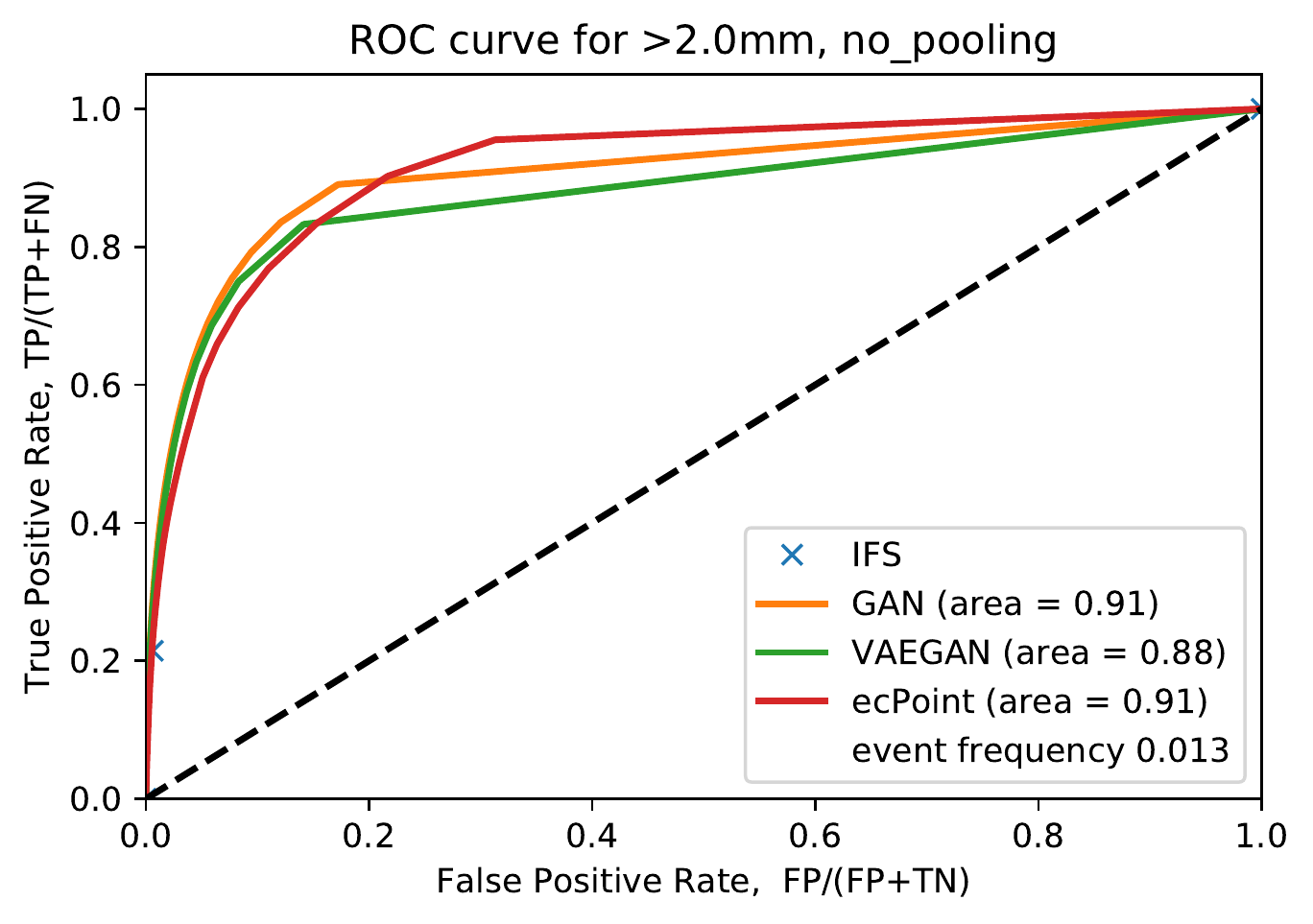}
\caption{ROC curves for the GAN, VAE-GAN and ecPoint models for a 2.0 mm/hr precipitation threshold, pixel-wise}
\label{fig:ROC-2.0-none}
\end{figure}

\begin{figure}
\centering
\noindent\includegraphics[width=0.75\textwidth]{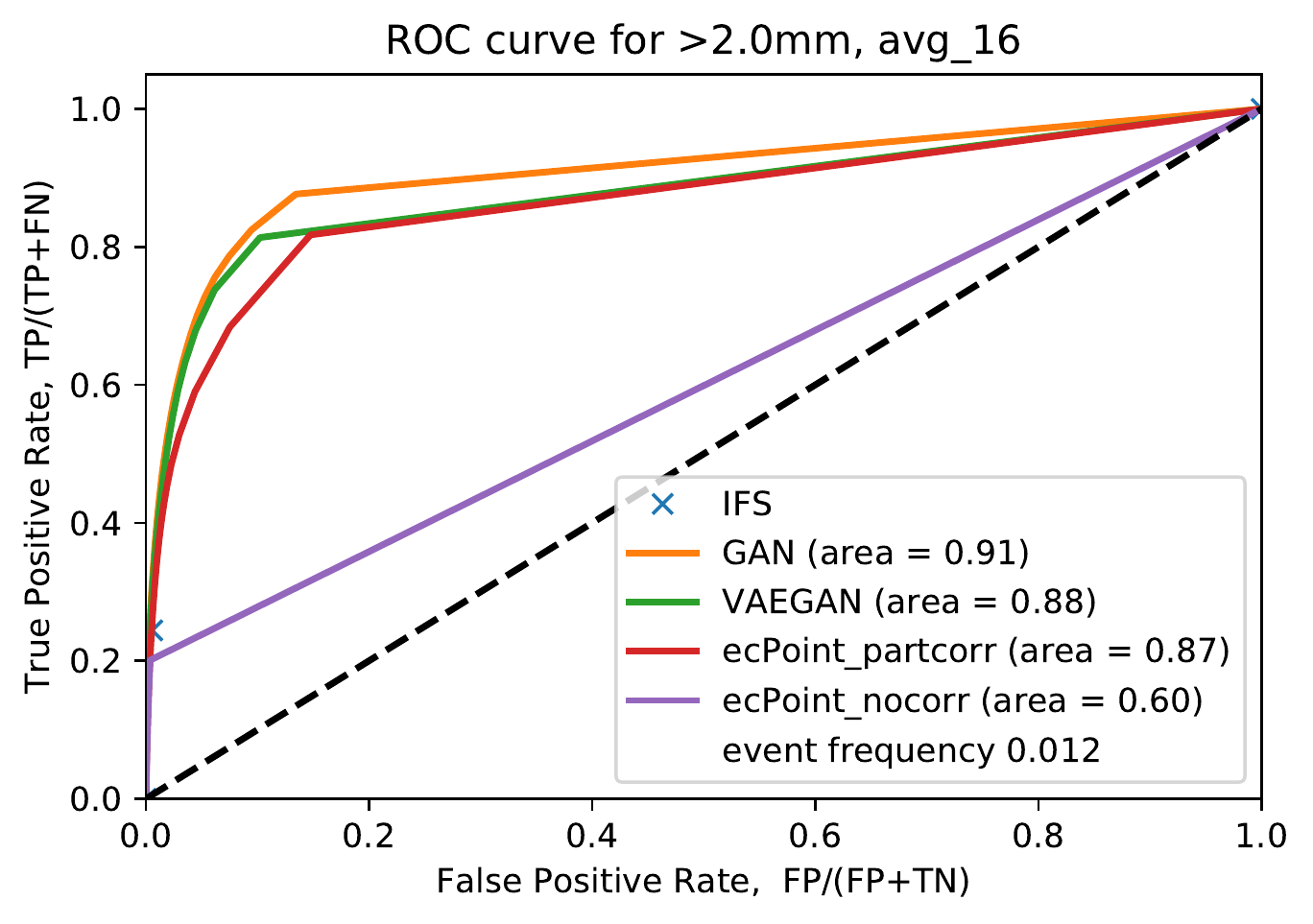}
\caption{ROC curves for the GAN, VAE-GAN and ecPoint models for a 2.0 mm/hr precipitation threshold, average pooling (16$\times$16 pixels)}
\label{fig:ROC-2.0-avg}
\end{figure}

\begin{figure}
\centering
\noindent\includegraphics[width=0.75\textwidth]{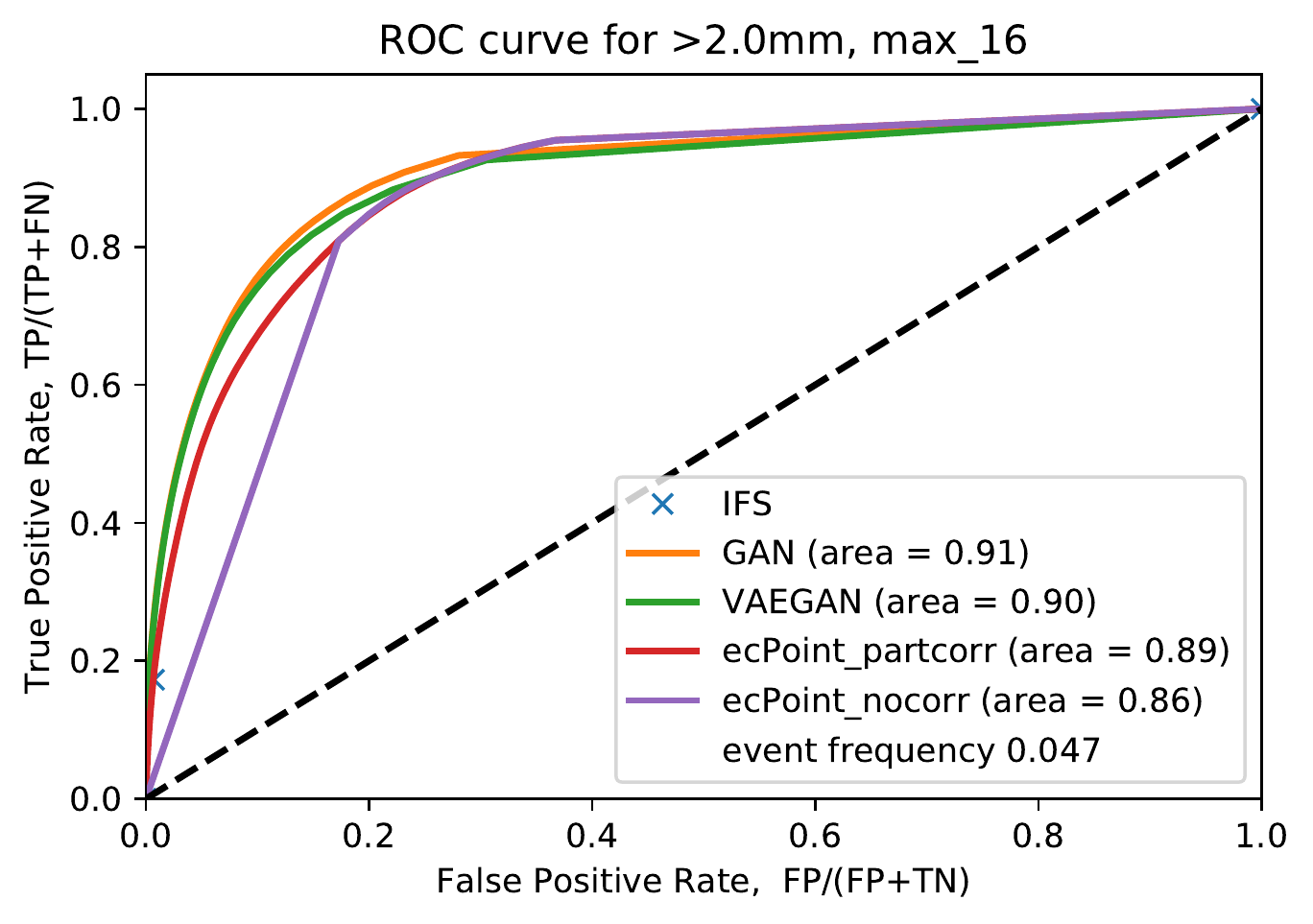}
\caption{ROC curves for the GAN, VAE-GAN and ecPoint models for a 2.0 mm/hr precipitation threshold, max pooling (16$\times$16 pixels)}
\label{fig:ROC-2.0-max}
\end{figure}

\begin{figure}
\centering
\noindent\includegraphics[width=0.75\textwidth]{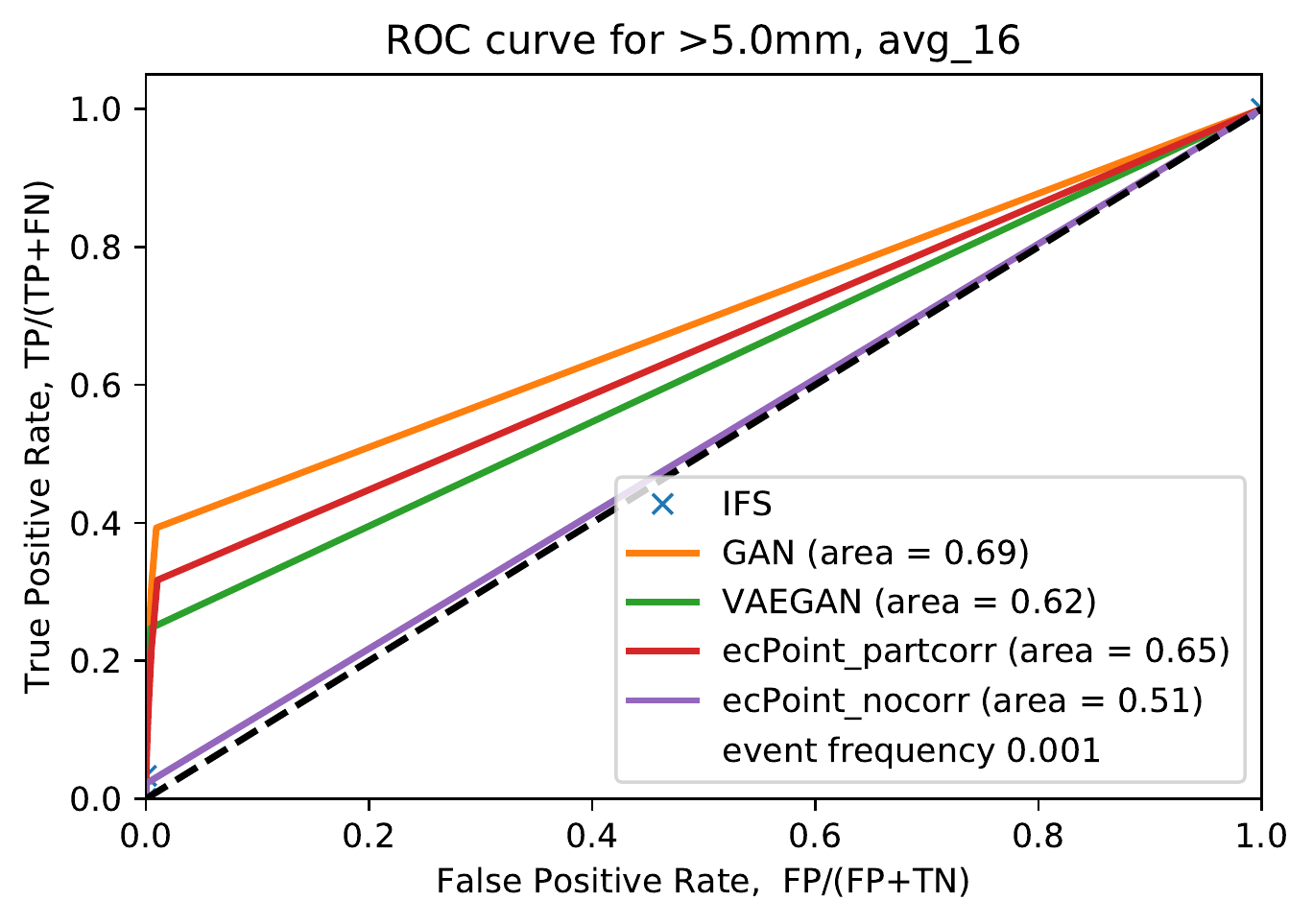}
\caption{ROC curves for the GAN, VAE-GAN and ecPoint models for a 5.0 mm/hr precipitation threshold, average pooling (16$\times$16 pixels)}
\label{fig:ROC-5.0-avg}
\end{figure}

\begin{figure}
\centering
\noindent\includegraphics[width=0.75\textwidth]{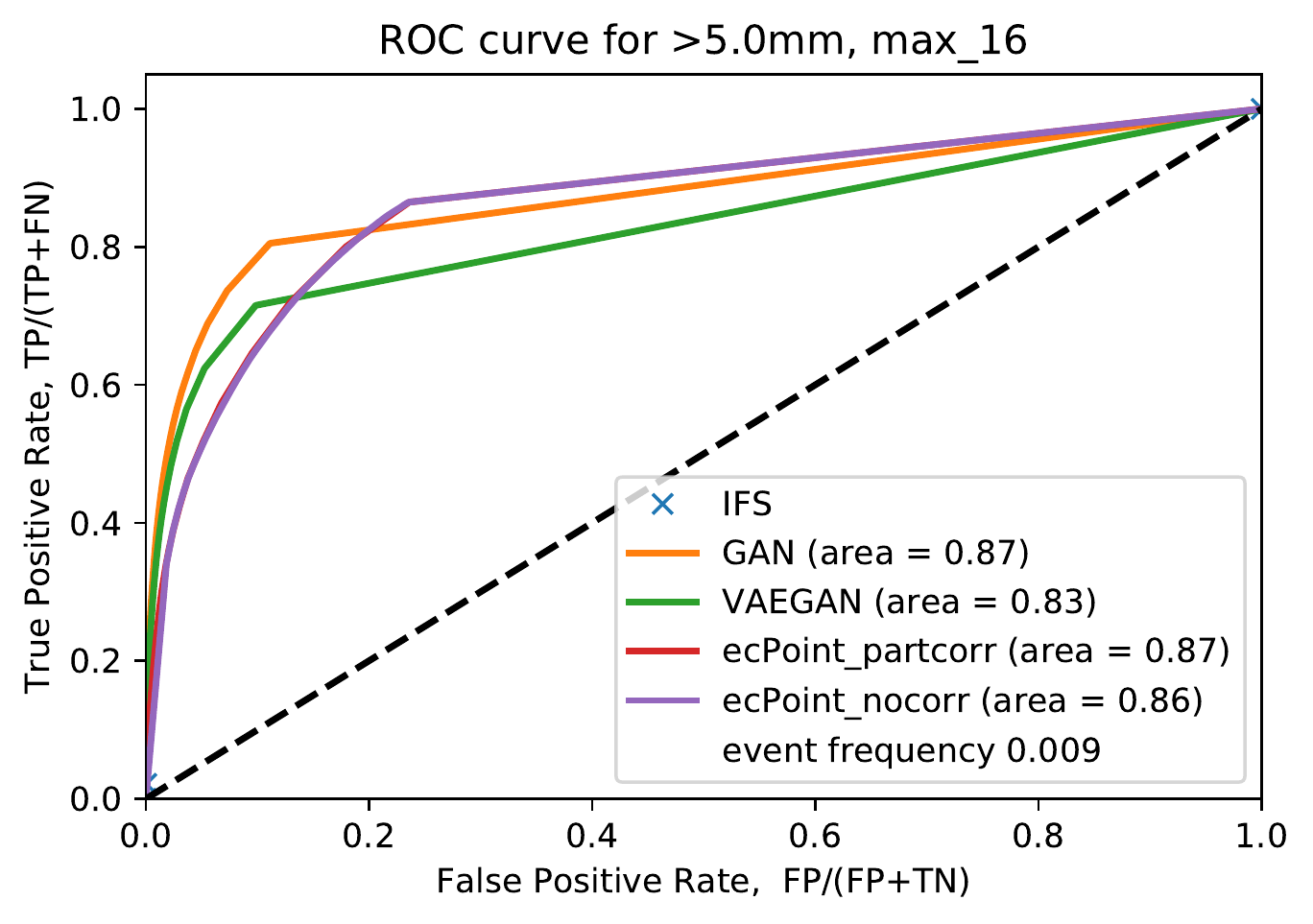}
\caption{ROC curves for the GAN, VAE-GAN and ecPoint models for a 5.0 mm/hr precipitation threshold, max pooling (16$\times$16 pixels)}
\label{fig:ROC-5.0-max}
\end{figure}

\begin{figure}
\centering
\noindent\includegraphics[width=0.7\textwidth]{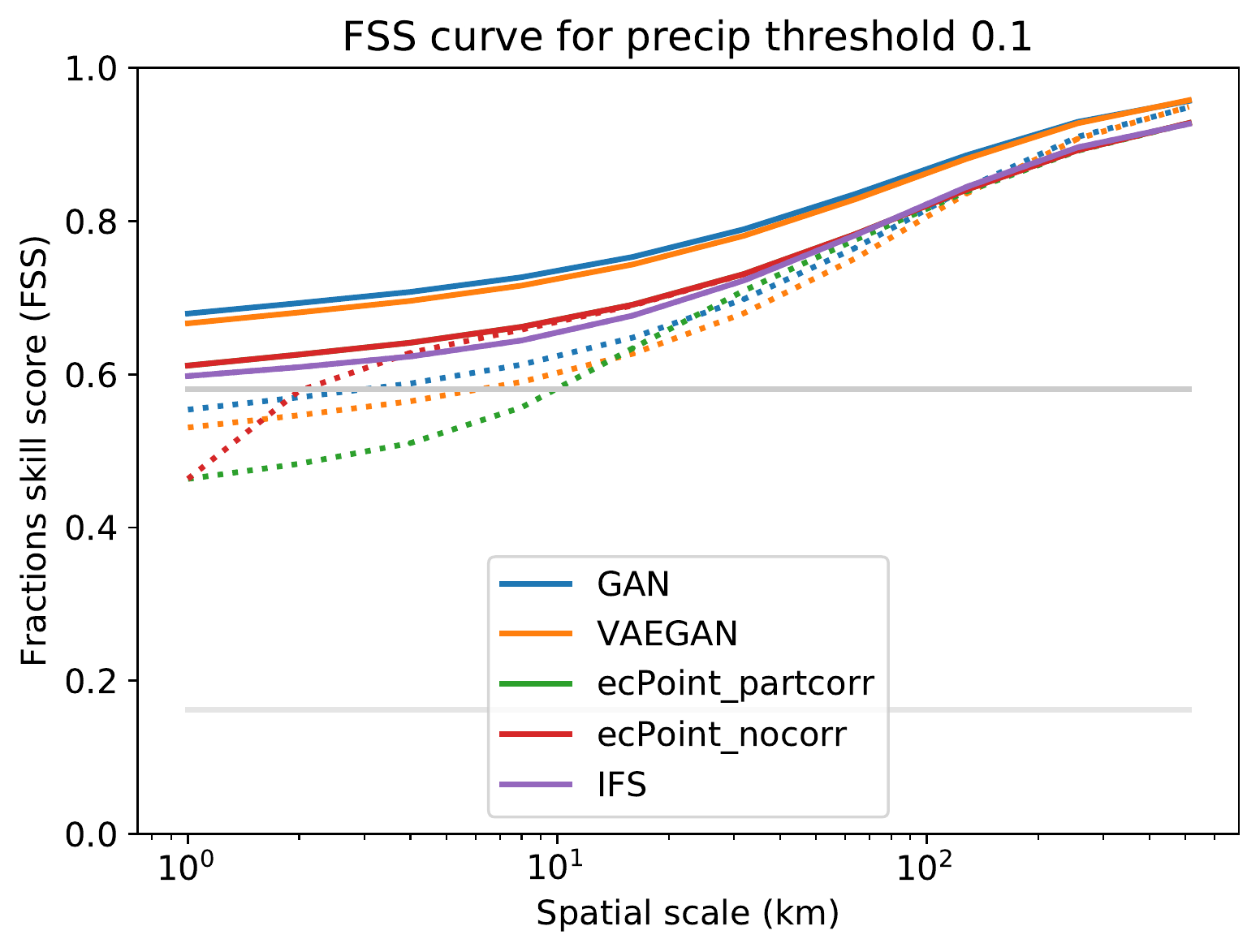}
\caption{FSS curves for the GAN, VAE-GAN and ecPoint models, 0.1 mm/hr precipitation threshold}
\label{fig:FSS-0.1}
\end{figure}

\begin{figure}
\centering
\noindent\includegraphics[width=0.7\textwidth]{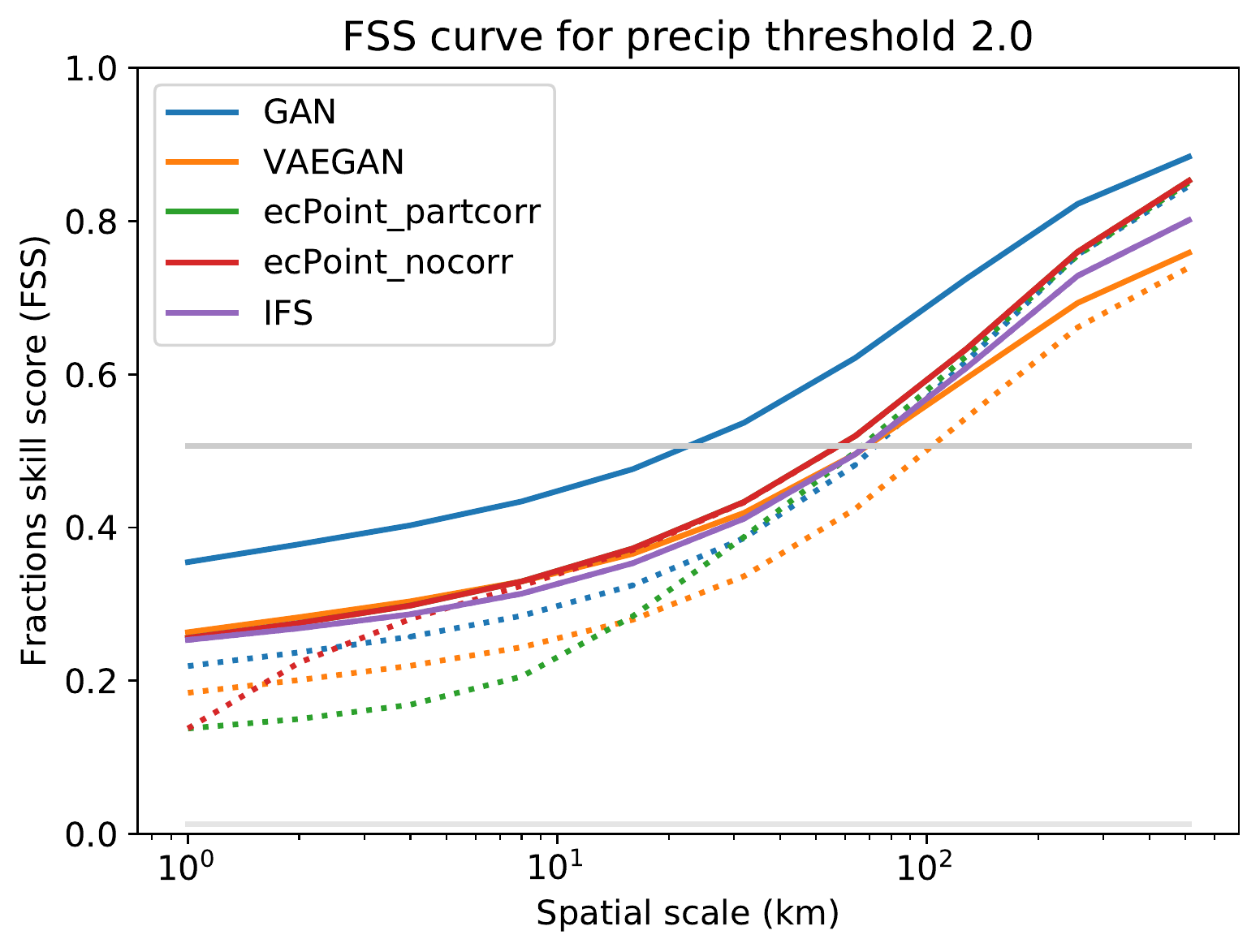}
\caption{FSS curves for the GAN, VAE-GAN and ecPoint models, 2.0 mm/hr precipitation threshold}
\label{fig:FSS-2.0}
\end{figure}

\begin{figure}
\centering
\noindent\includegraphics[width=0.75\textwidth]{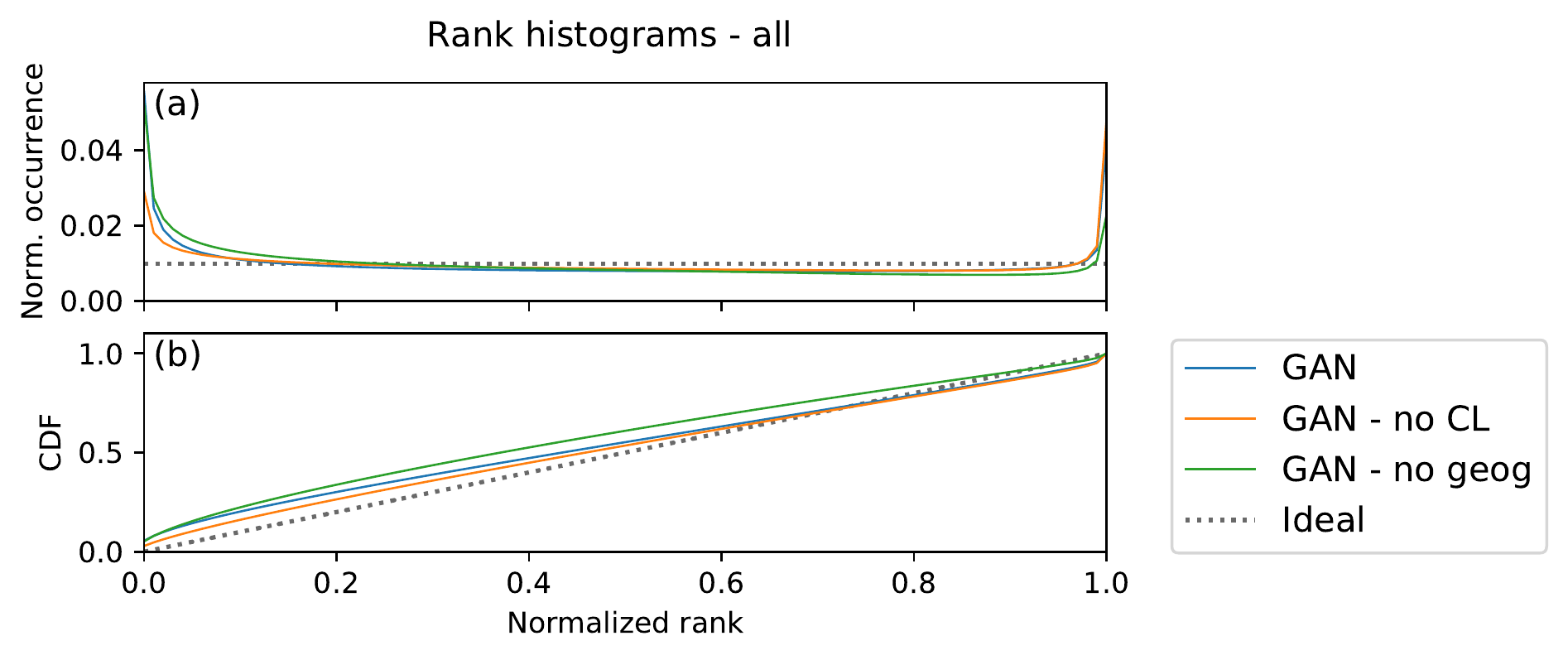}
    \caption{Calibration plot for various ablated GAN models: (a) shows the frequency of per-pixel normalised ranks for the trained models evaluated on the hold-out dataset (2020). The dotted grey line shows the ideal distribution for comparison. (b) shows the same as panel a, except displaying the CDFs of the distributions.}
    \label{fig:ranks-ablation}
\end{figure}

\begin{figure}
\centering
\noindent\includegraphics[width=0.75\textwidth]{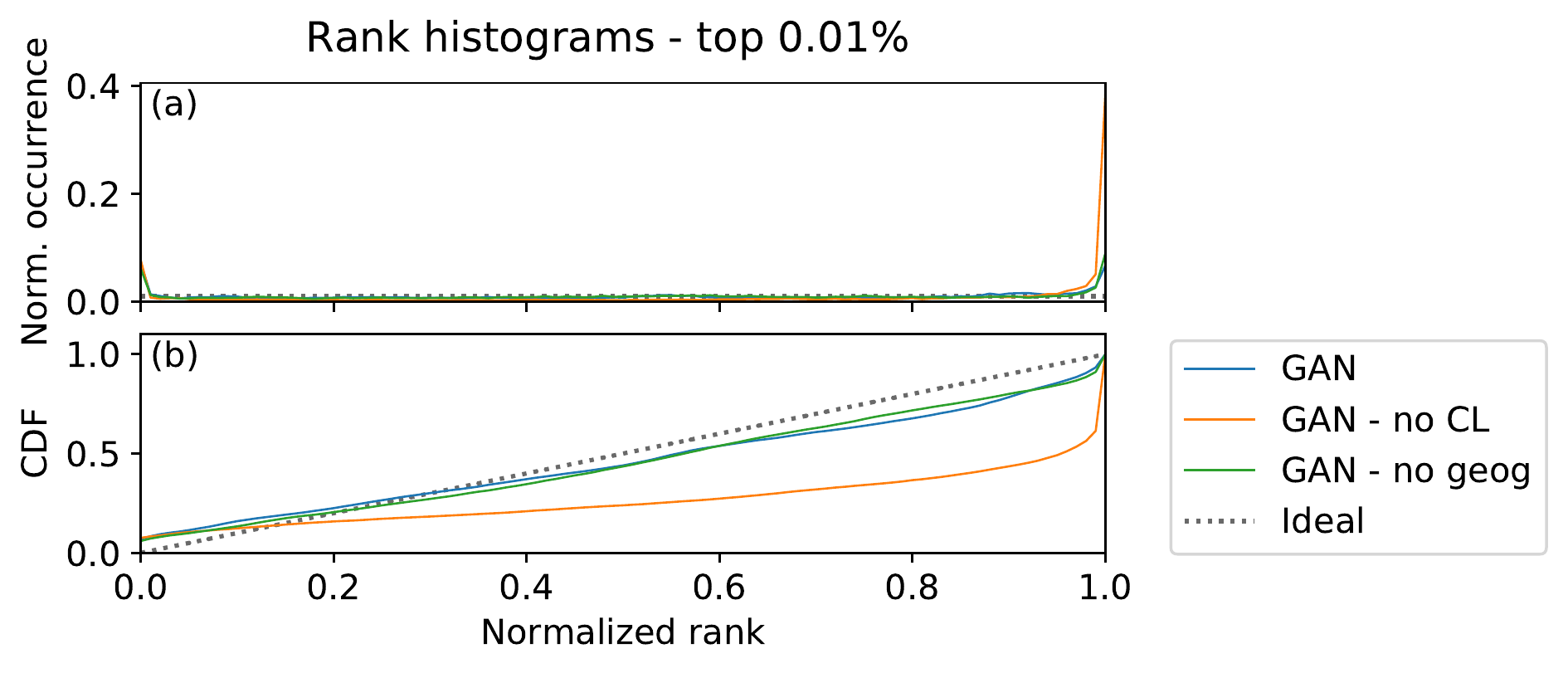}
    \caption{Thresholded calibration plot for various ablated GAN models: (a) shows the frequency of per-pixel normalised ranks over the 0.01\% threshold for the trained models evaluated on the hold-out dataset (2020). The dotted grey line shows the ideal distribution for comparison. (b) shows the same as panel a, except displaying the CDFs of the distributions.}
    \label{fig:ranks-ablation-thresh}
\end{figure}

\begin{figure}
\centering
\noindent\includegraphics[width=0.85\textwidth]{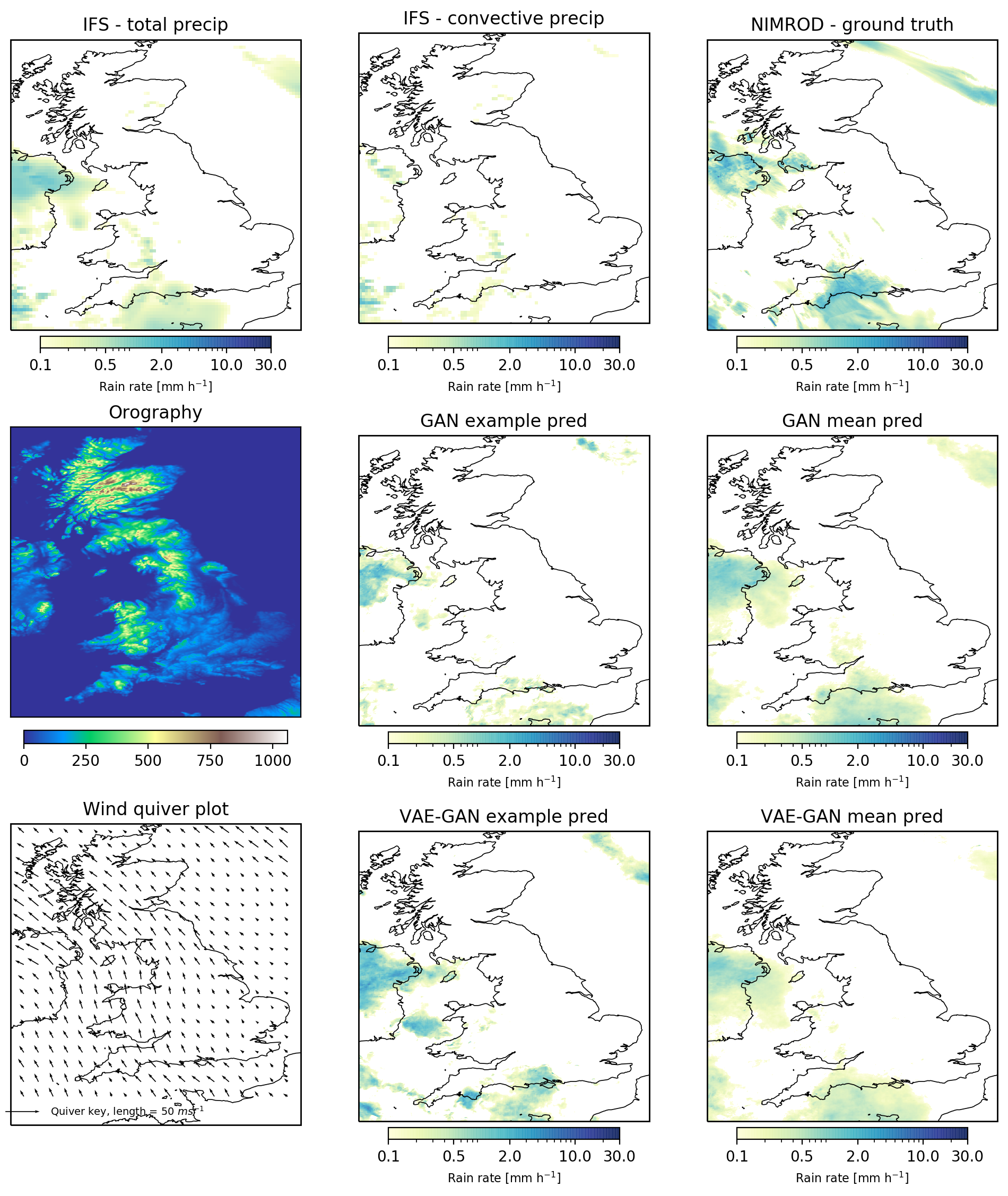}
\caption{Model prediction examples and means including input fields of total precipitation, wind speed and direction, and orography, 11:00-12:00 UTC, 05 April 2019}
\label{fig:pred-example-1}
\end{figure}

\begin{figure}
\centering
\noindent\includegraphics[width=0.85\textwidth]{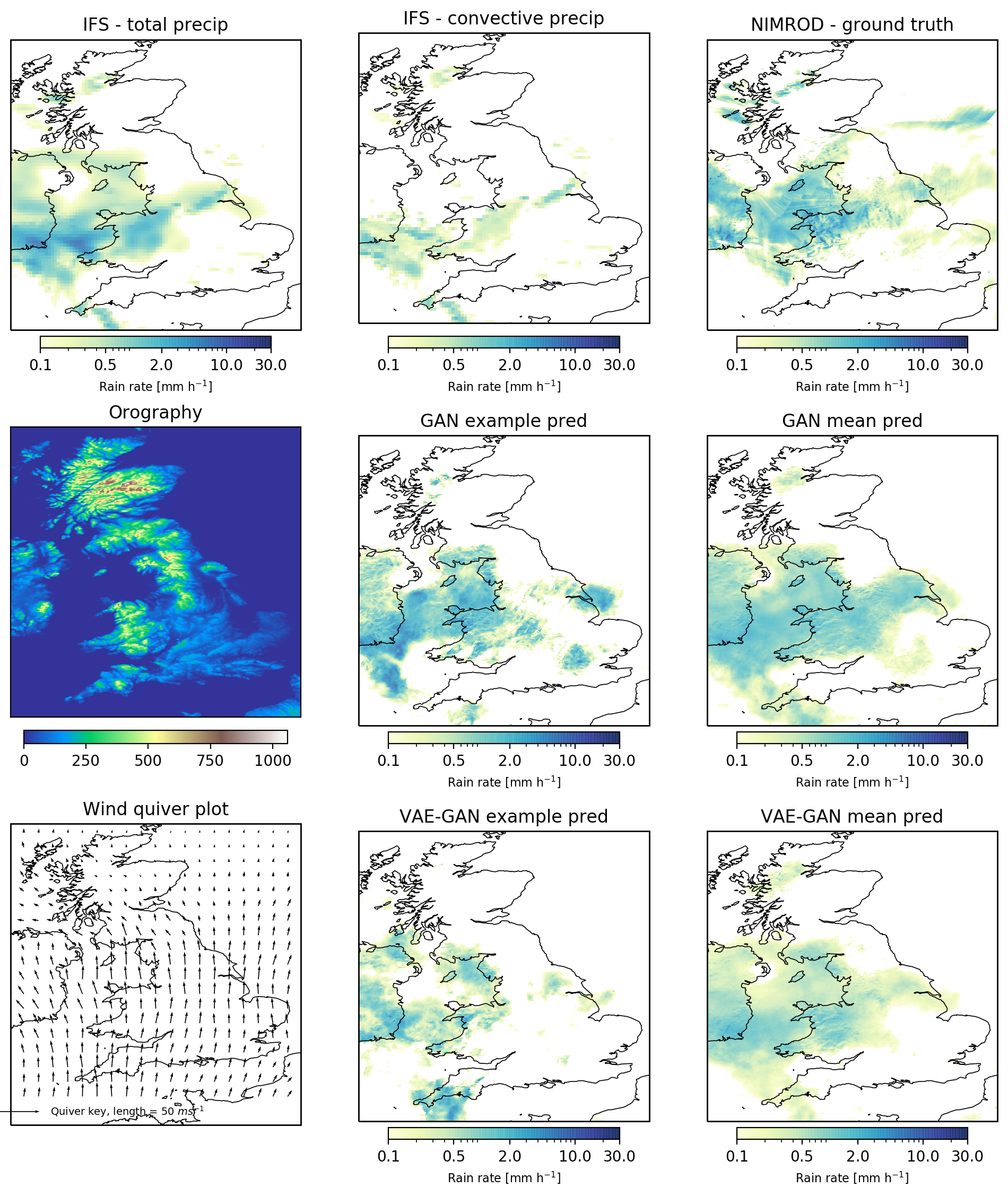}
\caption{Model prediction examples and means including input fields of total precipitation, wind speed and direction, and orography, 18:00-19:00 UTC, 23 June 2019}
\label{fig:pred-example-2}
\end{figure}

\begin{figure}
\centering
\noindent\includegraphics[width=0.85\textwidth]{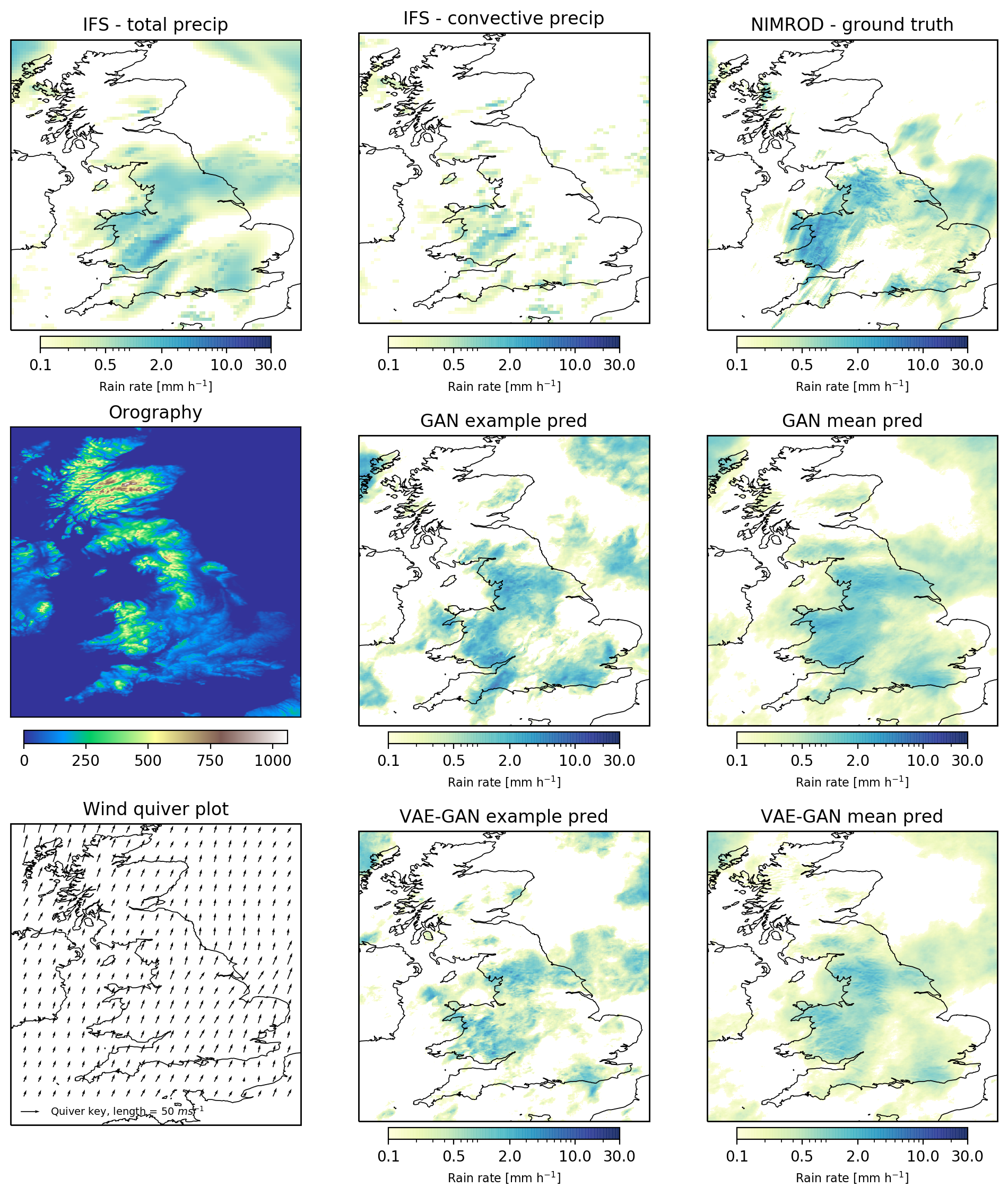}
\caption{Model prediction examples and means including input fields of total precipitation, wind speed and direction, and orography, 04:00-05:00 UTC, 19 Dec 2019}
\label{fig:pred-example-3}
\end{figure}

\begin{figure}
\centering
\noindent\includegraphics[width=0.85\textwidth]{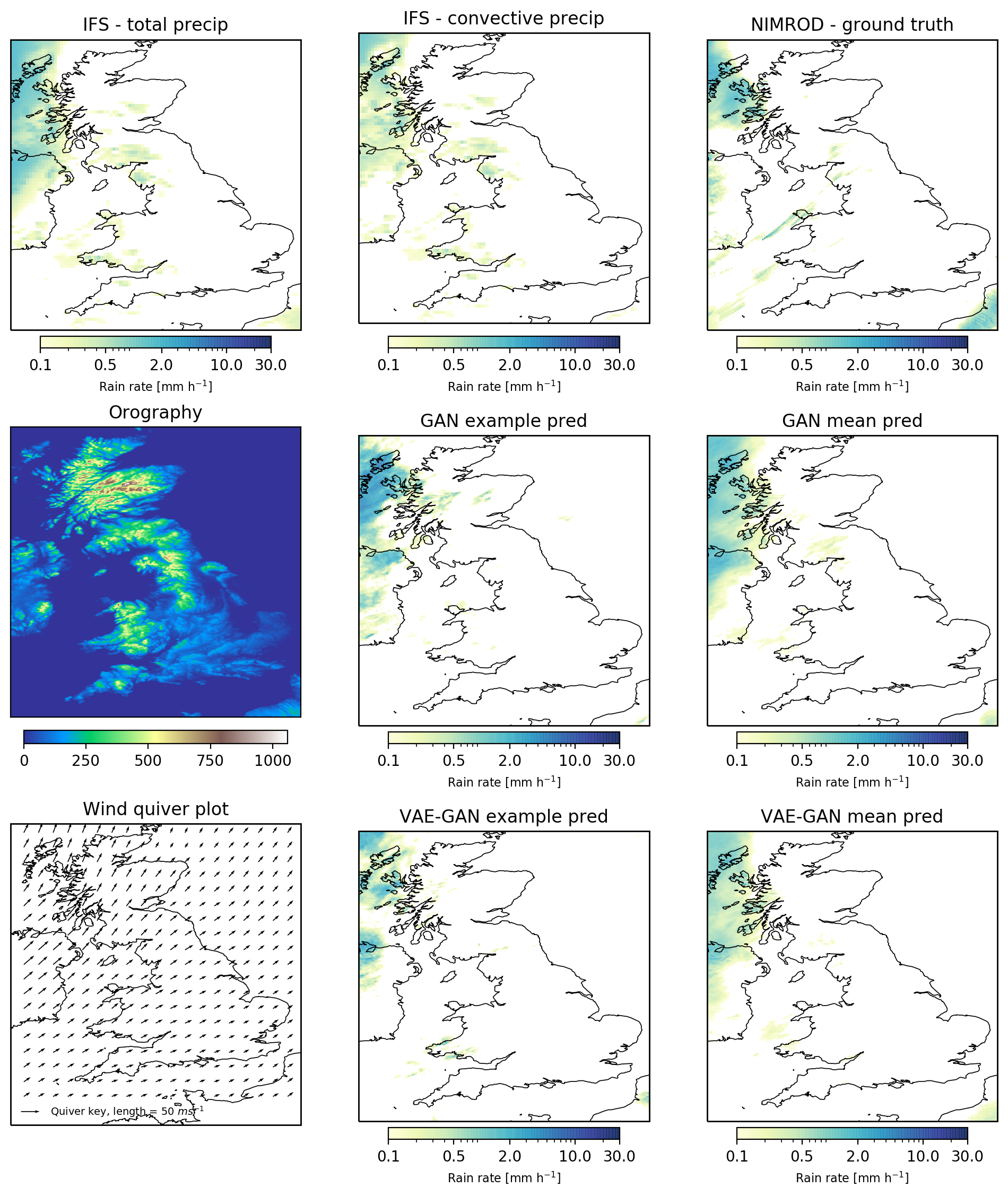}
\caption{Model prediction examples and means including input fields of total precipitation, wind speed and direction, and orography, 13:00-14:00 UTC, 08 Feb 2020}
\label{fig:pred-example-4}
\end{figure}

\begin{enumerate}
\item Introduction
\item Description of case studies
\item Figures~\ref{fig:ranks-GAN-lead-time} to~\ref{fig:pred-example-4}
\end{enumerate}

\noindent\section{\textbf{Introduction}}


{\noindent\subsection{Rank histogram plots for increasing lead times}

Figures~\ref{fig:ranks-GAN-lead-time} to \ref{fig:ranks-ecpoint-lead-time-thresh} show rank histogram plots for the GAN, VAE-GAN and ecPoint models for increasing lead time.

These plots are assessed on all available 2020 00Z forecasts, around 350 events in total. An ensemble size of 100 is used in each case. When evaluated on all events, all 3 methods (GAN, VAE-GAN, and ecPoint approach baseline) show only small differences between 24h, 48h and 72h lead-time.  However, all three methods show very slight worsening at increasing lead times. We would like to note that all 3 were trained on 7--17hr lead time data. When evaluated on the top 0.01\% of IFS predictions, all methods perform noticeably worse at 48h and 72h lead times than at 24h lead time. Our neural network approaches are under-dispersive, whereas the ecPoint approach is over-dispersive.}

\noindent\subsection{Precision-Recall curves}

Precision-recall curves (PRC) for the GAN, VAE-GAN and ecPoint part-correlation and no-correlation models are plotted here. These were generated by evaluating over 256 different images, with an ensemble size of 100.

Precision is the number of true positives divided by the sum of the true positives and false positives. Recall is calculated as the ratio of the number of true positives to the sum of the true positives plus false negatives. Recall is the same as sensitivity. P-R curves are useful in cases where there is an imbalance in the observations between the two classes, such as in our dataset where there are many more light-rain events than heavy-rain events.

Whilst the baseline is fixed with ROC curves, the baseline of P-R curves is determined by the ratio of positives (P) and negatives (N) as $y = \frac{P}{P+N}$, indicated by the dashed line. A model with perfect skill is depicted as a point at (1,1). A skilful model is represented by a curve that tends towards $(1,1)$ above the flat line of no skill. The area under each line is given, as a proxy for overall performance. However, this should be interpreted with caution, particularly in cases where the straight-line portion between $(0, 1)$ or $(1, \frac{P}{P+N})$ and the initial/final data values contribute significantly to the area.

\noindent\subsection{Further RAPSD plot examples}

\begin{figure}
\centering
\noindent\includegraphics[width=0.75\textwidth]{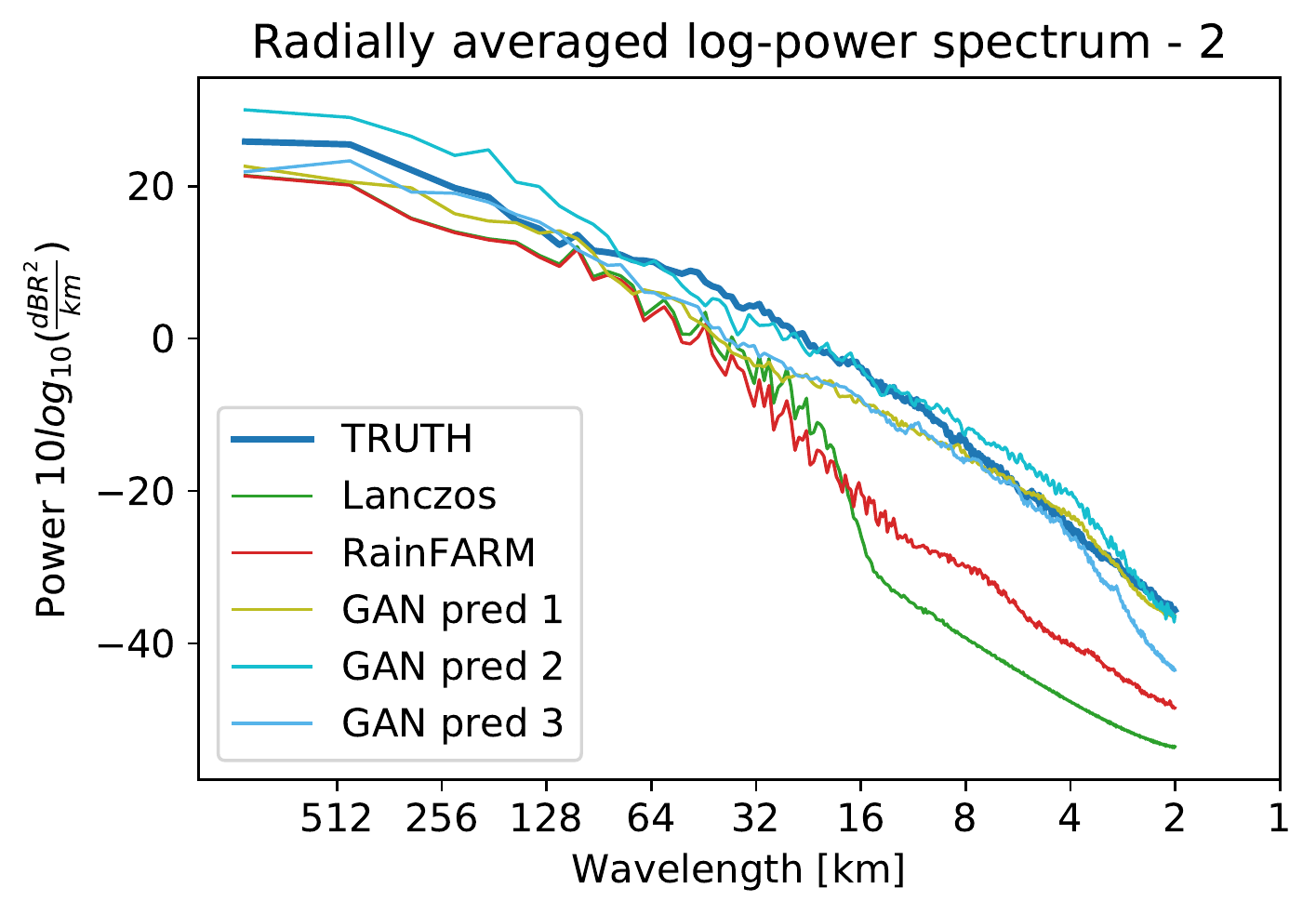}
\caption{Plot displaying the radially-averaged log power spectra of example image 2 comparing our GAN model to both Lanczos interpolation and the RainFARM method}
\label{fig:GAN-RAPSD-2}
\end{figure}

\begin{figure}
\centering
\noindent\includegraphics[width=0.75\textwidth]{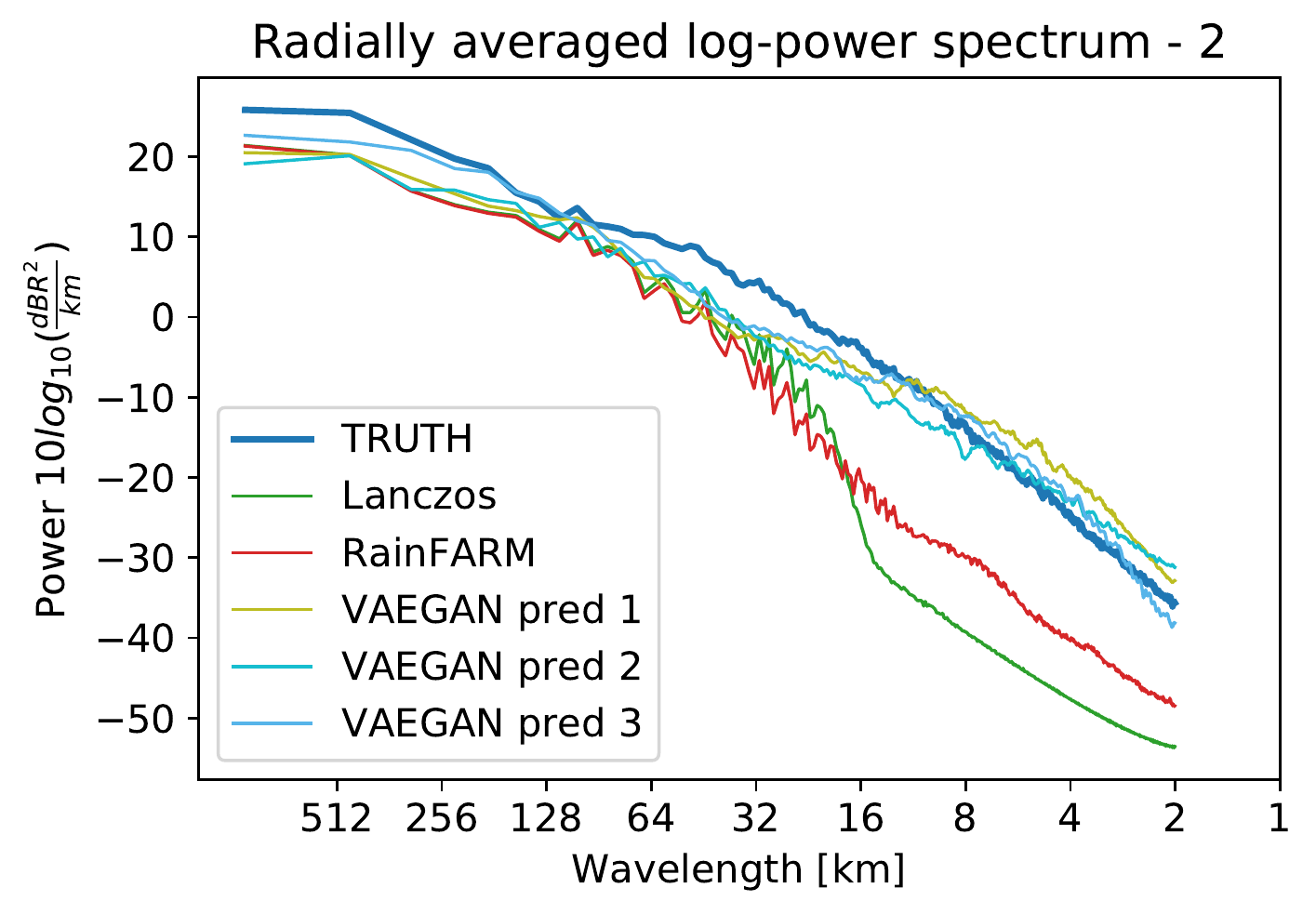}
\caption{Plot displaying the radially-averaged log power spectra of example image 2 comparing our VAE-GAN model to both Lanczos interpolation and the RainFARM method}
\label{fig:VAEGAN-RAPSD-2}
\end{figure}

\begin{figure}
\centering
\noindent\includegraphics[width=0.75\textwidth]{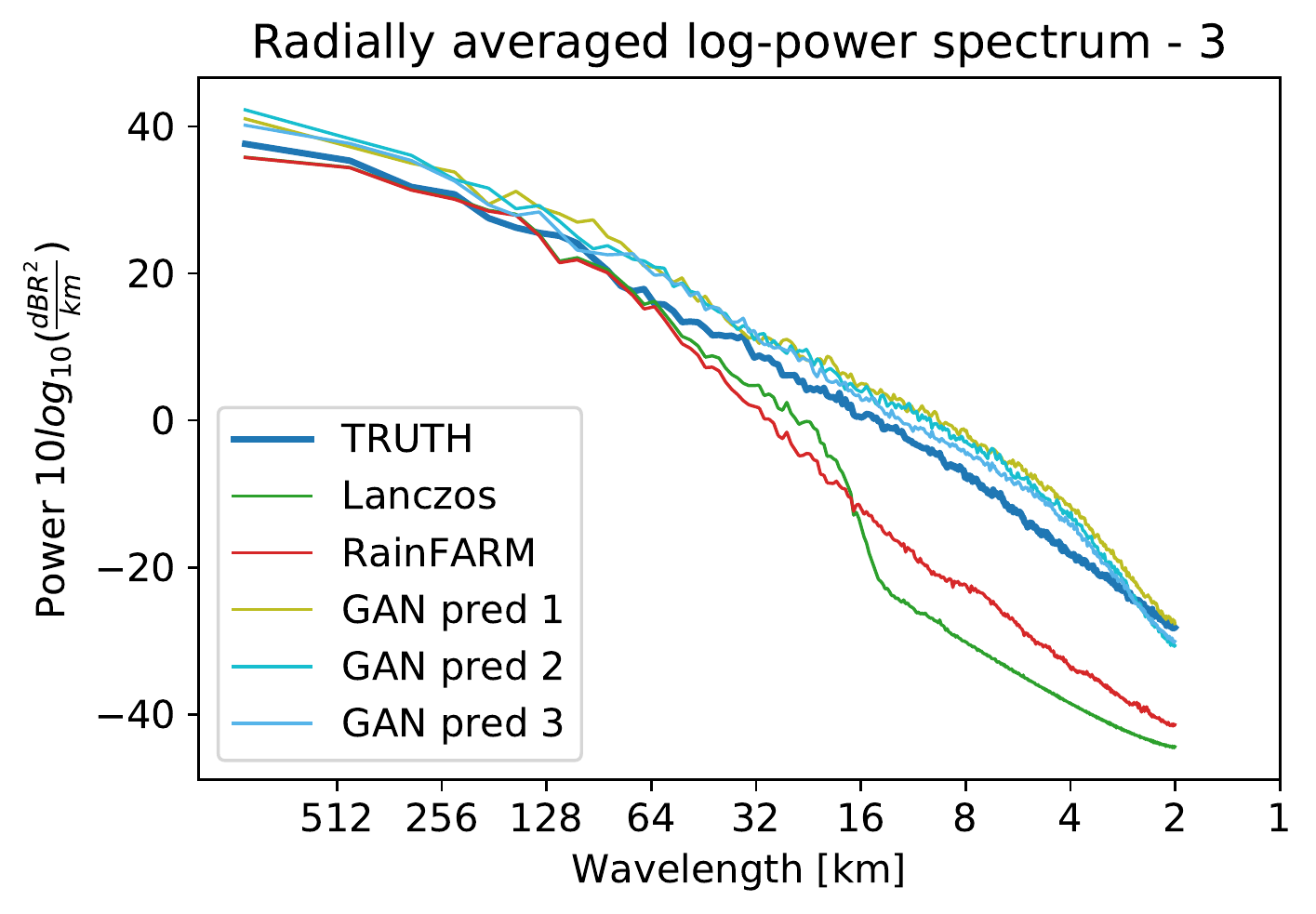}
\caption{Plot displaying the radially-averaged log power spectra of example image 3 comparing our GAN model to both Lanczos interpolation and the RainFARM method}
\label{fig:GAN-RAPSD-3}
\end{figure}

\begin{figure}
\centering
\noindent\includegraphics[width=0.75\textwidth]{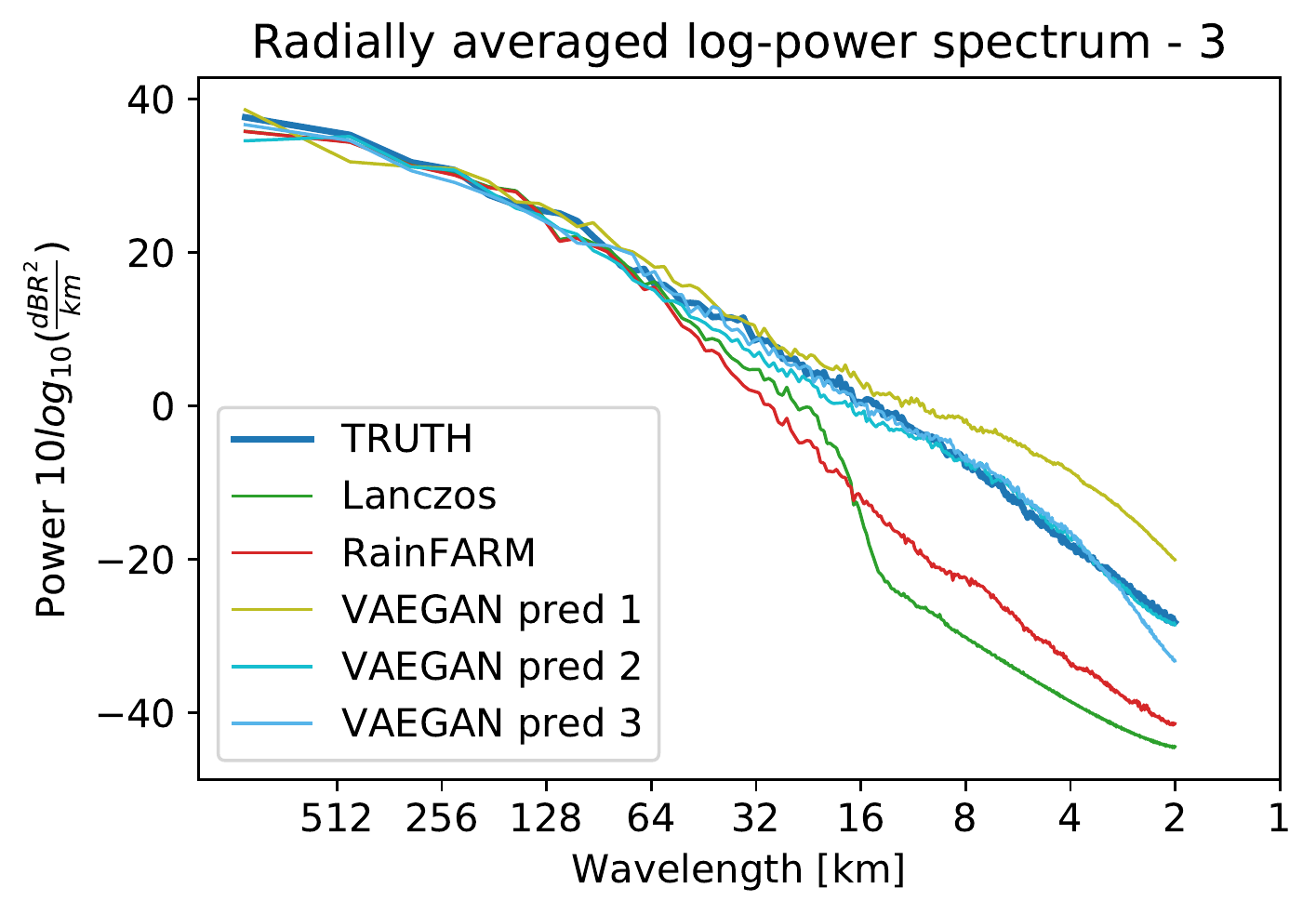}
\caption{Plot displaying the radially-averaged log power spectra of example image 3 comparing our VAE-GAN model to both Lanczos interpolation and the RainFARM method}
\label{fig:VAEGAN-RAPSD-3}
\end{figure}

Figures~\ref{fig:GAN-RAPSD-2} to \ref{fig:VAEGAN-RAPSD-3} show further RAPSD plot examples for the GAN and the VAE-GAN. These plots are included here to show that the RAPSD results are consistently good across all the example images. Figures~\ref{fig:GAN-RAPSD-2} and~\ref{fig:VAEGAN-RAPSD-2} show plots for the second standard example image case for the GAN and the VAE-GAN, respectively, and Figures~\ref{fig:GAN-RAPSD-3} and~\ref{fig:VAEGAN-RAPSD-3} show RAPSD plots for the third example image. The GAN plots also include the Lanczos and RainFARM model RAPSD for comparison. The RAPSD plots are generally fairly consistent: both the GAN and VAE-GAN models clearly outperform the Lanczos and RainFARM methods. These plots clearly demonstrate the added value of using the GAN and VAE-GAN models over interpolation of the IFS forecast (e.g. using Lanczos interpolation). It is also interesting to note the spread between individual model ensemble members in both the GAN and the VAE-GAN.

\noindent\subsection{Further ROC curve examples}

The ROC curves for the GAN and VAE-GAN models shown here were generated by evaluating over 256 different images, with a batch size of 1 (for memory reasons) and an ensemble size of 100. A full set of ROC curves are shown here for 0.1 mm/hr and 2 mm/hr in the appendix, in addition to the average pooling and max pooling plots for the 0.5 mm/hr and 5 mm/hr thresholds shown in the main body of the paper.

\noindent\subsection{Further FSS plot examples}
FSS curves for the GAN and VAE-GAN models are plotted in Figures~\ref{fig:FSS-0.1} and \ref{fig:FSS-2.0}. These were generated by evaluating over 256 different images, with an ensemble size of 100. FSS curves are shown here for the additional 0.1 mm/hr and 2 mm/hr thresholds.

\noindent\subsection{Rank histogram plots for ablation studies}
Figures~\ref{fig:ranks-ablation} and \ref{fig:ranks-ablation-thresh} show rank histograms for the ``no content-loss'' and ``no geographic fields'' ablation studies, compared with the original GAN, on all events and on the top 0.01\% of forecast predictions.

\noindent\subsection{Further model prediction examples}

These figures show the GAN and VAE-GAN model predictions for a further four different weather scenarios, on top of the four used in the main paper, and are included here to demonstrate the wide-ranging capabilities of the models.

\noindent\section{\textbf{Description of case studies used in the main paper}}

The first example is from 21:00-22:00 UTC of the 7\textsuperscript{th} June 2019, and shows a highly-structured pattern of rainfall, with a curved band of moderate precipitation across the centre of the image surrounded by lighter rainfall, with additional structure in the bottom right-hand corner.
All GAN and VAE-GAN model predictions produce a similar banded structure to the NIMROD ground truth image, with locally sharply varying structure within the rainfall band. The models also make bold predictions of extremely localised intense rainfall, whereas the IFS forecast fails to capture the peak intensities of the rainfall. There is fine-scale variation between the different predictions, but the large-scale structure remains consistent, suggesting a high level of confidence in this forecast.

The second example is from 18:00-19:00 UTC of the 1\textsuperscript{st} May 2019, and shows lighter, scattered rainfall across the country. The IFS forecast does not capture the full extent of the rainfall and under-predicts the intensity in places, whilst over-predicting the area of light rainfall over Northern Ireland. The GAN and VAE-GAN predictions show a much more realistic, finely detailed structure within the precipitation patches, whilst maintaining the same, broadly correct, large-scale structure. The models successfully remove the over-prediction of light rain over Northern Ireland. There is significant variation between the different predictions, corresponding to significant uncertainty in the scenario.

The third example is from 01:00-02:00 UTC of the 5\textsuperscript{th} August 2019, and shows two distinct bands of rain across the image, with the large-scale structure captured reasonably well by the IFS prediction. However, the western band of rain has a much more defined spatial structure than displayed in the IFS forecast, which predicts light rainfall over a greater area than in reality. The north-eastern band of rain has a fairly simple structure, but the intensity is significantly under-predicted by the IFS forecast. Firstly, and perhaps most importantly, the GAN model predictions capture the full intensity of the rainfall. The VAE-GAN corrects the intensity somewhat, but not enough. Secondly, unlike the IFS forecast, the GAN and VAE-GAN do not over-predict light rainfall for a large area for the western rainfall band. There is significant variation in the location of the precipitation due to the inherent uncertainty, but there is evidence the models are using the orographic information to produce sensible predictions, as the rainfall is consistently predicted over the Highlands, the Brecon Beacons, and the western Pennines.

The fourth example is from 04:00-05:00 UTC of the 27\textsuperscript{th} January 2019, and shows scattered light rainfall clustered into two groups, one in the north and one in the south. The NIMROD data shows lots of fine-scale structure and some medium-intensity rainfall over the Highlands. The IFS forecast only predicts light rainfall, and whilst the large-scale structure is broadly correct, there is an area of rainfall forecast to the north-east of the UK over the North Sea that is not present in the NIMROD data. Two of the three GAN model predictions also show this area of rainfall forecast, with the third showing some rainfall forecast in the same area.  The VAE-GAN predictions seem to largely eliminate this spurious prediction.  However, overall, the GAN and VAE-GAN predictions closely mirror the NIMROD data, showing patchy, light rainfall across much of the country with a more realistic-looking structure than the blurrier IFS forecast. The patch of rain over the south of the UK is well-calibrated with location and intensity matching the NIMROD data closely for all three example GAN model predictions, while the VAE-GAN predictions perhaps under-predict this slightly. There is again evidence of the orography informing the model, with a consistent concentration of rainfall centred on the Highlands. The NIMROD data shows bands of rain aligned north-west to south-east with the wind. We are pleased to see similar features existing in the GAN predictions.

\end{article}
\clearpage

